\documentclass[twocolumn,showpacs,amsmath,amssymb,floatfix]{revtex4}

\usepackage{float}
\usepackage{epsfig}
\usepackage{color}
\usepackage{times}
\usepackage{pst-all}
\usepackage{rotate}

\begin{document}

\title{Effect of extreme data loss on long-range correlated and
  anti-correlated signals quantified by detrended fluctuation
  analysis}

\author{Qianli~D.Y.~Ma$^{1, 2}$, Ronny~P.~Bartsch$^{1}$, Pedro~Bernaola-Galv\'an$^3$, Mitsuru~Yoneyama$^4$, and Plamen~Ch.~Ivanov$^{1, 3, 5}\footnote{Corresponding author: plamen@buphy.bu.edu}$}

\affiliation{$^1$ Harvard Medical School and Division of Sleep Medicine, Brigham and Women's Hospital, Boston, MA 02115, USA\\
$^2$ College of Geography and Biological Information, Nanjing University of Posts and Telecommunications, Nanjing 210003, China\\
$^3$ Departamento de F\'isica Aplicada II, Universidad de M\'alaga, 29071 M\'alaga, Spain\\
$^4$ Mitsubishi Chemical Group, Science and Technology Research Center Inc., Yokohama 227-8502, Japan\\
$^5$ Center for Polymer Studies and Department of Physics, Boston University, Boston, MA 02215, USA}

\begin{abstract}

  Detrended fluctuation analysis (DFA) is an improved method of
  classical fluctuation analysis for nonstationary signals where
  embedded polynomial trends mask the intrinsic correlation properties
  of the fluctuations. To better identify the intrinsic correlation
  properties of real-world signals where a large amount of data is
  missing or removed due to artifacts, we investigate how extreme data
  loss affects the scaling behavior of long-range power-law correlated
  and anti-correlated signals.  We introduce a new segmentation
  approach to generate surrogate signals by randomly removing data
  segments from stationary signals with different types of long-range
  correlations. The surrogate signals we generate are characterized by
  four parameters: (i) the DFA scaling exponent $\alpha$ of the
  original correlated signal $u(i)$, (ii) the percentage $p$ of the
  data removed from $u(i)$, (iii) the average length $\mu$ of the
  removed (or remaining) data segments, and (iv) the functional form
  $P(l)$ of the distribution of the length $l$ of the removed (or
  remaining) data segments. We find that the global scaling exponent
  of positively correlated signals remains practically unchanged even
  for extreme data loss of up to 90\%. In contrast, the global scaling
  of anti-correlated signals changes to uncorrelated behavior even
  when a very small fraction of the data is lost. These observations
  are confirmed on two examples of real-world signals: human gait and
  commodity price fluctuations. We further systematically study the
  local scaling behavior of surrogate signals with missing data to
  reveal subtle deviations across scales. We find that for
  anti-correlated signals even 10\% of data loss leads to significant
  monotonic deviations in the local scaling at large scales from the
  original anti-correlated towards uncorrelated behavior. In contrast,
  positively correlated signals show no observable changes in the
  local scaling for up to 65\% of data loss, while for larger
  percentage of data loss, the local scaling shows overestimated
  regions (with higher local exponent) at small scales, followed by
  underestimated regions (with lower local exponent) at large scales.
  Finally, we investigate how the scaling is affected by the average
  length, probability distribution and percentage of the remaining
  data segments in comparison to the removed segments. We find that
  the average length $\mu_r$ of the remaining segments is the key
  parameter which determines the scales at which the local scaling
  exponent has a maximum deviation from its original
  value. Interestingly, the scales where the maximum deviation occurs
  follow a power-law relationship with $\mu_r$. Whereas the percentage
  of data loss determines the extent of the deviation. The results
  presented in this paper are useful to correctly interpret the
  scaling properties obtained from signals with extreme data loss.
\end{abstract}

\maketitle

\section{Introduction} \label{secintr}

In real-world signals data can be missing or unavailable to a very
large extent, especially in archaeological, geological and
physiological recordings which often once recorded in the past can not
be generated again. Knowing the effects which data loss may have on
the correlations and other dynamical properties of the output signals
of a given system is instrumental in accurately quantifying and
modeling the underlying mechanisms driving the dynamics of the
system. Significant data loss can also be caused by failure of the
data collection equipment, as well as by the removal of artifacts or
noise-contaminated data segments. To correctly interpret results
obtained from correlated signals with missing data, it is important to
understand how the dynamical properties of such signals are affected
by the degree of data loss. Here we systematically investigate how
loss of data changes the scaling properties of various long-range
power-law anti-correlated and positively correlated
signals. Specifically, we develop a segmentation approach to generate
surrogate signals by randomly removing data segments from stationary
long-range power-law correlated signals, and we study how the
correlation properties are affected by (i) the percentage of removed
data, (ii) the average length of the removed (or remaining) data
segments and (iii) the functional form of the probability distribution
of the removed (remaining) segments. We utilize the detrended
fluctuation analysis (DFA) to quantify the effect of extreme data loss
on the scaling properties of long-range correlated signals.

Scaling (fractal) behavior was first encountered in a class of
physical systems~\cite{Stanley-1995,Shlesinger-1987,Liebovitch-1994}
which for a given ``critical'' value of their parameters, exhibit
complex organization among their individual components, leading to
correlated interactions over a broad range of scales. This class of
complex systems are typically characterized by (i) multi-component
nonlinear feedback interactions, (ii) non-equilibrium output dynamics,
and (iii) high susceptibility and responsiveness to
perturbations. Scaling behavior has been found in a diverse group of
systems --- ranging from earthquakes, to traffic jams and economic
crashes, to neuronal excitations as well as the dynamics of integrated
physiologic systems under neural control --- and has been associated
with the underlying mechanisms of regulation of these
systems~\cite{Ivanov-1998,Ashkenazy-2002}. The output signals of such
systems exhibit continuous fluctuations over multiple time and/or
space scales~\cite{Bassingthwaighte-1994,Malik-1995}, where the
amplitudes and temporal/spatial organization of the fluctuations are
characterized by absence of dominant scale, i.e., scale-invariant
behavior. Due to the nonlinear mechanisms controlling the underlying
interactions, the output signals of these systems are also typically
non-stationary, which masks the intrinsic correlations. Traditional
methods such as power-spectrum and auto-correlation
analysis~\cite{Hurst-1951,Mandelbrot-1969,Stratonovich-1981} are not
suitable for nonstationary signals.

DFA is a robust method suitable for detecting long-range power-law
correlations embedded in nonstationary
signals~\cite{Peng-1994,Taqqu-1995}. It has been successfully applied
to a variety of fields where scale-invariant behavior emerges, such as
DNA~\cite{Peng-1994,Peng-1992,Peng-1993,Buldyrev-1993,Ossadnik-1994,%
  Stanley-1994,Mantegna-1994,Havlin-1995a,Peng-1995b,Havlin-1995,%
  Mantegna-1996,Buldyrev-1998,Stanley-1999,Li-2003,Hackenberg-2005},
cardiac dynamics~\cite{Peng-1995,Ho-1997,%
  Barbi-1998,Ivanov-1999,Pikkujamsa-1999,%
  Ashkenazy-1999,Maekikallio-1999,Absil-1999,%
  Toweill-2000,Bunde-2000,Laitio-2000,Ashkenazy-2000,%
  Ashkenazy-2001,Ivanov-2001,Kantelhardt-2002,Karasik-2002,%
  Ivanov-2004,Bartsch-2005,Schmitt-2007,Schmitt-2009}, human
locomotion~\cite{Hausdorff-1995,Ashkenazy-2002,Bartsch-2007,Ivanov-2009},
circadian rhythm~\cite{Hu-2004,Ivanov-2007,Ivanov-2007a,Hu-2007},
neural receptors in biological systems \cite{Bahar-2001},
seismology~\cite{Varotsos-2003,Varotsos-2009},
meteorology~\cite{Ivanova-1999}, climate temperature
fluctuations~\cite{Koscielny-Bunde-1998,Koscielny-Bunde-1998a,%
Talkner-2000,Bunde-2001,Monetti-2003,Bunde-2005},
river flow and discharge \cite{Montanari-2000,Matsoukas-2000}, and
economics~\cite{Liu-1997,Vandewalle-1997,Vandewalle-1998,Liu-1999,Janosi-1999,%
  Ausloos-1999,Roberto-1999,Vandewalle-1999a,Grau-Carles-2000,%
  Ausloos-2000,Ausloos-2000a,Ausloos-2001,Ausloos-2001a,Ivanov-2004a}.
The DFA method may also help identify different states of the same
system exhibiting different scaling behavior --- e.g., the DFA scaling
exponent $\alpha$ for heart-beat intervals is significantly different
for healthy and sick
individuals~\cite{Peng-1995,Ashkenazy-1999,Schmitt-2007} as well as
for wake and sleep
states~\cite{Ivanov-1999,Bunde-2000,Kantelhardt-2002,Ivanov-2007a,Schmitt-2009}.

Elucidating the intrinsic mechanisms of a given system requires an
accurate analysis and proper interpretation of the dynamical (scaling)
properties of its output signals. It is often the case that the
scaling exponent quantifying the temporal (spatial) organization of
the systems' dynamics across scales is not always the same, but
depends on the scale of observation, leading to distinct crossovers
--- i.e., the value of the scaling exponent may be different for
smaller compared to larger scales. Such behavior has been observed for
diverse systems, for example: (i) the spontaneous motion of microbeads
bound to the cytoskeleton of living cells as quantified by the
mean-square displacement does not exhibit a Brownian motion but
instead undergoes a transition from subdiffusive to superdiffusive
behavior with time~\cite{Metzner-2007}; (ii) cardiac dynamics of
healthy subjects during sleep are characterized by fluctuations in the
heartbeat intervals exhibiting a crossover from a higher scaling
exponent (stronger correlations) at small time scales (from seconds up
to a minute) to a lower scaling exponent (weaker correlations) at
large time scales (from minutes to hours), associated with changes in
neural autonomic control during
sleep~\cite{Ivanov-1999,Kantelhardt-2003}; and (iii) stock market
dynamics where both absolute price returns and intertrade times
exhibit a crossover from a lower scaling exponent at small time scales
(up to a trading day) to much higher exponent at large time scales
(from a trading day to many months), a behavior consistent for all
companies on the market~\cite{Liu-1999,Ivanov-2004a}. However,
crossovers may also be a result of various types of nonstationarities
and artifacts present in the output signals, which, if not carefully
investigated, may lead to incorrect interpretation and modeling of the
underlying mechanisms regulating the dynamics of a given
system~\cite{Schmitt-2007}.

In previous studies, we have systematically investigated the effects
of various types of nonstationarities, data pre-processing filters and
data artifacts on the scaling behavior of long-range power-law
correlated signals as measured by the DFA
method~\cite{Hu-2001,Chen-2002,Chen-2005}.  In particular, we studied
a type of nonstationarity which is caused by the presence of
discontinuities (gaps) in the signal, i.e., how randomly removing data
segments of fixed length affects the scaling properties of long-range
power-law correlated signals~\cite{Chen-2002}.  Such discontinuities
may arise from the nature of the recordings --- e.g., stock exchange
data are not recorded during the nights, weekends and holidays
\cite{Liu-1997,Vandewalle-1997,%
  Vandewalle-1998,Liu-1999,Janosi-1999,Ausloos-1999,Roberto-1999,%
  Vandewalle-1999a}. In these situations, discontinuities correspond
to segments of fixed size.

Alternatively, discontinuities may be caused by the fact that (i) part
of the data is lost due to various reasons, and/or (ii) some noisy and
unreliable portions of continuous recordings (e.g., measurement
artifacts) are discarded prior to analysis~\cite{Peng-1995,Ho-1997,%
  Barbi-1998,Ivanov-1999,Pikkujamsa-1999,%
  Ashkenazy-1999,Maekikallio-1999,Absil-1999,%
  Toweill-2000,Bunde-2000,Laitio-2000,Ashkenazy-2000,%
  Ashkenazy-2001,Ivanov-2001,Schmitt-2009}.  In these cases, the
lengths of the lost or removed data segments are random, and may
follow a certain type of distribution which can often be related to
the process responsible for the removal or loss of data --- e.g., a
data acquisition device which fails randomly with a given probability
$p$ will result in a geometric distribution $P(l)=(1-p)^lp$ with mean
$\mu=1/p$, where $l$ is the length of the data lost segments. Thus,
investigating the effect of data loss is essential to determine the
true correlation properties of the signal output of a given system.

To address this question, we propose a new segmentation algorithm to
generate surrogate signals by randomly removing data segments from
long-range power-law correlated signals with a-priori known scaling
properties, and we investigate the effects of the percentage of the
removed data, different average lengths and different distributions of
removed data segments. We compare the scaling behavior of the original
signals with the scaling of the surrogate signals by systematically
studying changes in the DFA scaling exponent. We utilize local scaling
exponents to reveal subtle deviations and to characterize changes in
the scaling behavior at different scales in signals with segment
removed. We note, that in our investigation we consider the effect of
data loss on signals where the scaling behavior remains constant for
the duration of the observations. Signals comprised of segments
characterized by different scaling exponents have been considered
elsewhere~\cite{Chen-2002}.

This paper is structured as follows: in Sec.~\ref{secdfa}, we briefly
describe the DFA method.  In Sec.~\ref{secproc} we describe how to
generate stationary long-range power-law correlated signals. In
Sec.~\ref{secsegm} we introduce an algorithm for randomly removing
data segments from these signals to test the effects of data loss on
the scaling behavior. In Sec.~\ref{secresultglobal}, we study the
effect of data loss on the global scaling of positively correlated and
anti-correlated artifically generated signals with different length,
and we show examples on two different sets of empirical data. In
Sec.~\ref{secresultgap} we compare the local scaling properties of
correlated signals before and after data removal by considering the
effect of several parameters of the removed segments. In
Sec.~\ref{secresultseg} we consider the inverse situation --- instead
of focusing on the properties of the removed segments we investigate
how the correlations/scaling of the signal depend on the properties of
the remaining data segments.  We summarize and discuss our findings in
Sec.~\ref{secconclusion}.

\section{Methods}\label{secmethod}
\subsection{Detrended fluctuation analysis (DFA)}\label{secdfa}

The DFA is a random walk based method~\cite{Peng-1994}. It is an
improvement of the classical fluctuation analysis (FA) for
nonstationary signals where embedded polynomial trends mask the
intrinsic correlation properties in the
fluctuations~\cite{Peng-1994}. The performance of DFA for signals with
different types of nonstationarities and artifacts has been
extensively studied and compared to other methods of correlation
analysis~\cite{Taqqu-1995,Hu-2001,Chen-2002,Chen-2005,Xu-2005,%
Alvarez-Ramirez-2005,Nagarajan-2006,Bashan-2008}. The DFA methods
involves the following steps \cite{Peng-1994}:

(i) A given signal $u(i)$ ($i=1,..,N$, where $N$ is the length of
the signal) is integrated to obtain the random walk profile
$y(k)\equiv\sum_{i=1}^{k}[u(i)-\langle u \rangle]$, where $\langle u
\rangle$ is the mean of $u(i)$.
 
(ii) The integrated signal $y(k)$ is divided into boxes of equal
length $n$.
 
(iii) In each box of length $n$ we fit $y(k)$ using a polynomial
function of order $\ell$ which represents the {\it trend\/} in that
box. The $y$ coordinate of the fit curve in each box is denoted by
$y_n(k)$. When a polynomial fit of order $\ell$ is used, we denote the
algorithm as DFA-$\ell$. Note that, due to the integration procedure
in step (i), DFA-$\ell$ removes polynomial trends of order $\ell-1$
in the original signal $u(i)$.

(iv) The integrated profile $y(k)$ is detrended by subtracting the
local trend $y_n(k)$ in each box of length $n$:
\begin{equation}
 Y(k)\equiv~y(k)-y_n(k).
\label{F2}
\end{equation} 
 
(v) For a given box length $n$, the root-mean-square (rms) fluctuation
function for this integrated and detrended signal is calculated:
\begin{equation}
 F(n)\equiv\sqrt{{1\over {N}}\sum_{k=1}^{N}[Y(k)]^2}.
\label{F2}
\end{equation} 

(vi) The above computation is repeated for a broad range of box
lengths $n$ (where $n$ represents a specific space or time scale) to
provide a relationship between $F(n)$ and $n$.

A power-law relation between the root-mean-square fluctuation function
$F(n)$ and the box size $n$, i.e., $F(n) \sim n^{\alpha}$, indicates
the presence of scaling-invariant behavior embedded in the
fluctuations of the signal $u(i)$. The fluctuations can be
characterized by a scaling exponent $\alpha$, a self-similarity
parameter which represents the long-range power-law correlation
properties of the signal. If $\alpha=0.5$, there is no correlation and
the signal is uncorrelated (white noise); if $\alpha<0.5$, the signal
is anti-correlated; if $\alpha>0.5$, the signal is positively
correlated; and $\alpha=1.5$ indicates Brownian motion (integrated
white noise). For stationary signals with long-range power-law
correlations, the value of the scaling exponent $\alpha$ is related to
the exponent $\beta$ characterizing the power spectrum
$S(f)=f^{-\beta}$ of the signal, where $\beta=2\alpha-1$
\cite{Peng-1993}. Thus, the special case of $1/f$ noise, where
$\beta=1$, observed in various physiological and biological system
dynamics, corresponding to $\alpha=1$. Since the power spectrum of
stationary signals is the Fourier transform of the auto-correlation
function, for signals with scale-invariant long-range positive
correlation and $\alpha<1$, one can find the following relationship
between the auto-correlation exponent $\gamma$ and the power spectrum
exponent $\beta$ for signals with scale-invariant long-range
correlations: $\gamma=1-\beta=2-2\alpha$, where $\gamma$ is defined by
the auto-correlation function $C(\tau)=\tau^{-\gamma}$, and should
satisfy $0<\gamma<1$~\cite{Kantelhardt-2001}.

We note that for anti-correlated signals, the scaling exponent
$\alpha$ obtained from the DFA method overestimates the true
correlations at small scales $n$~\cite{Hu-2001}. To avoid this
problem, one needs first to integrate the original anti-correlated
signal and then apply the DFA method. The correct scaling exponent can
thus be obtained from the relation between $n$ and $F(n)/n$ [instead
of $F(n)$] (see Fig.~\ref{fig-dfa-corr}a). This procedure is applied
for all cases of anti-correlated signals in this study. In our
analysis in the following sections we apply DFA-2. The choice of DFA-2
is dictated by the fact that this order of DFA-$l$ can accurately
quantify the scaling behavior of signals with exponents in the range
$0<\alpha<3$~\cite{Xu-2005}, which covers practically all signals
generated by real world systems. Moreover, earlier investigations have
demonstrated that DFA-2 is sufficient to accurately quantify a broad
range of nonstationary signals generated by different physiologic
dynamics --- e.g., for heartbeat and gait dynamics the exponent
$\alpha$ obtained from higher order DFA-$l$ is not significantly
different compared to $\alpha$ obtained from
DFA-2~\cite{Ivanov-2009}. Further, deviations from scaling which
appear at small scale become more pronounced in higher order
DFA-$l$~\cite{Kantelhardt-2001}. In order to provide an accurate
estimate of $F(n)$, the largest box size $n$ we use is $n=N/8$, where
$N$ is the signal length.

\subsection{Procedure to generate stationary signals with long-range
  power-law correlations}\label{secproc}

We use a modified Fourier filtering technique \cite{Makse-1996} to
generate stationary long-range power-law correlated signals $u(i)$
($i=1,2,...,N$) with mean $\langle u(i) \rangle=0$ and standard
deviation $\sigma=1$. The correlations of $u(i)$ are characterized by
a Fourier power spectrum of a power-law form $S(f) \sim f^{-\beta}$,
where $f$ is the frequency. By manipulating the Fourier spectrum of
random Gaussian-distributed sequences, we generate signal $u(i)$ with
desired power-law correlations. This method consists of the following
steps:

(i) First, we generate a Gaussian-distributed sequence $\eta(i)$ with
mean $\langle \eta(i) \rangle=0$ and standard deviation
$\sigma_{\eta}=1$, and we calculate its Fourier transformation
$\hat{\eta}(f)$.

(ii) Next, we generate $\hat{u}(f)$ using the following
transformation:
\begin{equation}
\hat{u}(f)=\hat{\eta}(f)\cdot f^{-\beta/2},
\label{M2}
\end{equation}
where $\hat{u}(f)$ is the Fourier transform of the desired correlated
signal $u(i)$ characterized by a Fourier power spectrum of the form
\begin{equation}
S(f)=|\hat{u}(f)|^2\sim f^{-\beta},
\label{M1}
\end{equation}    

(iii) We calculate the inverse Fourier transform of {$\hat{u}(f)$} to
obtain $u(i)$. The generated stationary signal $u(i)$ is then
normalized to zero mean and unit standard deviation.

\subsection{Algorithm to generate surrogate signals with randomly
  removed segments}\label{secsegm}

We introduce a new segmentation approach to generate surrogate
nonstationary signals $\tilde{u}(i)$ by randomly removing data
segments from a stationary correlated signal $u(i)$ and stitching
together the remaining parts of $u(i)$. Such ``cutting'' procedure is
often used in the pre-processing of data prior to analysis in order to
eliminate, for example, segments of data artifacts. The proposed
segmentation approach allows the simulation of empirical data series
where data segments are lost or removed. The surrogate signals
$\tilde{u}(i)$ are characterized by four parameters: (i) the DFA
scaling exponent $\alpha$ of the original signal $u(i)$, (ii) the
percentage $p$ of the data removed, (iii) the average length $\mu$ of
the removed data segments as well as (iv) the functional form $P(l)$
of the distribution of the length $l$ of the removed data segments.

\begin{figure}
\centering{
\setlength{\unitlength}{7mm}
\begin{picture}(12,5)
\linethickness{0.15mm}
\put(0, 0.35){$\tilde{u}(i)$}
\put(1.6,0.35){$u(1)$}
\put(2.6,0.35){$u(2)$}
\put(3.6,0.35){$u(3)$}
\put(4.6,0.35){$u(6)$}
\put(5.6,0.35){$u(7)$}
\multiput(1.5,0)(1,0){6}{\line(0,1){1}}
\multiput(1.5,0)(0,1){2}{\line(1,0){5.5}}
\multiput(7.2,0)(0.4,0){5}{\line(1,0){0.2}}
\multiput(7.2,1)(0.4,0){5}{\line(1,0){0.2}}

\put(2.0, 2){\vector(0,-1){1}}
\put(3.0, 2){\vector(0,-1){1}}
\put(4.0, 2){\vector(0,-1){1}}
\put(7.0, 2){\vector(-2,-1){2}}
\put(8.0, 2){\vector(-2,-1){2}}

\put(0, 2.35){$g(i)$}
\put(1.9,2.35){$1$}
\put(2.9,2.35){$1$}
\put(3.9,2.35){$1$}
\put(4.9,2.35){$0$}
\put(5.9,2.35){$0$}
\put(6.9,2.35){$1$}
\put(7.9,2.35){$1$}
\put(8.9,2.35){$0$}
\multiput(1.5,2)(1,0){9}{\line(0,1){1}}
\multiput(1.5,2)(0,1){2}{\line(1,0){8.5}}
\multiput(10.2,2)(0.4,0){5}{\line(1,0){0.2}}
\multiput(10.2,3)(0.4,0){5}{\line(1,0){0.2}}

\put(2.0, 4){\vector(0,-1){1}}
\put(3.0, 4){\vector(0,-1){1}}
\put(4.0, 4){\vector(0,-1){1}}
\put(5.0, 4){\vector(0,-1){1}}
\put(6.0, 4){\vector(0,-1){1}}
\put(7.0, 4){\vector(0,-1){1}}
\put(8.0, 4){\vector(0,-1){1}}
\put(9.0, 4){\vector(0,-1){1}}

\put(0, 4.35){$u(i)$}
\put(1.6,4.35){$u(1)$}
\put(2.6,4.35){$u(2)$}
\put(3.6,4.35){$u(3)$}
\put(4.6,4.35){$u(4)$}
\put(5.6,4.35){$u(5)$}
\put(6.6,4.35){$u(6)$}
\put(7.6,4.35){$u(7)$}
\put(8.6,4.35){$u(8)$}
\multiput(1.5,4)(1,0){9}{\line(0,1){1}}
\multiput(1.5,4)(0,1){2}{\line(1,0){8.5}}
\multiput(10.2,4)(0.4,0){5}{\line(1,0){0.2}}
\multiput(10.2,5)(0.4,0){5}{\line(1,0){0.2}}

\end{picture}
}
\caption{Illustration of generating a surrogate signal $\tilde{u}(i)$
  by removing data points from the original signal $u(i)$ according to
  a binary series $g(i)$. The positions $i$ where $g(i)=0$ (or 1)
  correspond to the positions at which data points in $u(i)$ are
  removed (or preserved) to obtain $\tilde{u}(i)$.}
\label{fig-data-remove}
\end{figure}
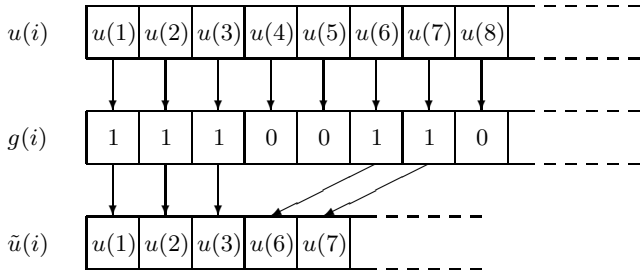

To generate a surrogate signal $\tilde{u}(i)$ from the original signal
$u(i)$, we first construct a binary sequence $g(i)$ with the same
length $N$ as $u(i)$. In our algorithm the positions $i$ where
$g(i)=0$ will correspond to the positions at which data points in
$u(i)$ are removed, while the positions where $g(i)=1$ will correspond
to the positions in $u(i)$ where data points are preserved
(Fig.~\ref{fig-data-remove}).

We developed the following method to construct the binary series $g(i)$:

(i) We generate the lengths $l_j$ ($j=1,2,...,M$) of the
segments that will be removed from the original signal $u(i)$ by
randomly drawing integer numbers from a given probability distribution
$P(l)$ with mean value $\mu$. Each integer number drawn from $P(l)$
represents the length of a segment removed from $u(i)$. The process
continues until the summation of the lengths of all removed segments
becomes equal or exceeds a predetermined amount $pN$ of data to be
removed, i.e.,
\begin{equation}
\sum_{j=1}^Ml_j~\ge~pN,
\label{equ-num-seg}
\end{equation}
where $M$ is the minimal number to fulfill Eq.~\ref{equ-num-seg}.
Eventually, we will cut the size of the last segment to obtain the
exact fraction $pN$ of the lost data.

(ii) We append a ``1'' to each element in the series $\{l_j\}$ which
will serve as a separator between two adjacent segments (see step
(iv)), and results in a new series $\{[l_j,1]\}$. Note that now the
summation over the series yields $pN+M$.

(iii) We append $N-(pN+M)$ ``1'' elements to the end of the series
$\{[l_j,1]\}$ to obtain an extended series where the sum of all
elements is $N$, equal to the length of the original series $u(i)$.
This extended series is then shuffled leading to a set of $M$ elements
$[l_j,1]$ randomly scattered in a ``sea'' of $N-(pN+M)$ ``1'' elements
(see Eq.~\ref{equ-shuffle}).

(iv) Next, we replace the numbers $l_j$ in Eq.~\ref{equ-shuffle} with
$l_j$ elements of zeros, to obtain a binary series $g(i)$ as shown in
Eq.~\ref{equ-expand}.
\begin{align}
\{\ldots,1, & \ \ \ \ [l_j,1], & 1,\ldots,1, & \ \ [l_{j+1},1], & [l_{j+2},1],\ \  & 1,\ldots\}\label{equ-shuffle}\\
\{\ldots,1, & \overbrace{0,\ldots,0,1}, & 1,\ldots,1, & \overbrace{0,\ldots,0,1}, & \overbrace{0,\ldots,0,1}, & 1,\ldots\}\label{equ-expand}
\end{align}

Note that, in step (iii) of our algorithm, the shuffling of the
extended series may lead to two or more $[l_j, 1]$ elements, which
represent removed data segments, to become direct neighbors
(Eq.~\ref{equ-shuffle}). Adding ``1'' to each element $\{l_j\}$ in
step (ii) thus ensures that adjacent $[l_j, 1]$ elements in the
shuffled extended series in Eq.~\ref{equ-shuffle} would not allow two
or more separate removed segments to be merged leading to the
formation of removed segments with longer average length $\mu$ and
different form of their probability distribution compared to the
original choice in step (i) of the algorithm.

Finally, the surrogate signal $\tilde{u}(i)$ is obtained by
simultaneously scanning the original signal $u(i)$ and the binary
series $g(i)$ from Eq.~\ref{equ-expand}, removing the $i$-th element
in $u(i)$ if $g(i)\equiv 0$ and concatenating the segments of the
remaining data (Fig.~\ref{fig-data-remove}).

In this study, we consider four different functional forms of the
probability distribution $P(l)$ of segment lengths $l$, i.e.,
exponential, Gaussian, $\delta$- and power-law distributions, and we
use the average length $\mu$ of the removed data segments as a common
parameter to compare the effect of removed data segments with
different distributions. For the exponential and
$\delta$-distribution, the average length $\mu$ is sufficient to
determine their probability distribution functions.  The Gaussian and
power-law distributions require additional parameters to be clearly
defined, and thus, we need to introduce boundary conditions, so that
these parameters can be related to the average length $\mu$.

The functional form of the Gaussian distribution is
\begin{equation}
P(l)=\frac{1}{\sqrt{2\pi\sigma^2}}exp\left[-\frac{(l-\mu)^2}{2\sigma^2}\right],
\label{equ-gaussian}
\end{equation}
where $\mu$ is the average and $\sigma$ is the standard deviation of
the segment lengths $l$. Since with a fixed small $\sigma$, the
Gaussian distribution is not much different from a
$\delta$-distribution, and with a fixed large $\sigma$, the Gaussian
distribution resembles an exponential distribution, we relate $\sigma$
with $\mu$ in such a way, as a boundary condition, that the smallest
segment ($l=1$) can only be obtained (statistically) once in each
realization, i.e., $P(l=1)~\equiv~1/pN$, where $N$ is the length of
the original signal, and $p$ is the percentage of data loss.

The functional form of a power-law distribution is given by
\begin{equation}
P(l)=al^k, l\in[1,l_{max}],
\label{equ-power-law}
\end{equation}
with $\int_1^{l_{max}}P(l)dl=1$ and the average length
$\mu=\int_1^{l_{max}}lP(l)dl$. Similar to the Gaussian distribution,
we set the probability of the largest segment to
$P(l=l_{max})~\equiv~1/pN$. With these three boundary conditions, we
can relate the three parameters $a$, $k$ and $l_{max}$ in
Eq.~\ref{equ-power-law} with the average length $\mu$. 

\begin{figure}
\centering{
\includegraphics[width=1.0\linewidth]{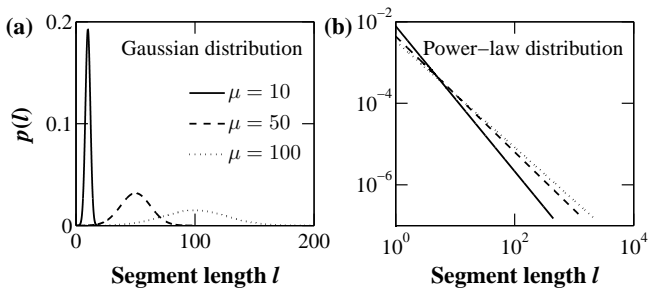}
}
\vspace{-0.5cm}
\caption{Examples of theoretical probability density for (a) Gaussian
  distribution and (b) power-law distribution used in our simulations
  of different situations of data loss. The parameters for the
  functional form of distributions are determined by the average
  length $\mu$ we chose for each simulation and by specific boundary
  conditions, i.e., for the Gaussian distribution, we set the
  probability of the smallest segment length $P(l=1)=1/pN$, and for
  the power-law distribution we set the probability of the largest
  segment length $P(l=l_{max})=1/pN$ (see text for details).}
\label{fig-pdf-gauss-pow-mu}
\end{figure}

\begin{figure}
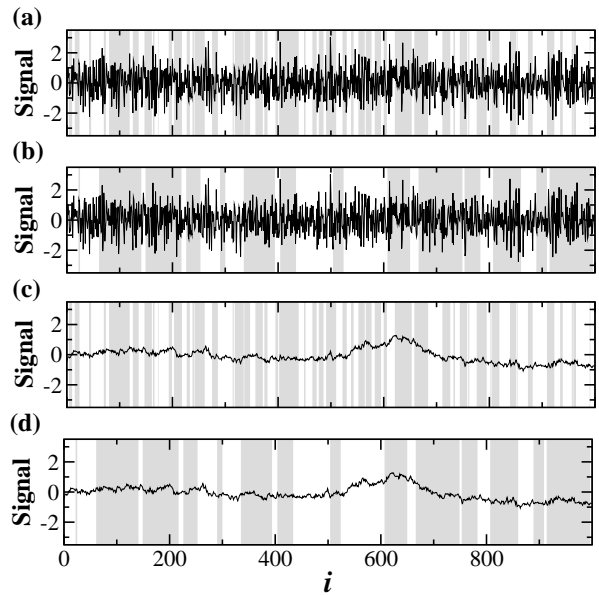

\centering{
\includegraphics[width=0.9\linewidth]{EPSfigures/signal_a03_gap_exp_t10_p50.eps}
}
\centering{
\includegraphics[width=0.9\linewidth]{EPSfigures/signal_a03_gap_exp_t50_p50.eps}
}
\centering{
\includegraphics[width=0.9\linewidth]{EPSfigures/signal_a13_gap_exp_t10_p50.eps}
}
\centering{
\includegraphics[width=0.9\linewidth]{EPSfigures/signal_a13_gap_exp_t50_p50.eps}
}
\caption{Illustration of data removal from stationary correlated
  signals.  Removed data segments (shaded regions) are randomly
  positioned within the original signal, and their lengths $l$ are
  drawn from an exponential distribution
  $P(l)~=~\frac{1}{\mu}\exp(-l/\mu)$ with average value $\mu$.  An
  average length $\mu=10$ is chosen for (a) the anti-correlated signal
  (DFA scaling exponent $\alpha=0.3$) and (b) the positively
  correlated signal ($\alpha=1.3$). Larger segments with $\mu=50$ are
  removed from (b) anti-correlated signal ($\alpha=0.3$) and (d)
  positively correlated signal ($\alpha=1.3$).}
\label{fig-signal-exp}
\end{figure}

In Fig.~\ref{fig-pdf-gauss-pow-mu}, we show examples of Gaussian and
power-law distributions with different average lengths $\mu$ based on
the criteria described above. Fig.~\ref{fig-signal-exp} shows examples
of our procedure of data removal. The lengths of the removed segments
were chosen to be exponentially distributed with different average
length.

\section{Results}\label{secresult}
\subsection{Effect of data loss on global
  scaling} \label{secresultglobal}

Previously, we have studied the effect of data loss on the scaling
behavior of long-range correlated signals by removing data segments
with fixed length~\cite{Chen-2002}. We have found that data loss in
anti-correlated signals substantially changes the scaling behavior
even when only 1\% of data are removed. In contrast, the scaling
behavior of (positively) correlated signals is practically not
affected even when up to 50\% of the data are removed. Data loss
generally causes a crossover in the scaling behavior of
anti-correlated signals.  At the scales larger than the crossover the
anti-correlated scaling behavior is completely destroyed and resembles
uncorrelated behavior.  This crossover is shifted to smaller scales
with increasing percentage of removed data or decreasing length of the
removed segments, indicating a stronger effect on the scaling
behavior.

In most cases, the length of data loss segments is not fixed but
random, and follows a certain distribution. How does the distribution
of data loss segments influence the scaling behavior of correlated
signals? In some cases, especially when archaeological data are
studied, the percentage of data loss can be extremely large (and can
reach up to 95\% !~\cite{Temin-2002}). Would the extreme data loss
affect also positively correlated signals? To address these questions,
in this section we study the effect of data loss caused by random
removal of data segments that follow a certain distribution.

\begin{figure}
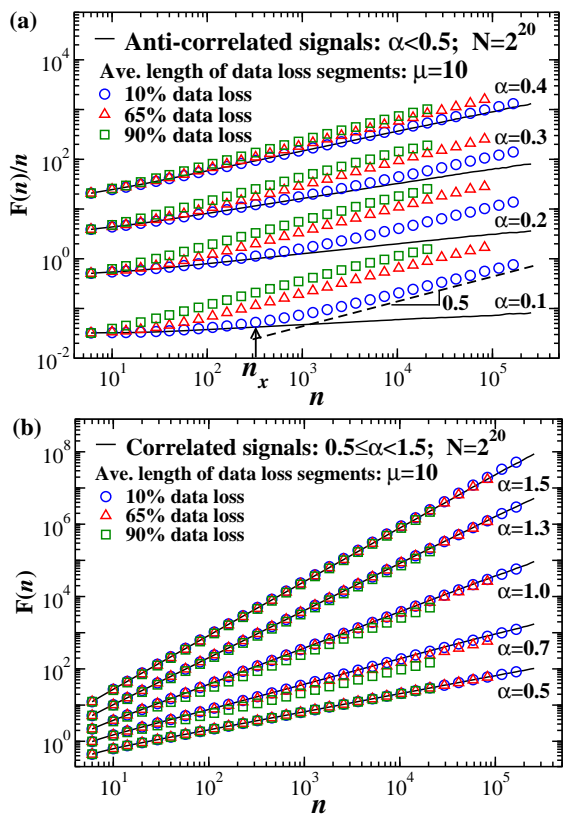

\centering{
\includegraphics[width=0.85\linewidth]{EPSfigures/dfa2_anticorr.exp.t10.ave.eps}
}
\vspace{0.1cm}
\centering{
\includegraphics[width=0.85\linewidth]{EPSfigures/dfa2_corr0515.exp.t10.ave.eps}
}
\caption{(Color online) Effect of data loss on the scaling behavior of
  long-range correlated signals with length $N=2^{20}$ (before data
  removal), zero mean and unity standard deviation. The lengths of the
  removed segments are drawn from an exponential distribution with
  mean $\mu=10$. (a) Scaling behavior of anti-correlated signals
  (scaling exponent $\alpha<0.5$) with a data loss of 10\% (blue
  circles), 65\% (red triangles) and 90\% (green squares). Note that,
  to obtain an accurate estimation of the DFA scaling exponent
  $\alpha$ for anti-correlated signals, we first integrate the signals
  and then we apply the DFA method. Thus, to obtain the correct
  scaling exponent for anti-correlated signals we divide $F(n)$ by $n$
  to account for the integration of the signals and next we plot
  $F(n)/n$ vs. the scale $n$ (see also Sec.~\ref{secdfa} and Fig.~15
  in~\cite{Hu-2001}). (b) Scaling behavior of positively correlated
  signals (scaling exponent $\alpha>0.5$) with 10\%, 65\% and 90\%
  data loss.  The scaling behavior of strongly anti-correlated data is
  dramatically changed even when only 10\% of the data are removed. A
  crossover at scale $n_x$ indicates a transition (arrow), due to loss
  of data in the signals, from the original anti-correlated behavior
  with $\alpha=0.1$ to an uncorrelated behavior with $\alpha=0.5$. In
  contrast, for positively correlated signals, i.e. $0.5<\alpha<1.5$
  only an extreme data loss of 90\% leads to small deviations from the
  original scaling behavior. This effect becomes weaker for increasing
  values of $\alpha$ . As expected, for $\alpha=0.5$ (white noise) and
  $\alpha=1.5$ (Brownian noise) data removal does not affect the
  scaling behavior.}
\label{fig-dfa-corr}
\end{figure}

First, we consider the case in which the lengths of data loss segments
are exponentially distributed. Following the approach introduced in
Sec.~\ref{secsegm}, we first generate stationary correlated signals
$u(i)$ with length $N=2^{20}$ and with scaling exponents $\alpha$
ranging from 0.1 to 1.5, and then randomly remove exponentially
distributed data segments from the original signal $u(i)$ to obtain
surrogate signals $\tilde{u}(i)$. As illustrated in
Fig.~\ref{fig-dfa-corr}, the rms fluctuation function $F(n)$ shows
similar changes in the scaling behavior as observed in
\cite{Chen-2002} where segments with a fixed length were removed from
the original signal. (i) The scaling behavior of surrogate signals
strongly depends on the scaling exponent $\alpha$ of the original
signals.  (ii) The anti-correlated signals substantially change their
scaling behavior even if only 10\% of the data are removed
(Fig.~\ref{fig-dfa-corr}(a)).  A crossover from anti-correlated to
uncorrelated ($\alpha=0.5$) behavior appears at scale $n_x$ due to
data loss, i.e., at the scales larger than $n_x$, the
anti-correlations in the original signals are completely
destroyed. The crossover scale $n_x$ is shifted to smaller scales with
increasing percentage of lost data. (iii) In contrast, positively
correlated signals show practically no changes for up to 65\% of data
loss (Fig.~\ref{fig-dfa-corr}(b)). Surprisingly, even with extreme
data loss of up to 90\% of the signal the scaling behavior is still
practically preserved, exhibiting a slightly lower exponent $\alpha$
(waker correlations) --- an effect which is less pronounced with
increasing values of $\alpha$ (see Fig.~\ref{fig-dfa-corr}(b)).

\begin{figure}
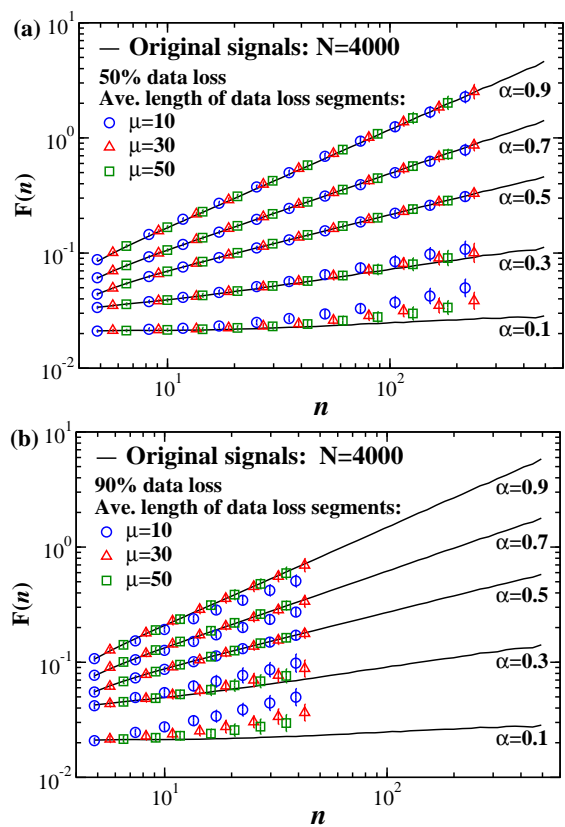

\centering{
\includegraphics[width=0.85\linewidth]{EPSfigures/dfa2_N4000_exp_p50_t10_t30_t50.eps}
}
\centering{
\includegraphics[width=0.85\linewidth]{EPSfigures/dfa2_N4000_exp_p90_t10_t30_t50.eps}
}
\caption{(Color online) Effect of data loss on the scaling behavior of
  {\it short} signals ($N=4000$ before data removal).  (a) Removing up
  to 50\% of the data (i.e., 2000 data points remain) does not have an
  observable effect on the scaling behavior of positively correlated
  signals and leads to small deviations from the original scaling
  behavior in anti-correlated signals. (b) Extreme data loss of 90\%
  (i.e., only 400 data points remain) leads to more pronounced
  deviations from the original scaling behavior. In general, the
  deviations are smaller with larger average length $\mu$ of removed
  segments.}
\label{fig-dfa-N4000}
\end{figure}

Next, we consider the case in which the length of the original signal
is much shorter ($N=4000$), as illustrated in
Fig.~\ref{fig-dfa-N4000}. We find that the scaling behavior of both
anti-correlated and positively correlated signals with extreme data
loss change in the same way as we observed in Fig.~\ref{fig-dfa-corr}
(where $N=2^{20}$). In addition, we find (see
Fig.~\ref{fig-dfa-N4000}) that when increasing the average length
$\mu$ of the data loss segments, the scaling behavior of the surrogate
signals deviates less from the original scaling behavior. Thus,
removing the same percentage of the data using longer (and fewer)
segments has a lesser impact on the scaling behavior of both
positively correlated and anti-correlated signals compared to removing
segments with smaller average length $\mu$.

\begin{figure}
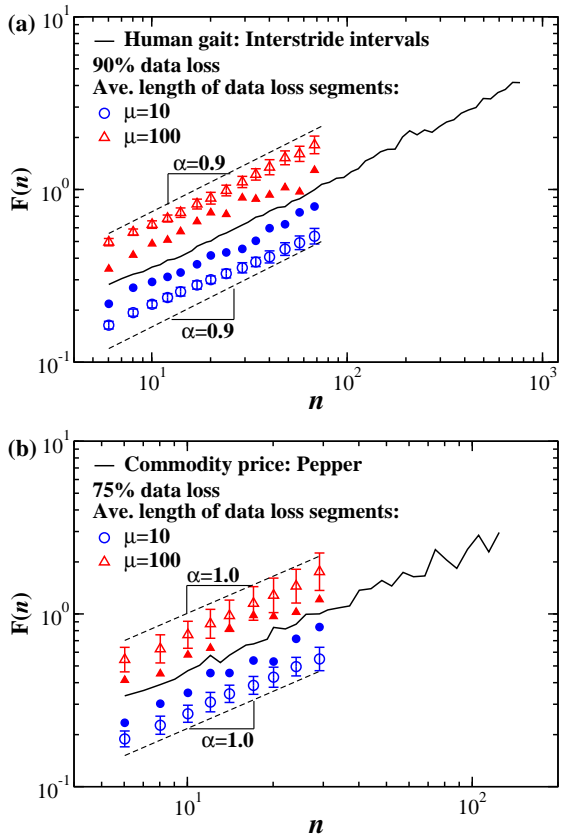

\begin{center}
\includegraphics[width=0.85\linewidth]{EPSfigures/dfa2_gaptest_gait.eps}
\end{center}
\vspace{-0.5cm}
\begin{center}
\includegraphics[width=0.85\linewidth]{EPSfigures/dfa2_gaptest_price.eps}
\end{center}
\vspace{-0.5cm}
\caption{ (Color online) Two examples of the effect of extreme data
  loss: (a) interstride intervals of human gait, and (b) annual prices
  of pepper in England in the period 1209--1914. Removing up to 90\%
  of the gait intervals and up to 75\% of the commodity data using
  segments of different average length $\mu$ does not significantly
  affect the global scaling behavior. Closed symbols represent a
  single realization and open symbols indicate the mean and standard
  deviations obtained from 100 realizations of randomly removing data
  segments. The lengths of the removed data segments are drawn from an
  exponential distribution.}
\label{fig-dfa-gait-price}
\end{figure}

To show how missing data segments affect correlations in real-world
signals, we consider two examples of complex scale-invariant dynamics:
(i) human gait as a representative of integrated physiologic systems
under neural control with multiple-component feedback interactions
(Fig.~\ref{fig-dfa-gait-price}a), and (ii) commodity price
fluctuations from England across several centuries reflecting complex
economic and social interactions (Fig.~\ref{fig-dfa-gait-price}b). In
agreement with our tests on surrogate signals shown in
Fig.~\ref{fig-dfa-corr} and Fig.~\ref{fig-dfa-N4000}, our analyses of
real data confirm the observation that even extreme data loss of up to
90\% does not significantly affect the global scaling behavior of
positively correlated ($\alpha>0.5$) signals.

\subsection{Properties of removed data segments: Effect of data loss
  on local scaling}\label{secresultgap}

To reveal in greater detail the effect of data loss, we investigate 
the local scaling behavior of the $F(n)$ curves by fitting $F(n)$ 
locally in a window of size $w=3log2$. We determine the local
scaling exponent $\alpha_{loc}$ at different scales $n$ by moving
the window $w$ in small steps of size $\Delta=\frac{1}{4}log2$ starting 
at $n=4$.

\begin{figure*}
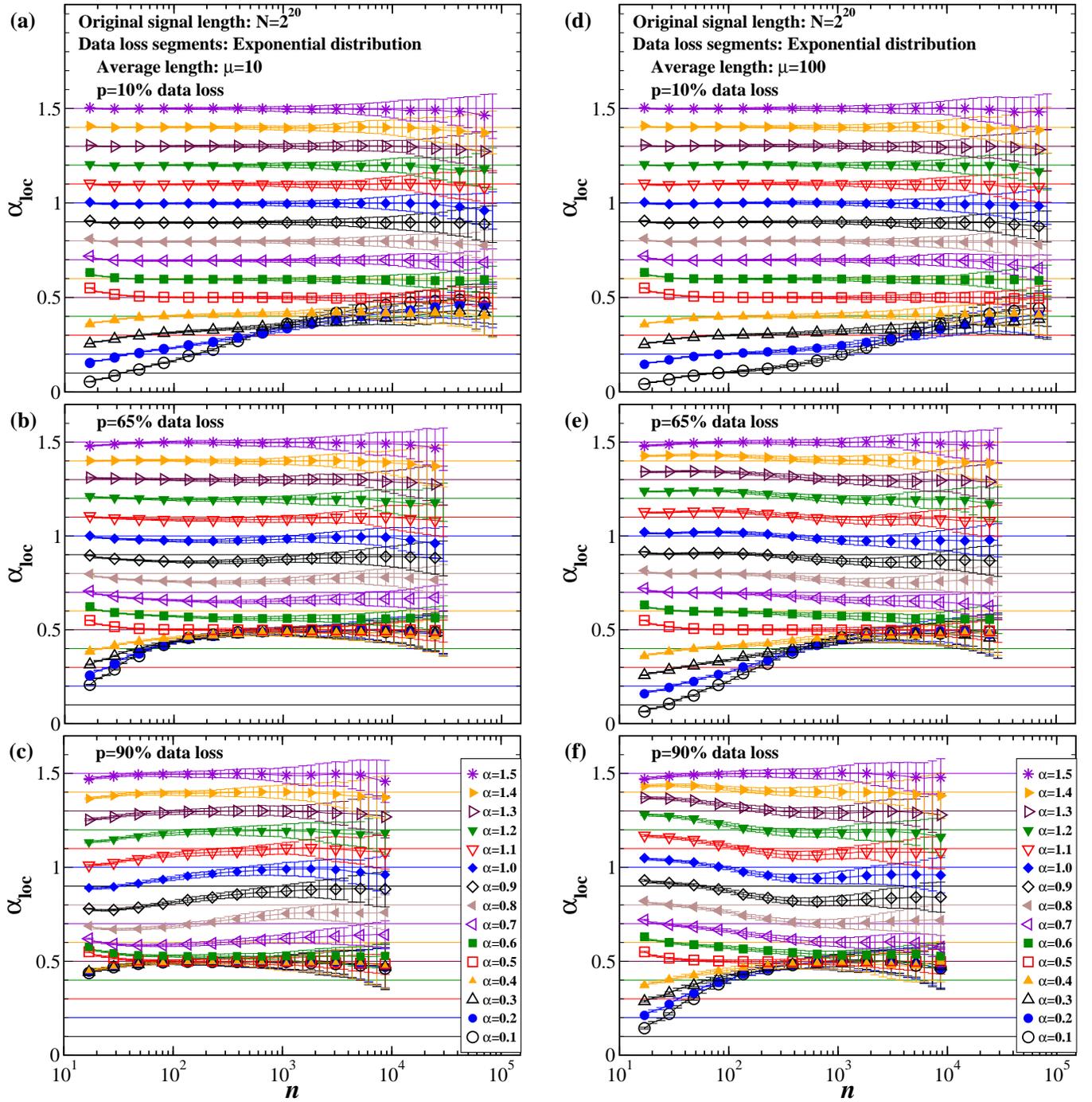

\centering{
\includegraphics[width=0.48\linewidth]{EPSfigures/aloc_exp_t10_p10.eps}
\hspace{0.5cm}
\includegraphics[width=0.48\linewidth]{EPSfigures/aloc_exp_t100_p10.eps}
}
\centering{
\includegraphics[width=0.48\linewidth]{EPSfigures/aloc_exp_t10_p65.eps}
\hspace{0.5cm}
\includegraphics[width=0.48\linewidth]{EPSfigures/aloc_exp_t100_p65.eps}
}
\centering{
\includegraphics[width=0.481\linewidth]{EPSfigures/aloc_exp_t10_p90.eps}
\hspace{0.5cm}
\includegraphics[width=0.481\linewidth]{EPSfigures/aloc_exp_t100_p90.eps}
}
\caption{(Color online) Effect of data loss on the local scaling
  behavior (quantified by local scaling exponent $\alpha_{loc}$) of
  long-range power-law correlated signals. The symbols indicate
  average $\alpha_{loc}$ values obtained from 100 different
  realizations of surrogate signals with the same correlation exponent
  $\alpha$, and the error bars show the standard deviations. The more
  data are removed, the more the scaling exponent deviates from the
  original exponent. The data loss segments are exponentially
  distributed with average length $\mu=10$ ((a)-(c)) and $\mu=100$
  ((d)-(f)). For anti-correlated signals, the removal of larger
  segments ($\mu=100$) has less effect on the scaling behavior. For
  positively correlated signals, the deviations vary across scales,
  showing both overestimated and underestimated regions.}
\label{fig-aloc-exp-t10}
\end{figure*}

In Fig.~\ref{fig-aloc-exp-t10}, we show $\alpha_{loc}$ for 10\%, 65\%
and 90\% of data loss, and the average length of the data loss
segments is $\mu=10$ (cp. Fig.~\ref{fig-dfa-corr}).  The scaling
behavior of anti-correlated signals shows systematic deviations from
the original behavior: the stronger the anti-correlations, the faster
is the decay of $\alpha_{loc}$ towards 0.5 (uncorrelated
behavior). The deviations are stronger when more data were removed
from the original signal. Note that when 90\% of the data are removed,
the correlation properties of originally anti-correlated signals are
completely destroyed (Fig.~\ref{fig-aloc-exp-t10}(c)), because there
are practically no consecutive data points of the original signals
preserved in the surrogates when $\mu=10$ and $p=90\%$ (see
Sec.~\ref{secresultseg} and Eq.~\ref{equ-mu}). When increasing the
average length of the removed segments from $\mu=10$ to $\mu=100$
(Fig.~\ref{fig-aloc-exp-t10}), the scaling behavior of anti-correlated
signals is less affected and $\alpha_{loc}=0.5$ is reached at larger
scales.

For positively correlated signals ($0.5<\alpha<1.5$), the effect of
data loss is more complex. The local scaling exponents show
significant and systematic deviations from the original scaling
behavior not observed in the rms fluctuation functions $F(n)$ in
Fig.~\ref{fig-dfa-corr}(b).  The deviations from the original scaling
behavior are more pronounced for a higher percentage of data loss and
vary across scales. For small average length ($\mu=10$,
Fig.~\ref{fig-aloc-exp-t10}a-c), the local scaling exponent is
underestimated at small scales and gradually recovers to the original
scaling behavior at larger scales. For a larger average length of
removal data segments ($\mu=100$, Fig.~\ref{fig-aloc-exp-t10}d-f), we
find overestimated regions at small scales and underestimated regions
at large scales. The overestimation of the local scaling behavior is
more pronounced for stronger positively correlated signals, while the
underestimation is more pronounced for weaker positively correlated
signals.

An interesting phenomenon seen in Fig.~\ref{fig-aloc-exp-t10} is that
for anti-correlated signals the scale at which $\alpha_{loc}$ reaches
0.5 (uncorrelated behavior) is shifted towards smaller scales with
increasing percentage of data loss. Similarly, for positively
correlated signals, the overestimated and underestimated regions are
also shifted towards smaller scales, when a higher percentage of data
is removed. This phenomenon occurs in both cases $\mu=10$ and
$\mu=100$.

\begin{figure*}
\centering{
\includegraphics[width=0.465\linewidth]{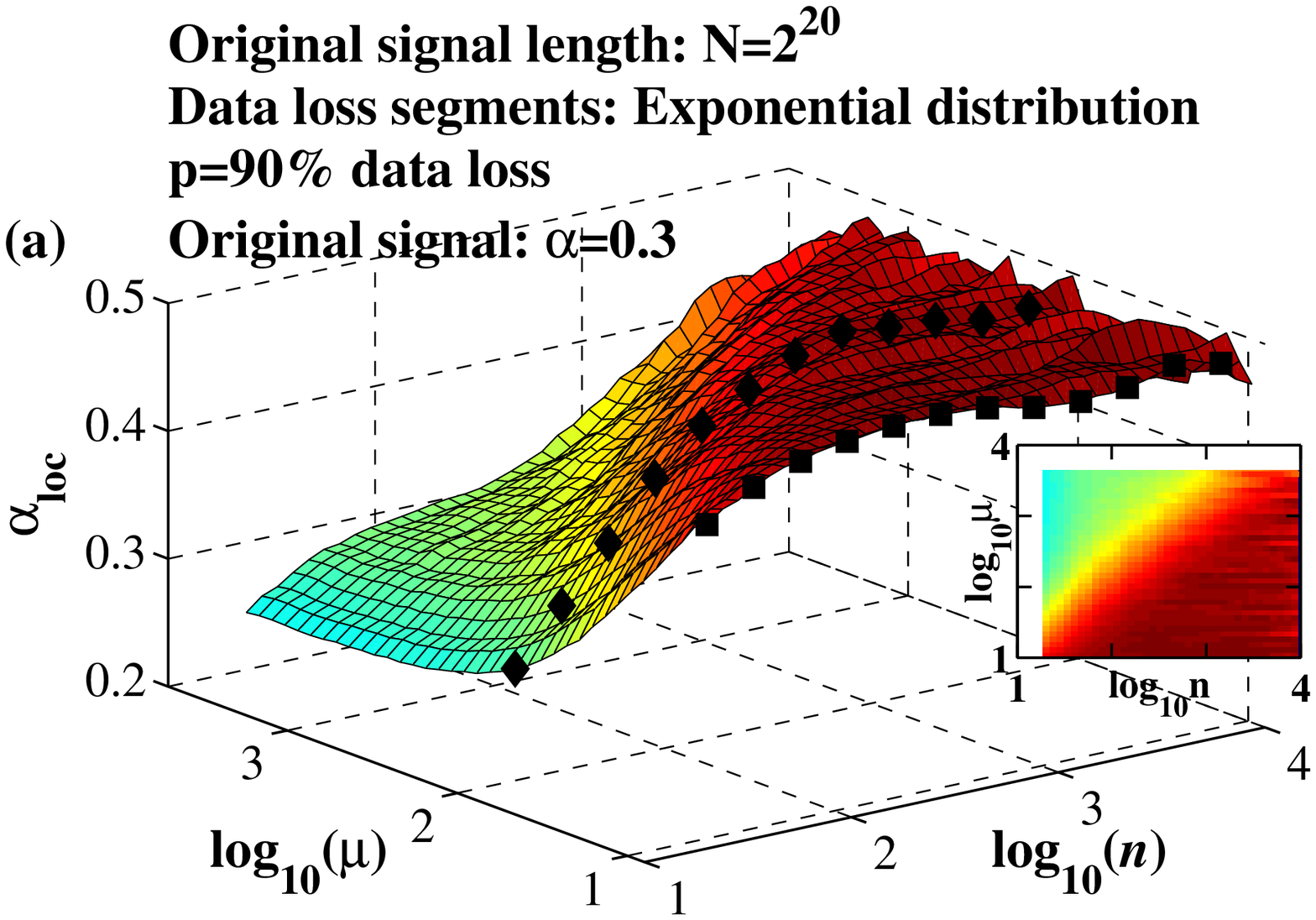}
\hspace{0.5cm}
\includegraphics[width=0.465\linewidth]{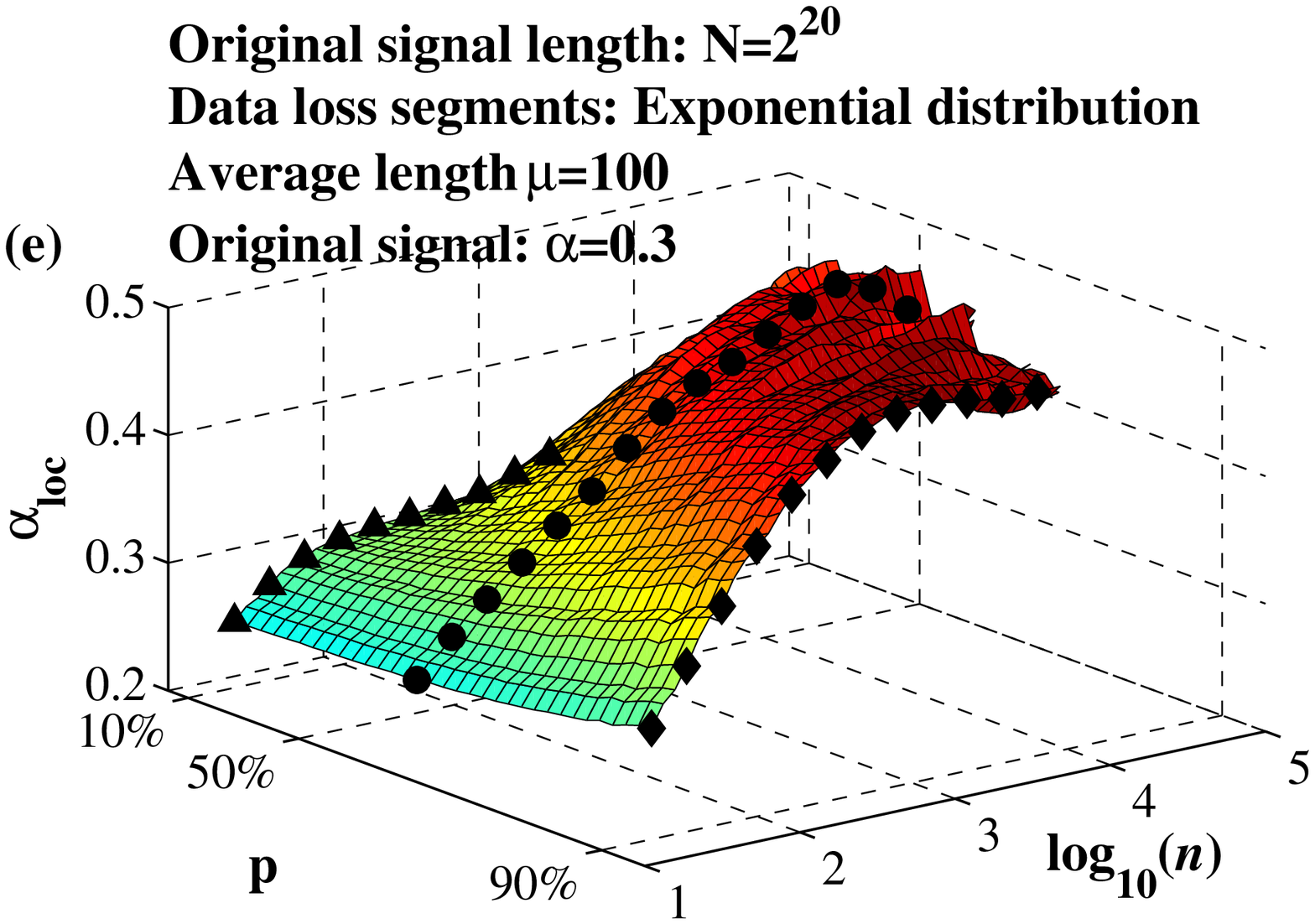}
}
\centering{
\includegraphics[width=0.465\linewidth]{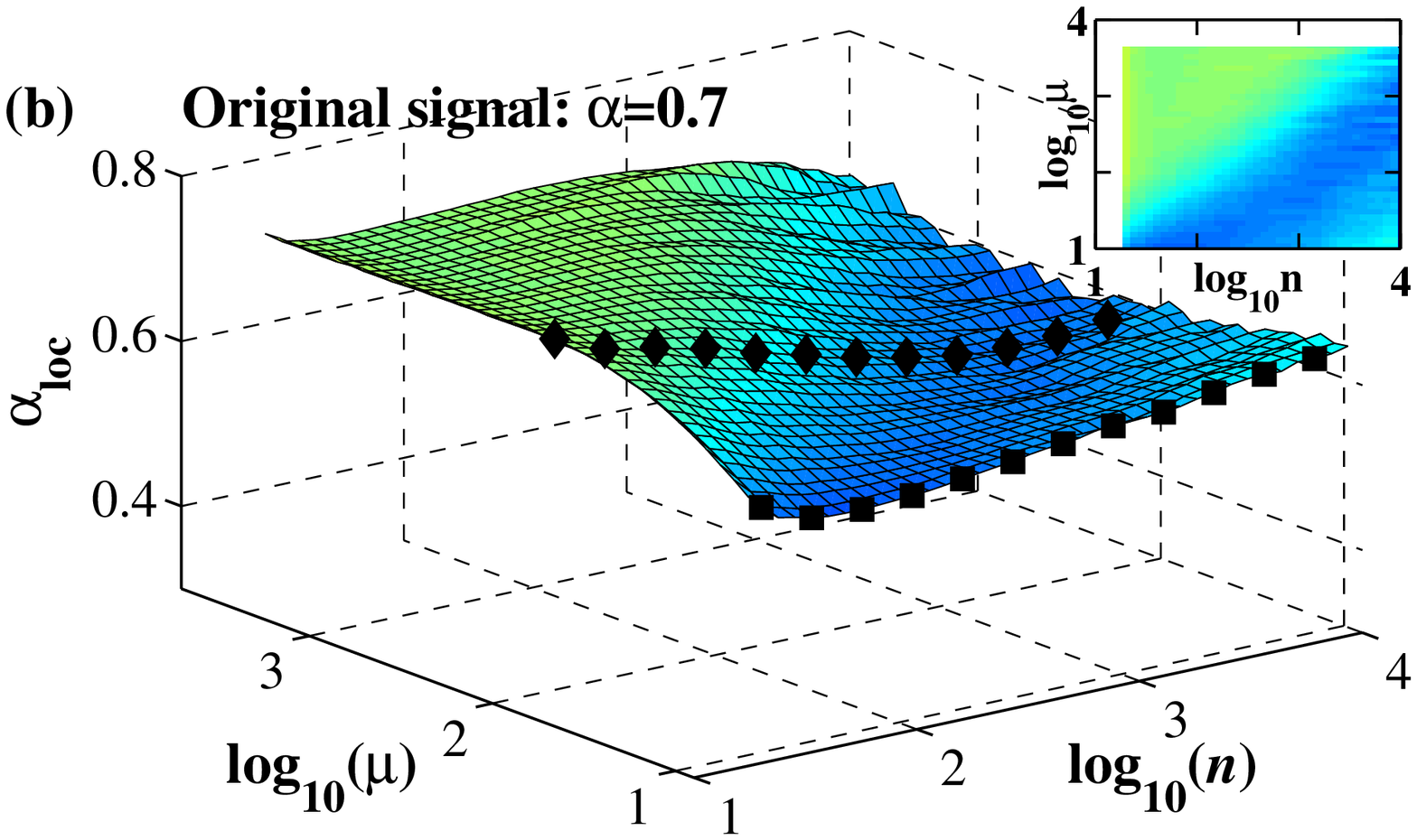}
\hspace{0.5cm}
\includegraphics[width=0.465\linewidth]{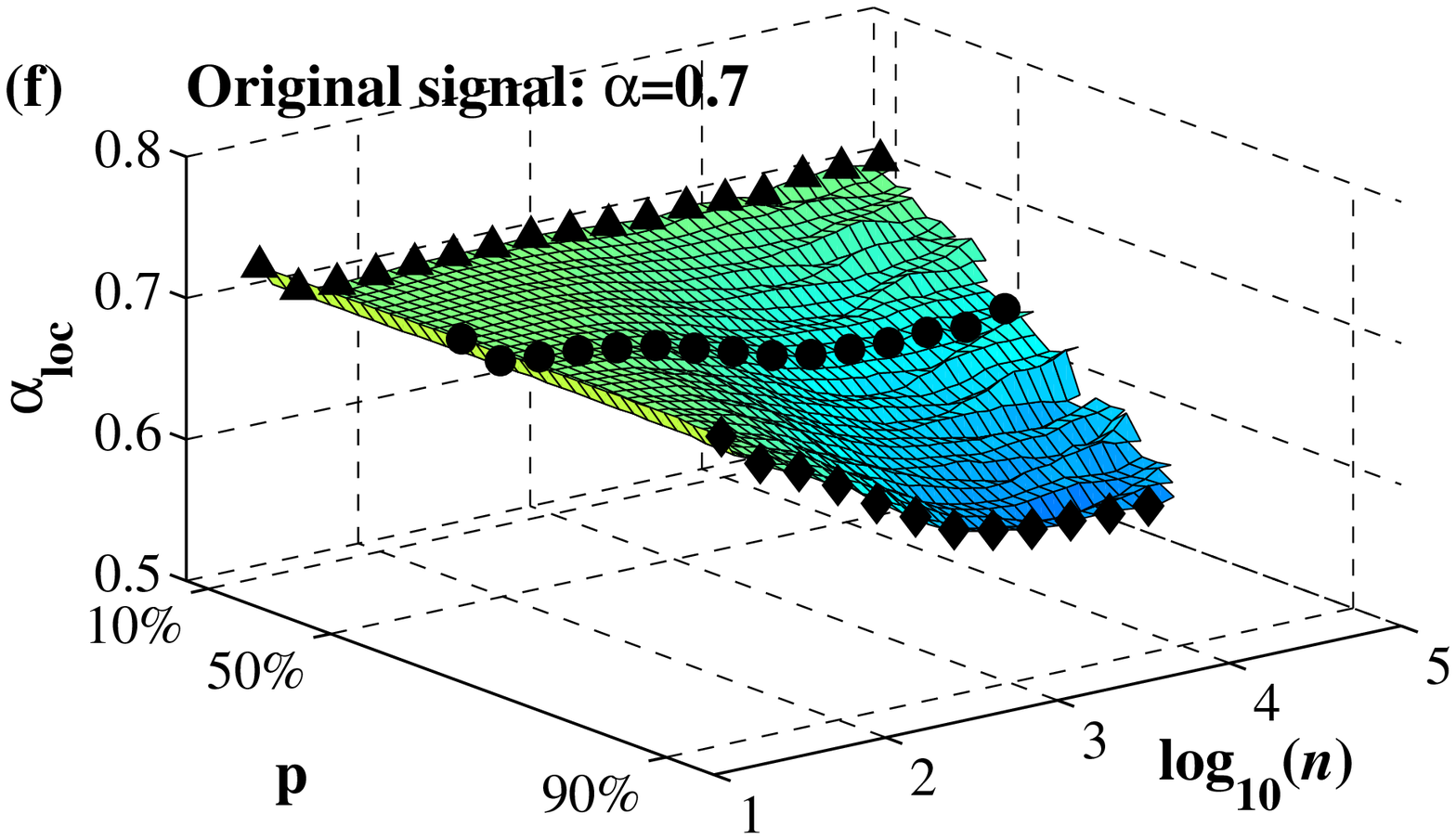}
}
\centering{
\includegraphics[width=0.465\linewidth]{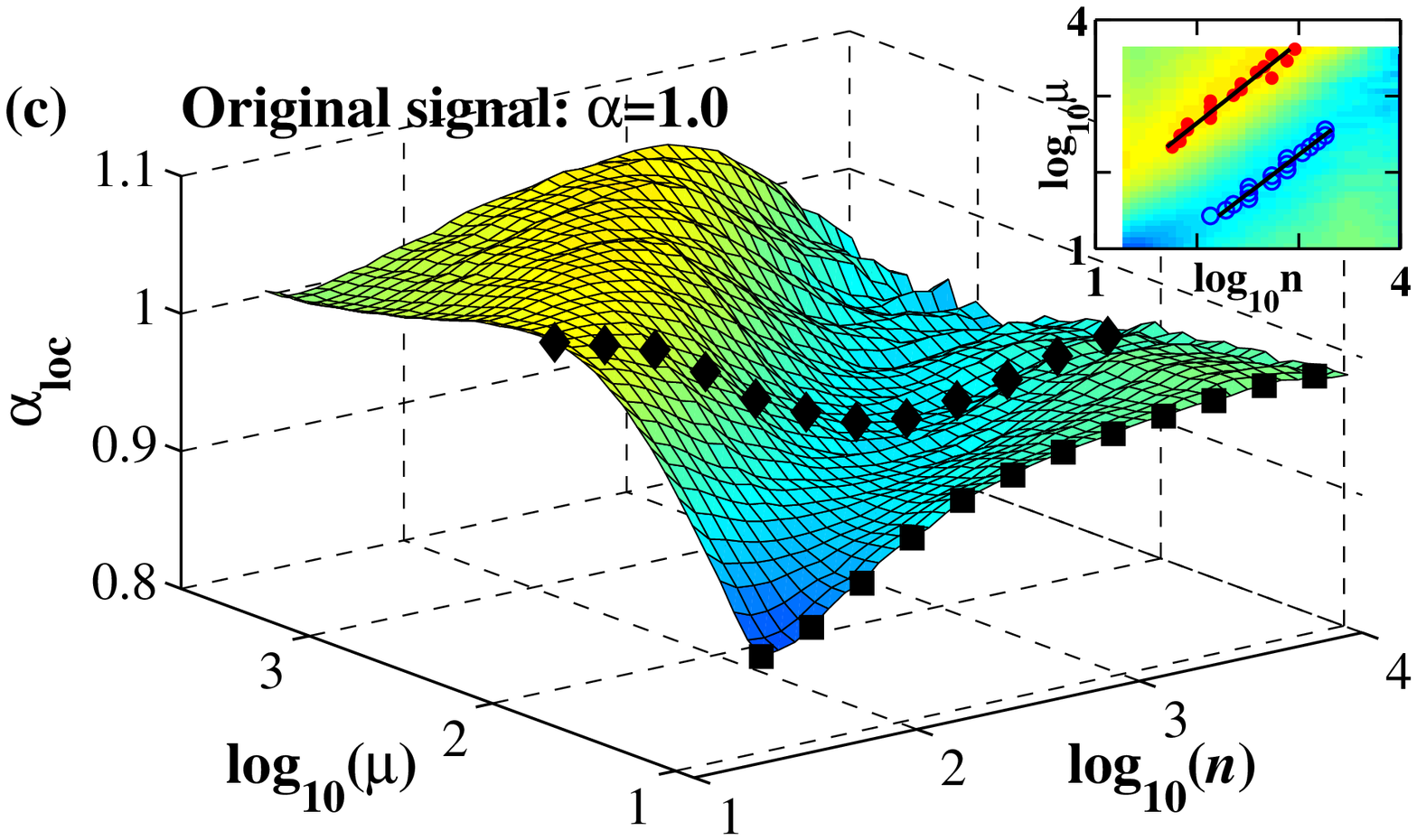}
\hspace{0.5cm}
\includegraphics[width=0.465\linewidth]{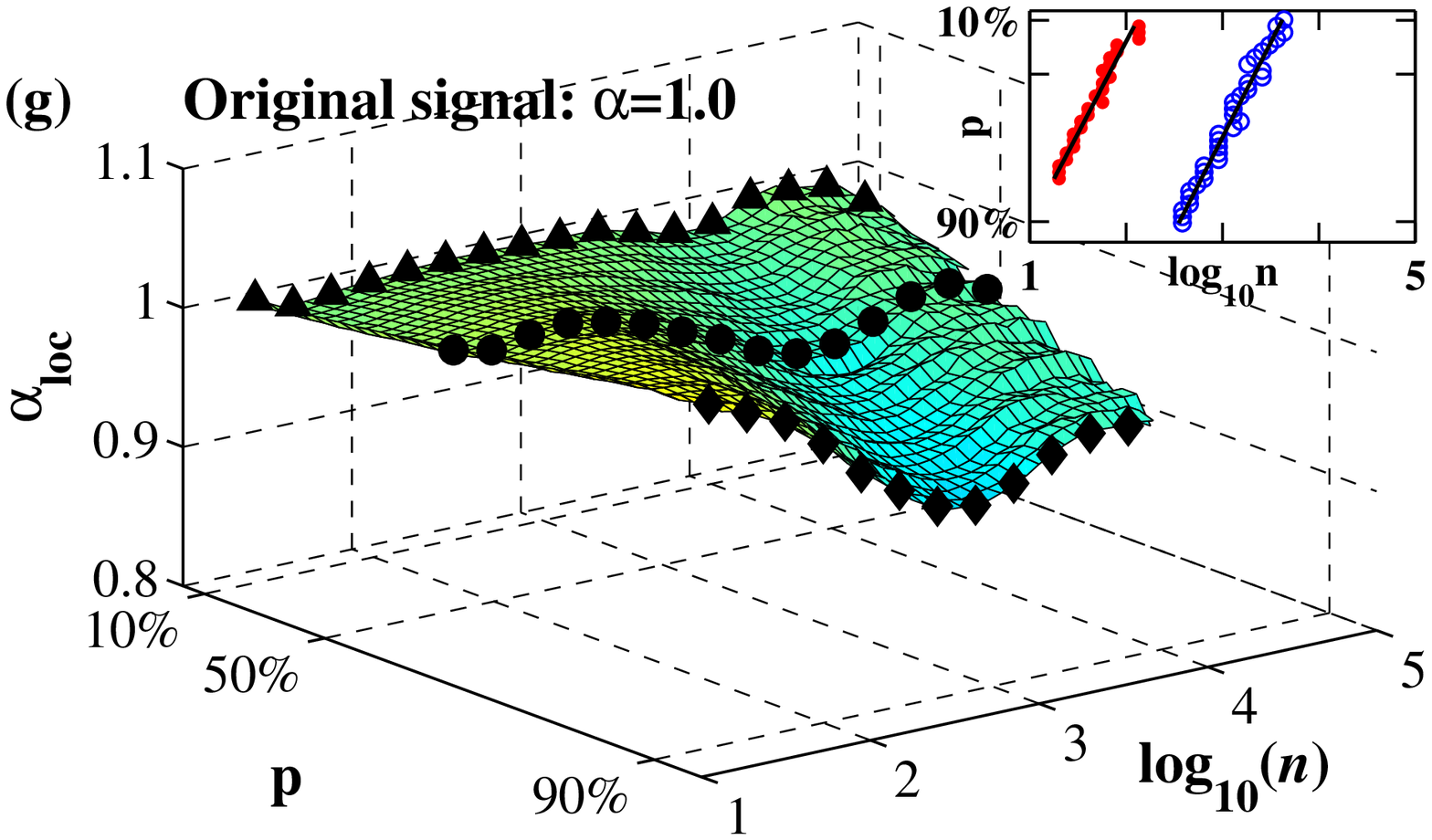}
}
\centering{
\includegraphics[width=0.465\linewidth]{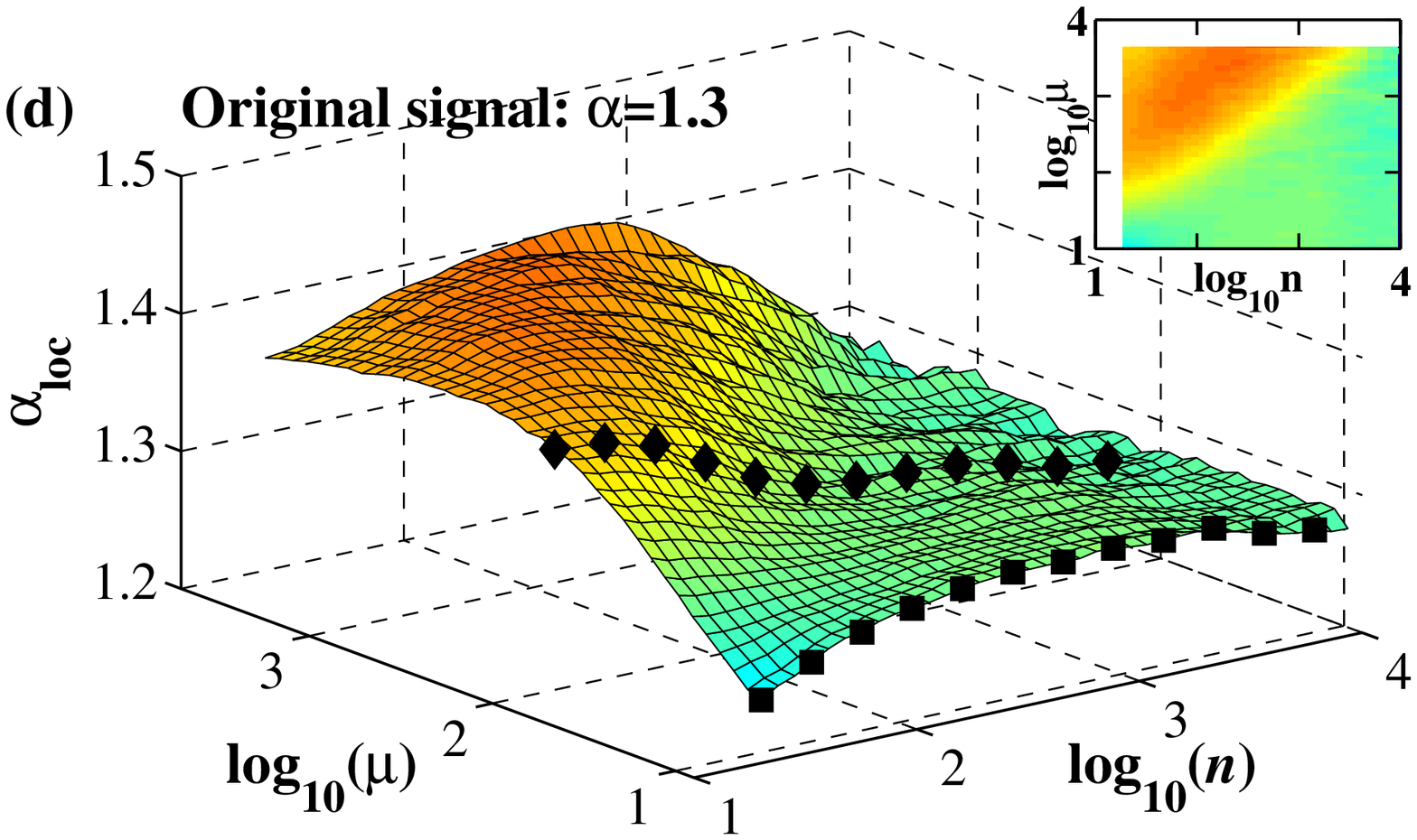}
\hspace{0.5cm}
\includegraphics[width=0.465\linewidth]{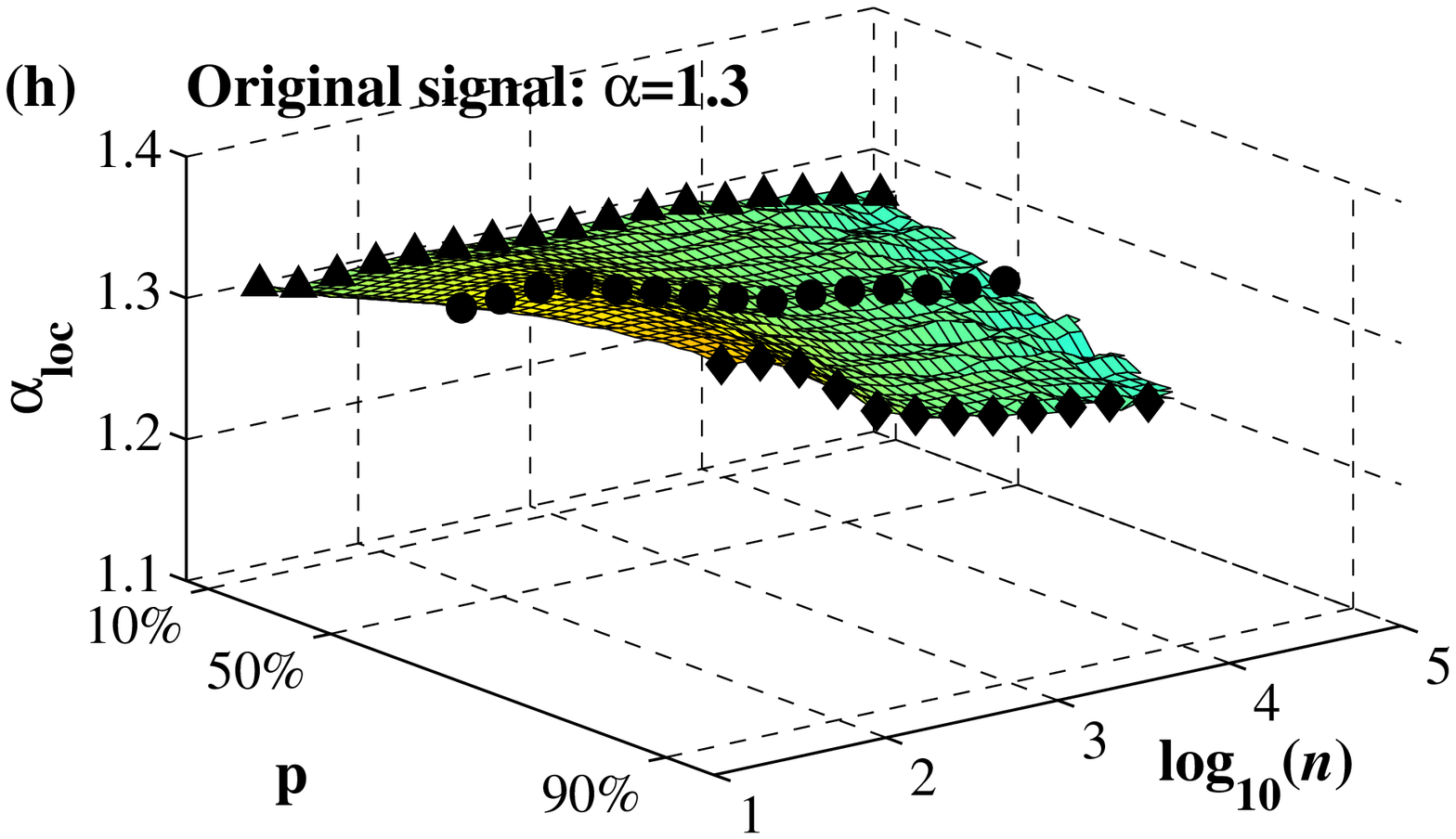}
}
\caption{(Color online) Effect of the average length $\mu$ of data
  loss segments (a)-(d) and effect of the percentage $p$ of data
  loss (e)-(h) on the local scaling behavior in anti-correlated
  signals [(a), (e): $\alpha=0.3$] and positively correlated signals
  [(b), (f): $\alpha=0.7$; (c), (g): $\alpha=1.0$; (d), (h):
  $\alpha=1.3$]. For (a)-(d), $p=90\%$ of data are removed, and for
  (e)-(h), the average length of removed segments $\mu=100$. In all
  the cases, the removed segments are exponentially distributed, and
  the length of the original signals $N=2^{20}$. To clearly see the
  power-law relation between the average length $\mu$ of removed
  segments and the scale $n$ at which $\alpha_{loc}$ achieves the same
  value, the $\alpha_{loc}$ values are projected into the
  $\log_{10}\mu$--$\log_{10}n$ plane (see color-coded insets in
  figures (a)--(d)).  The symbols in the inset figures in (c) and (g)
  indicate the positions where $\alpha_{loc}$ values reach a maximum
  (red closed circle) or a minimum (blue open circle), and depict the
  shift of the overestimated and underestimated regions to large
  scales with increasing $\mu$ and decreasing $p$. The local scaling
  curves highlighted by black symbols correspond to the curves shown
  in Fig.~\ref{fig-aloc-exp-t10} (rectangle: $\mu=10$, $p=90\%$;
  diamond: $\mu=100$, $p=90\%$; circle: $\mu=100$, $p=65\%$; triangle:
  $\mu=100$; $p=10\%$).}
\label{fig-aloc3d-exp-p90}
\end{figure*}

To understand precisely how the two parameters --- the average length
$\mu$ of the data loss segments and the percentage $p$ of data
loss --- influence changes in the local scaling behavior, in
Fig.~\ref{fig-aloc3d-exp-p90}a-d we show how $\alpha_{loc}$ changes with
the average length $\mu$ of the removed segments. For anti-correlated
signals, the scale at which $\alpha_{loc}$ reaches 0.5 monotonically
increases and shows a power-law relationship with $\mu$
(Fig.~\ref{fig-aloc3d-exp-p90}a). For positively correlated signals,
as shown in Fig.~\ref{fig-aloc3d-exp-p90}b-d, the overestimated
regions at small scales as well as the underestimated regions at large
scales are shifted to higher scales with increasing $\mu$. This shift
in the local scaling behavior also follows a power-law with average
length $\mu$ (Fig.~\ref{fig-aloc3d-exp-p90}c, inset).

In Fig.~\ref{fig-aloc3d-exp-p90}e-h, we show how the percentage $p$ of
data loss influence changes in the local scaling behavior. For a fixed
average length $\mu=100$, we find that the deviation from the original
scaling behavior is more pronounced for higher values of $p$ in both
anti-correlated and positively correlated signals, as also observed in
Fig.~\ref{fig-aloc-exp-t10}. The scaling behavior of positively
correlated signals also shows overestimated regions at small scales
and underestimated regions at large scales
(Fig.~\ref{fig-aloc3d-exp-p90}f-h), although not as clear as in
Fig.~\ref{fig-aloc3d-exp-p90}b-d. Both regions are shifted to larger
scales with decreasing percentage of data loss as illustrated in the
inset in Fig.~\ref{fig-aloc3d-exp-p90}g.

\begin{figure}
\centering{
\includegraphics[width=1.0\linewidth]{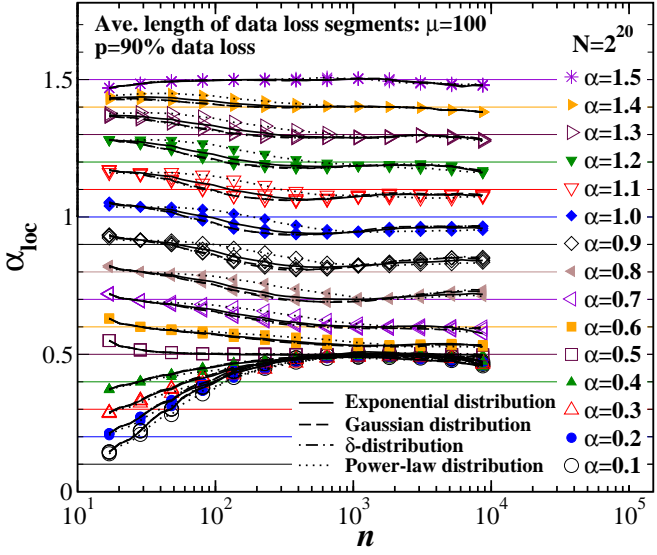}
}
\vspace{-0.5cm}
\caption{(Color online) Effect of different kinds of distributions of
  data loss segments on the local scaling behavior. The power-law
  distributed data loss segments lead to higher values of
  $\alpha_{loc}$ for positively correlated signals and lower values
  for anti-correlated signals compared to the other
  distributions. There is no difference between Gaussian and
  $\delta$-distributed segments which yield slightly lower
  $\alpha_{loc}$ values than exponentially distributed signals. For
  anti-correlated signals, exponentially, Gaussian and
  $\delta$-distributed segments lead to identical $\alpha_{loc}$
  values whereas the power-law distribution yields slightly lower
  local scaling exponents.}
\label{fig-aloc-vary-t100-p90}
\end{figure}

\begin{figure}
\centering{
\includegraphics[width=0.93\linewidth]{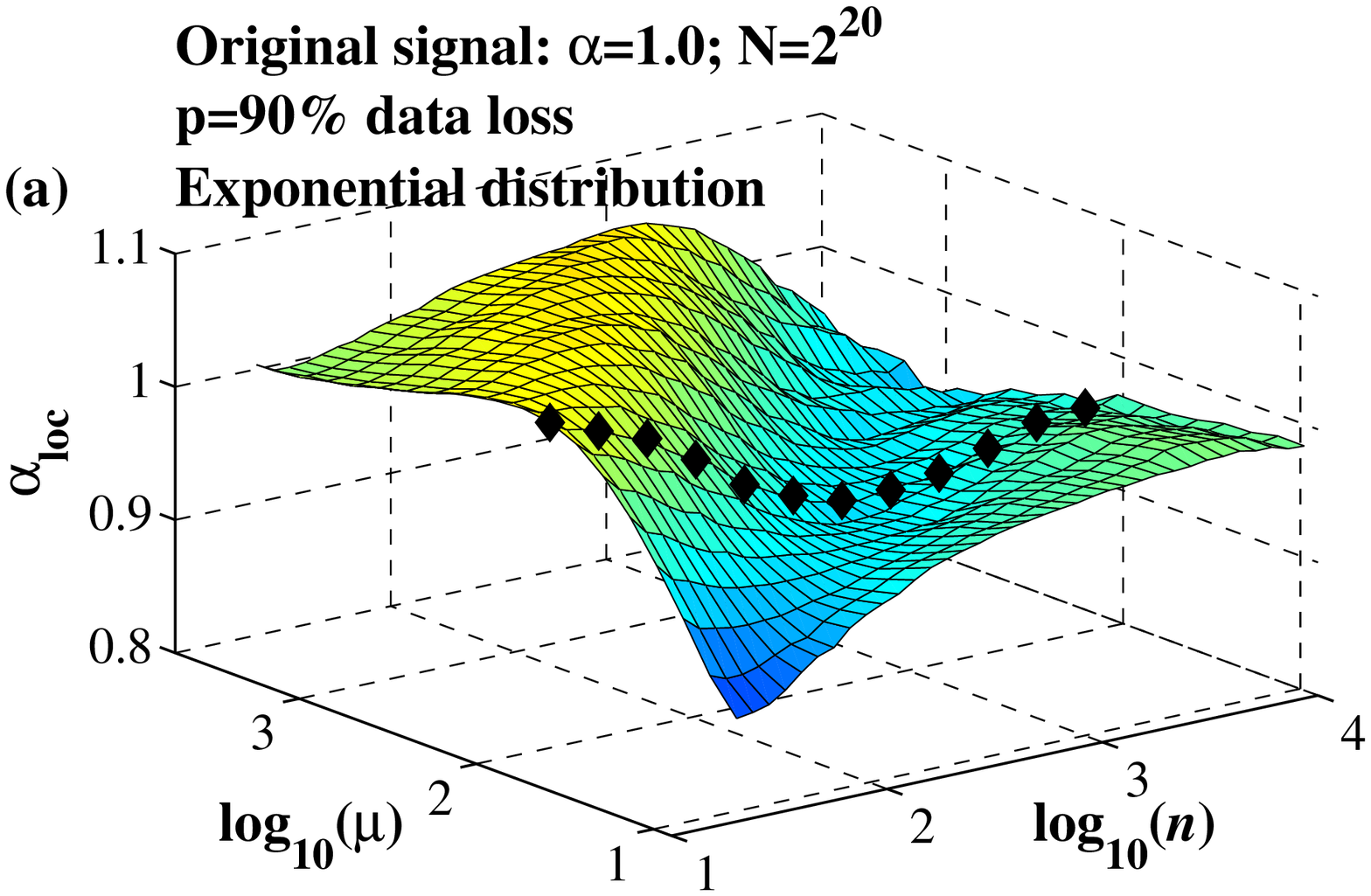}
}
\centering{
\includegraphics[width=0.93\linewidth]{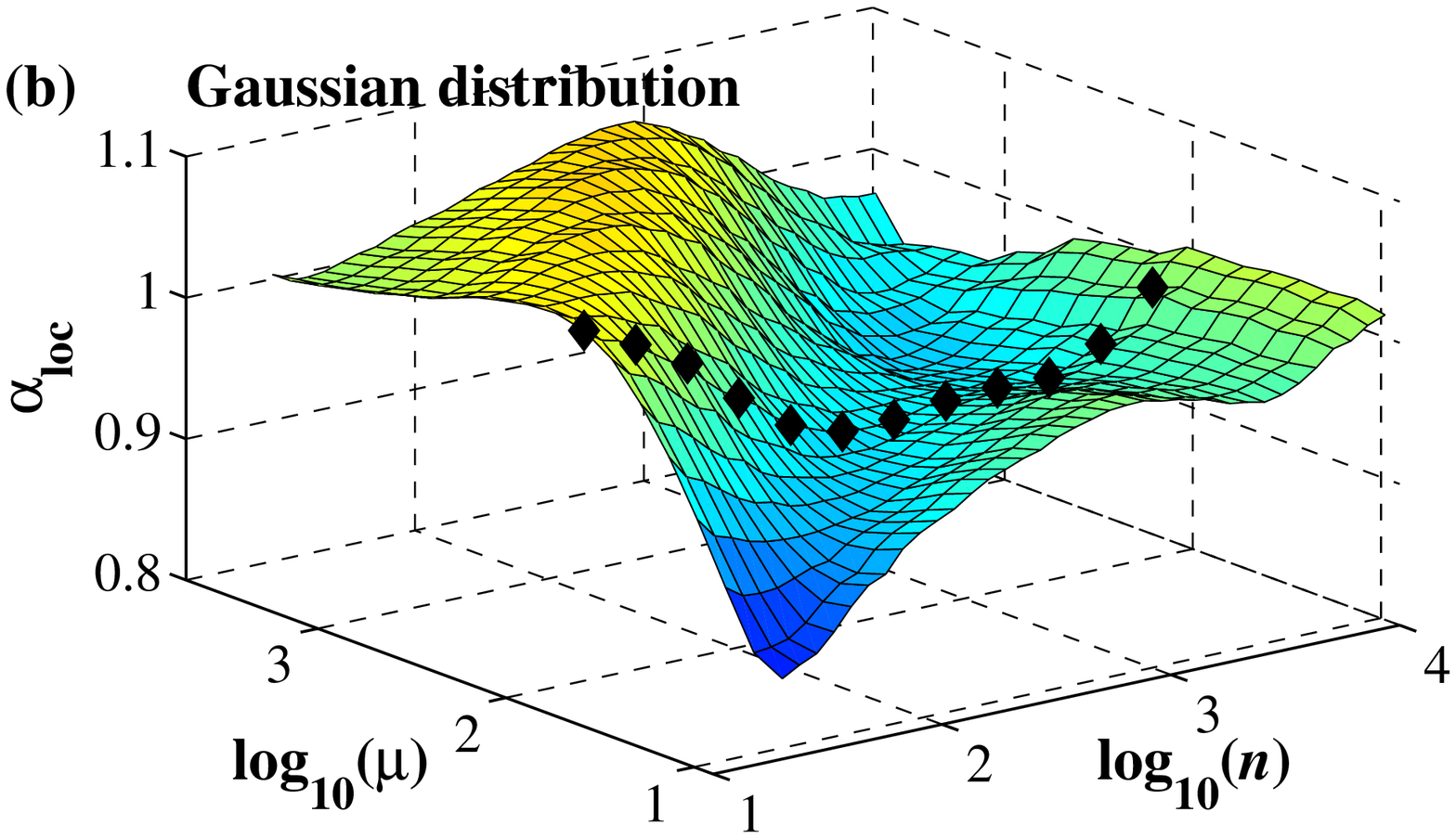}
}
\centering{
\includegraphics[width=0.93\linewidth]{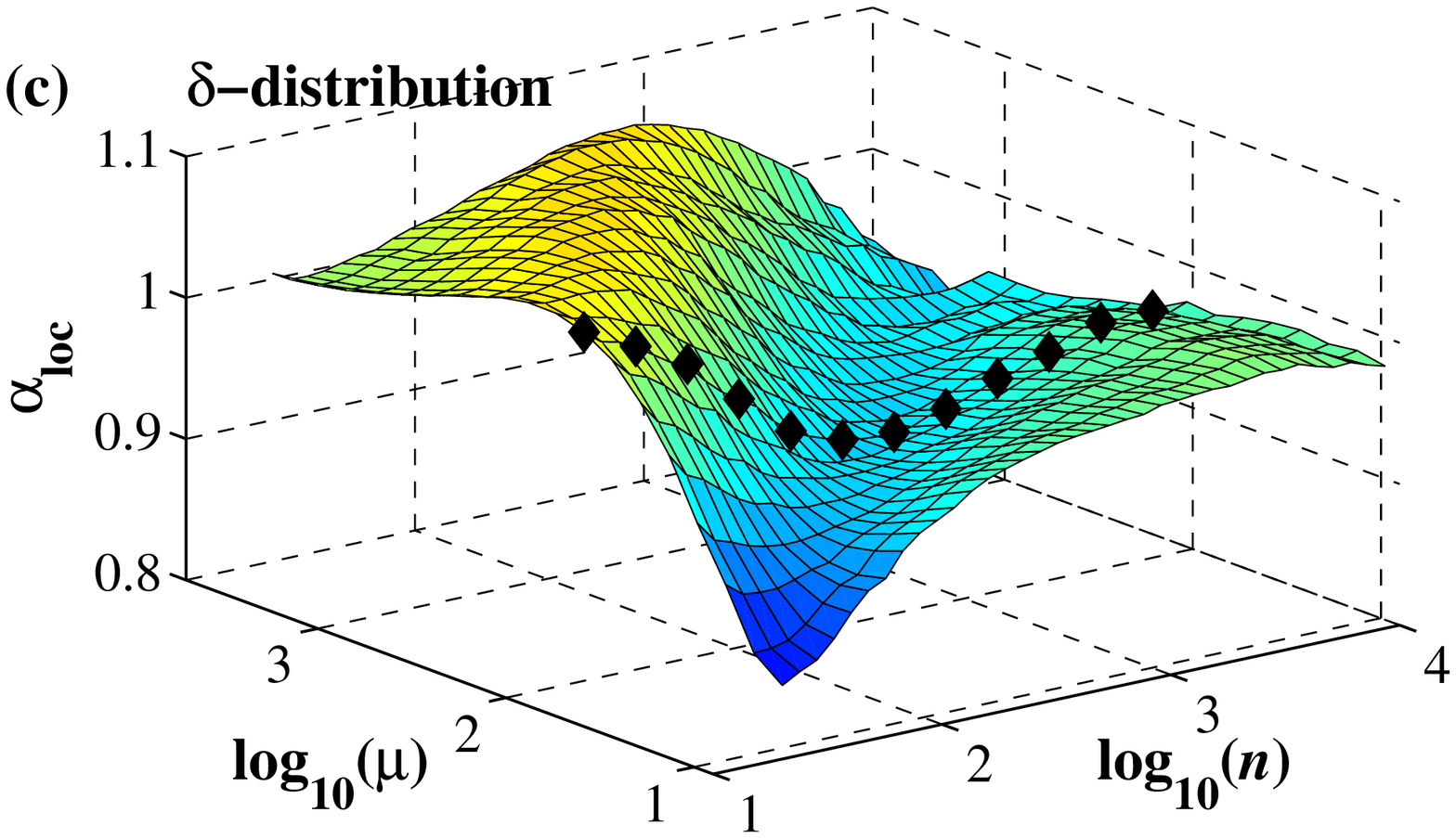}
}
\centering{
\includegraphics[width=0.93\linewidth]{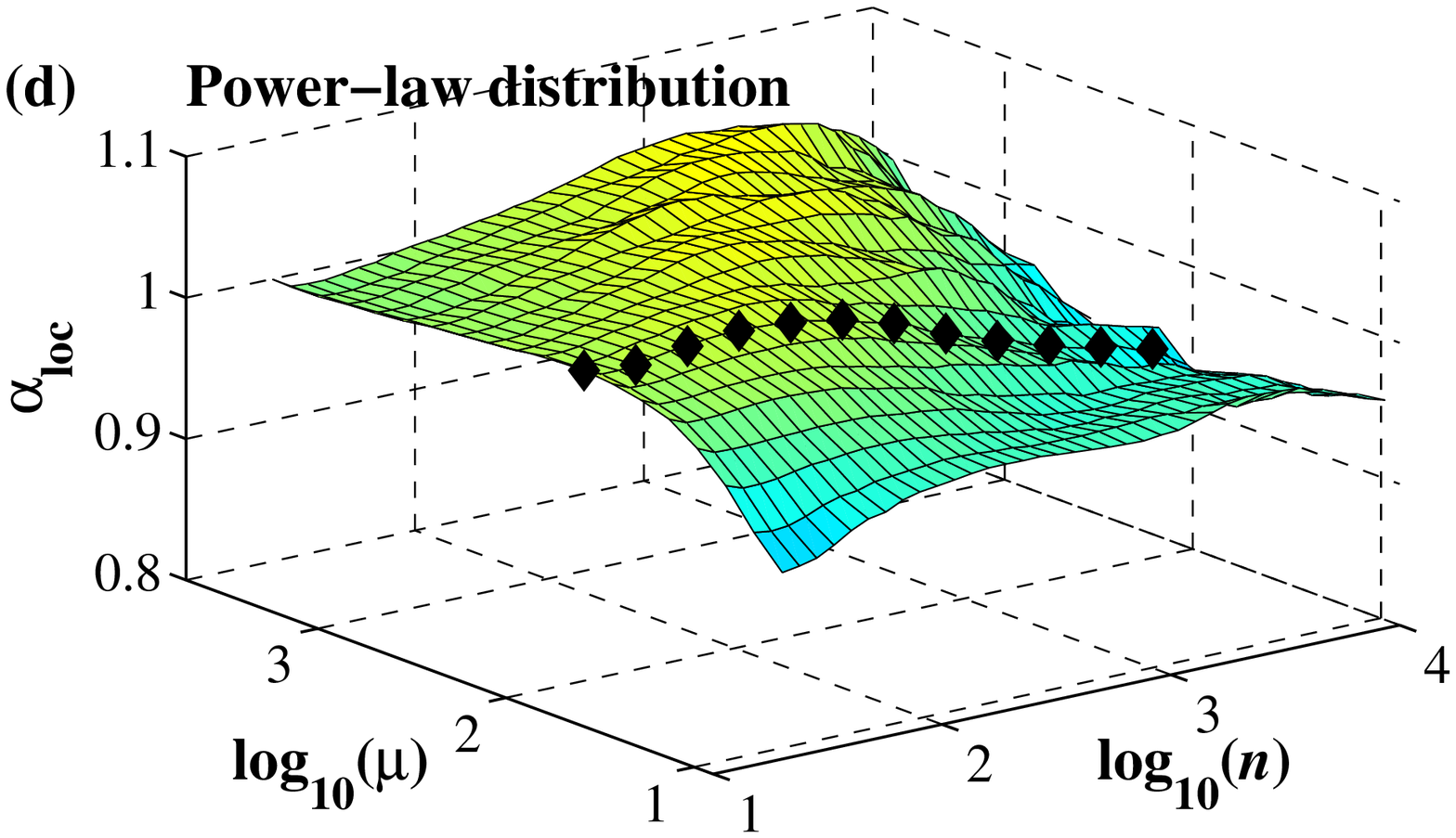}
}
\caption{(Color online) Effect of the average length $\mu$ of data
  loss segments on the local scaling behavior in long-range correlated
  signal with $\alpha=1.0$. The length of the data loss segments are
  (a) exponentially distributed, (b) Gaussian distributed, (c)
  $\delta$-distributed and (d) power-law distributed. In all the
  cases, $p=90\%$ of data are removed, and the length of the original
  signals $N=2^{20}$. The behavior of how $\alpha_{loc}$ changes with
  $\mu$ is similar for exponential, Gaussian and
  $\delta$-distribution, while the power-law distribution shows less
  variations. The local scaling curves highlighted by black symbols
  correspond to the curves shown in
  Fig.~\ref{fig-aloc-vary-t100-p90}.}
\label{fig-aloc3d-noise10-gap-p90}
\end{figure}

\begin{figure}
\centering{
\includegraphics[width=0.93\linewidth]{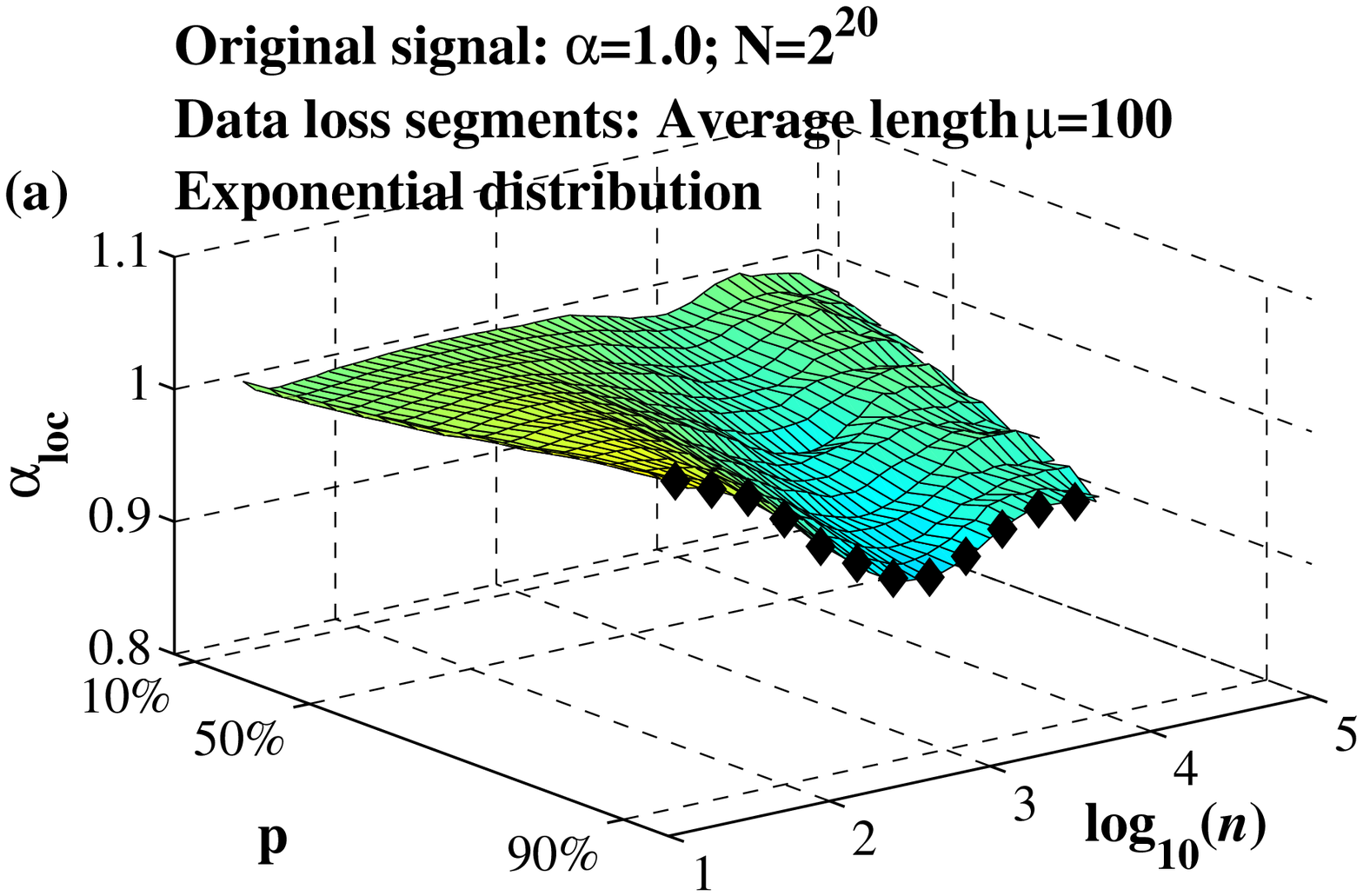}
}
\centering{
\includegraphics[width=0.93\linewidth]{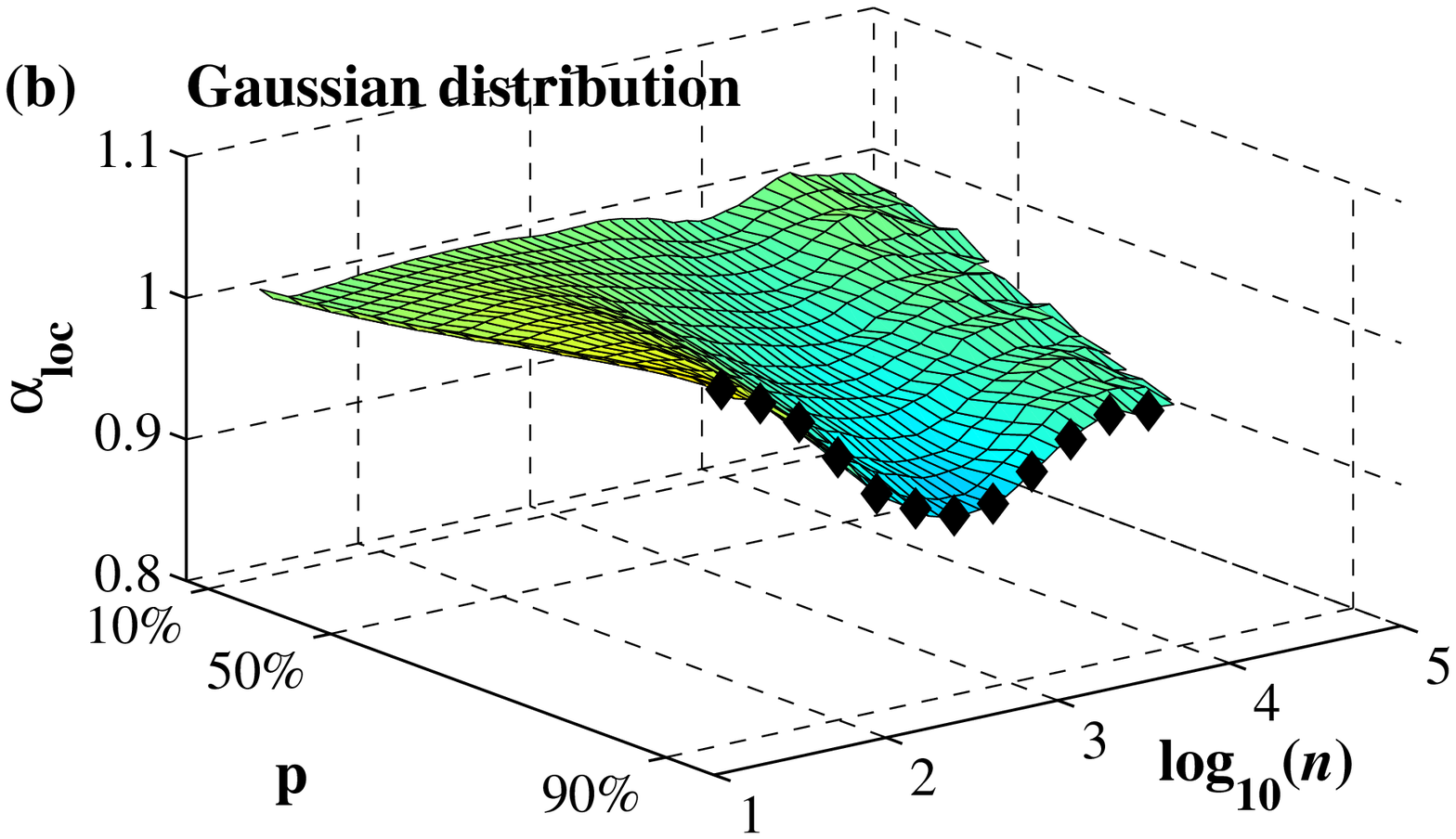}
}
\centering{
\includegraphics[width=0.93\linewidth]{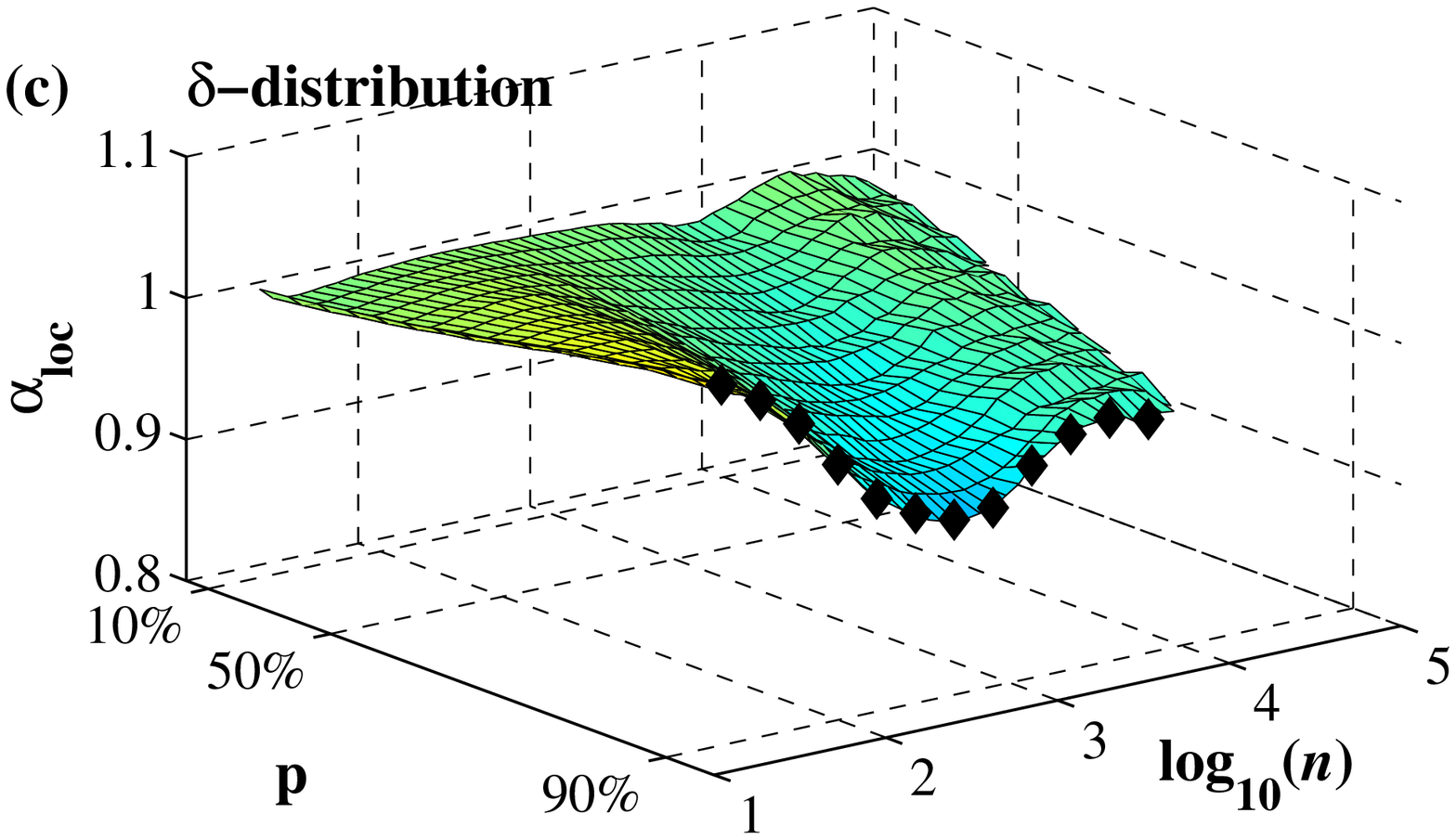}
}
\centering{
\includegraphics[width=0.93\linewidth]{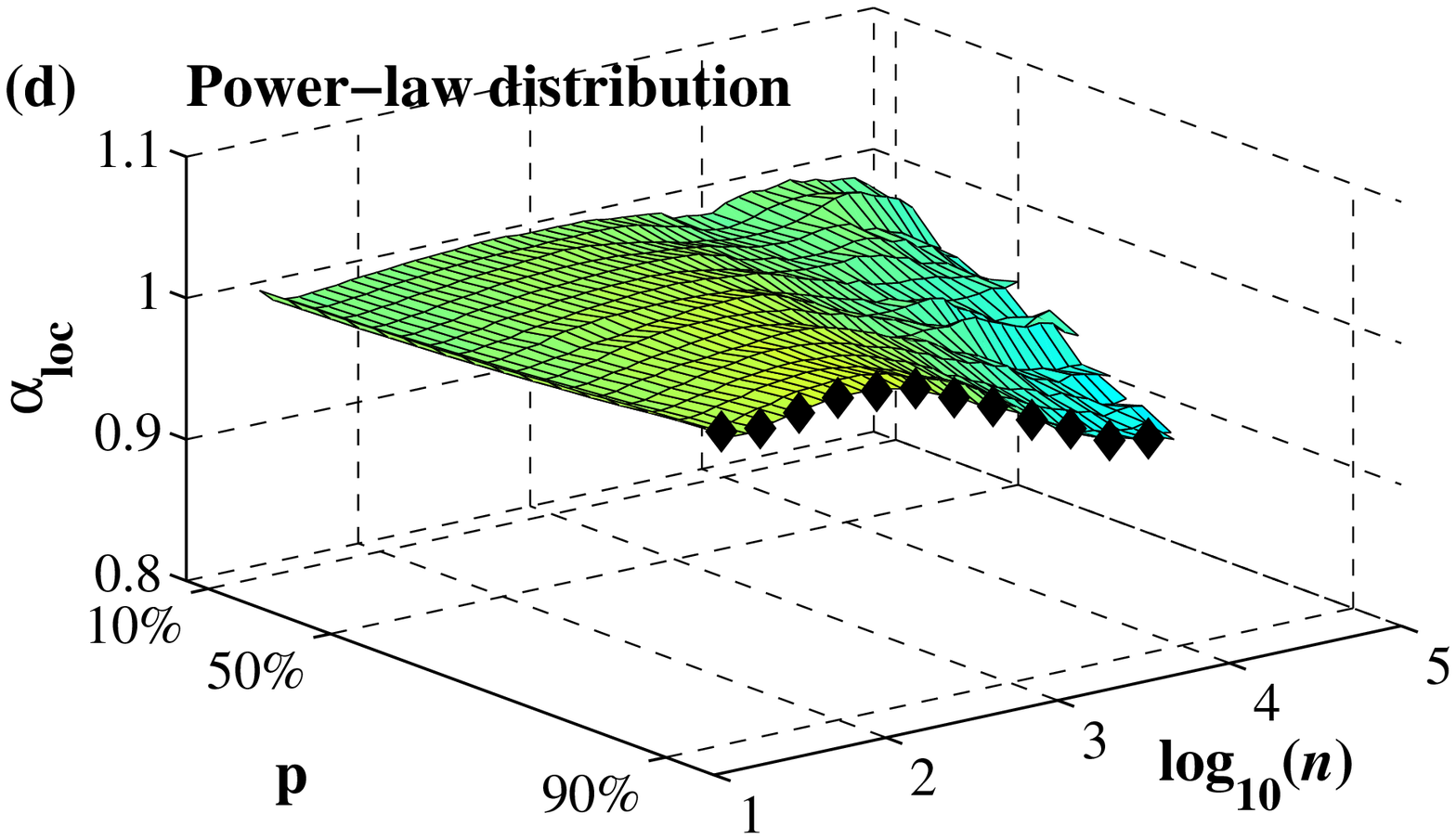}
}
\caption{(Color online) Effect of the percentage $p$ of data
  loss on the local scaling behavior in long-range correlated signal
  with $\alpha=1.0$. The length of the data loss segments are (a)
  exponentially distributed, (b) Gaussian distributed, (c)
  $\delta$-distributed and (d) power-law distributed. In all the
  cases, the average length of removed segments $\mu=100$, and the
  length of the original signals $N=2^{20}$. Similar to
  Fig.~\ref{fig-aloc3d-noise10-gap-p90}, the exponential, Gaussian and
  $\delta$-distributions show similar changes in $\alpha_{loc}$ with
  $p$, while the power-law distribution shows less variations. The
  local scaling curves highlighted by black symbols correspond to the
  curves shown in Fig.~\ref{fig-aloc-vary-t100-p90}.}
\label{fig-aloc3d-noise10-gap-t100}
\end{figure}

To understand whether different functional forms of distributions of
data loss segments have different effects on the scaling behavior, we
repeated the same tests with three other kinds of distributions: a
Gaussian distribution, a $\delta$-distribution (i.e., segments have
fixed length) and a power-law distribution. We find that all three
kinds of distributions show similar deviations from the original local
scaling behavior as reported above for exponentially distributed data
loss segments. However, for power-law distributed segments lengths,
the estimated local scaling exponents are generally higher (lower)
across scales for positively (anti-) correlated signals
(Fig.~\ref{fig-aloc-vary-t100-p90}). When increasing the average
length $\mu$ of the removed data segments or increasing the percentage
$p$ of data loss, the power-law distribution shows less variations
than the other three kinds of distributions
(Fig.~\ref{fig-aloc3d-noise10-gap-p90} and
Fig.~\ref{fig-aloc3d-noise10-gap-t100}).

\subsection{Properties of remaining data segments: Effect of data loss
  on local scaling}\label{secresultseg}

In the previous section, we tested the effect of data loss by
specifying the distribution and average length of {\it removed}
segments. In this section, we study the effect of data loss by
specifying the distribution and average length of {\it remaining} data
segments. The results obtained by focusing on the properties of
remaining data segments are different from what was shown above
and will lead to a better understanding of the effect of data loss
on the scaling behavior of long-range correlated signals.

\begin{figure}
\centering{
\includegraphics[width=1\linewidth]{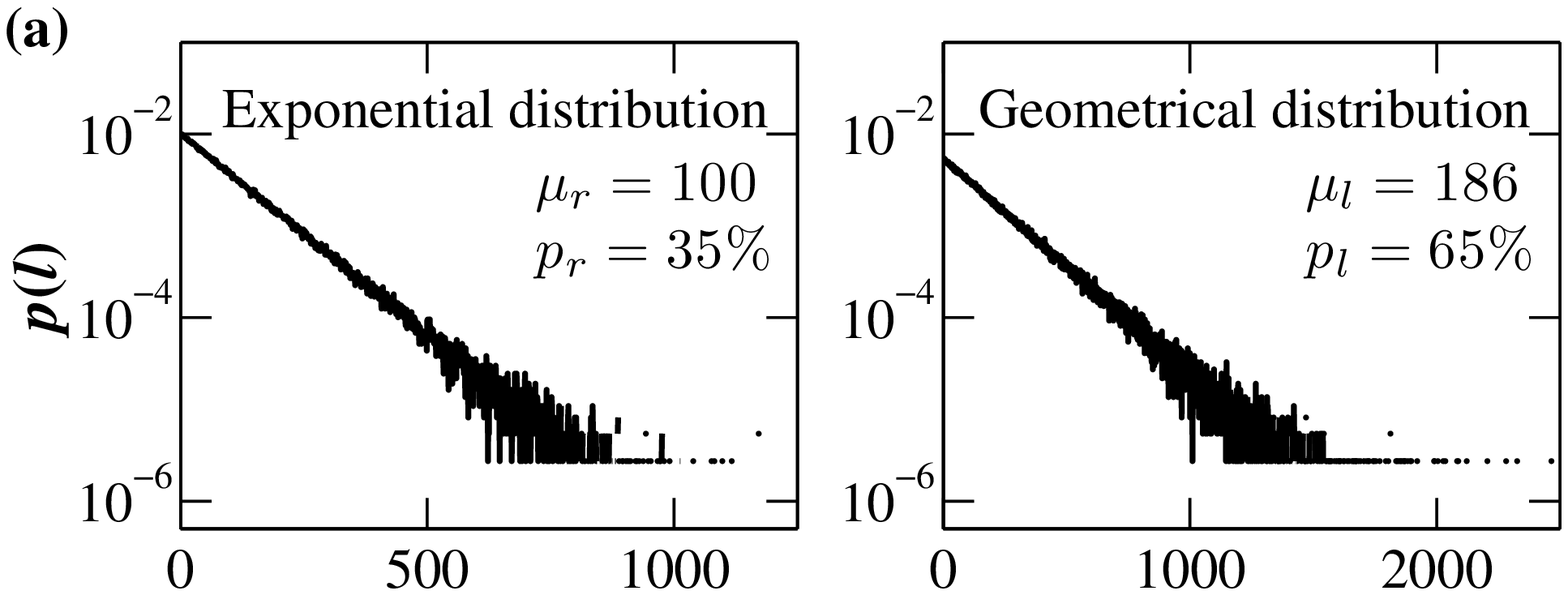}
}
\centering{
\includegraphics[width=1\linewidth]{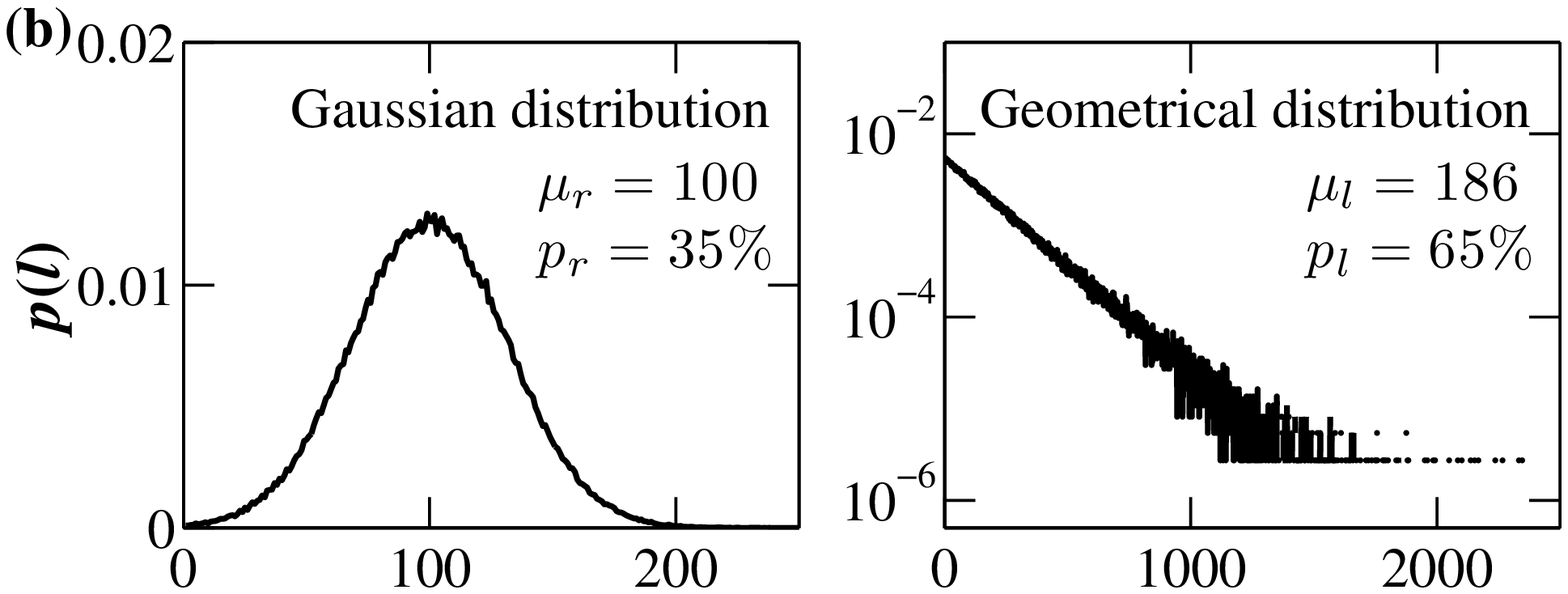}
}
\centering{
\includegraphics[width=1\linewidth]{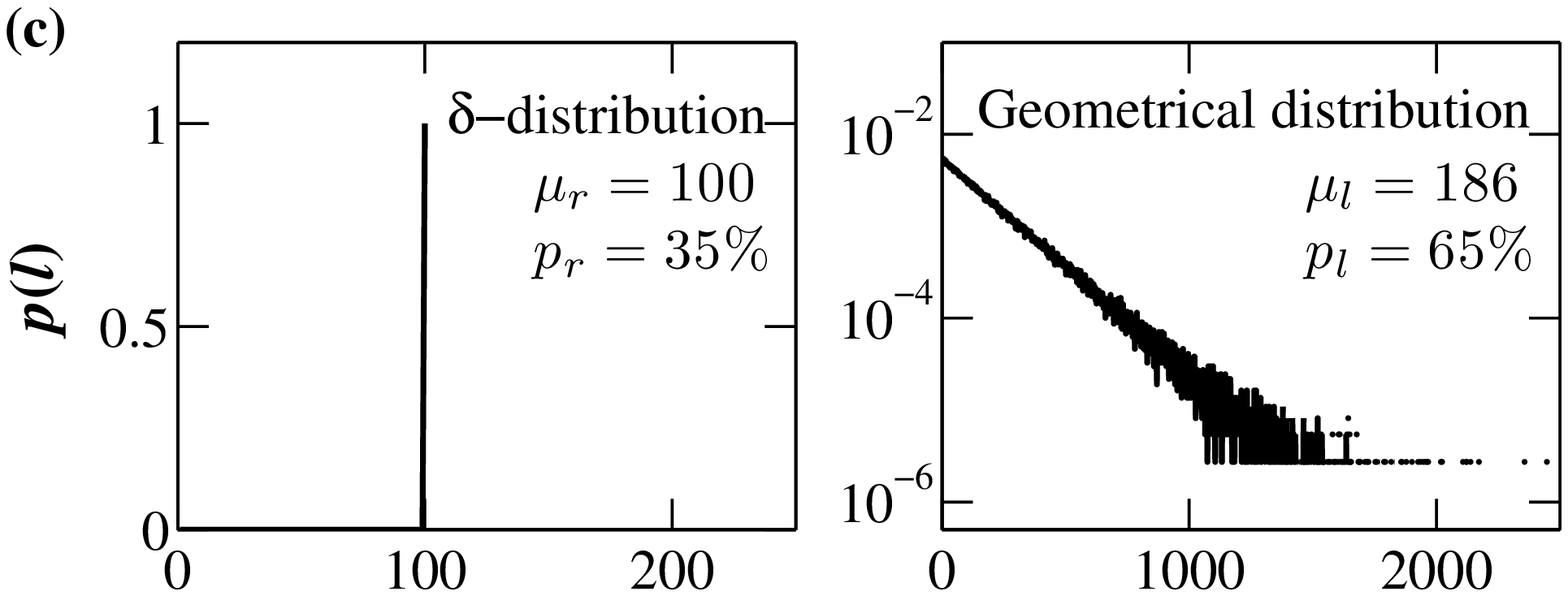}
}
\centering{
\includegraphics[width=1\linewidth]{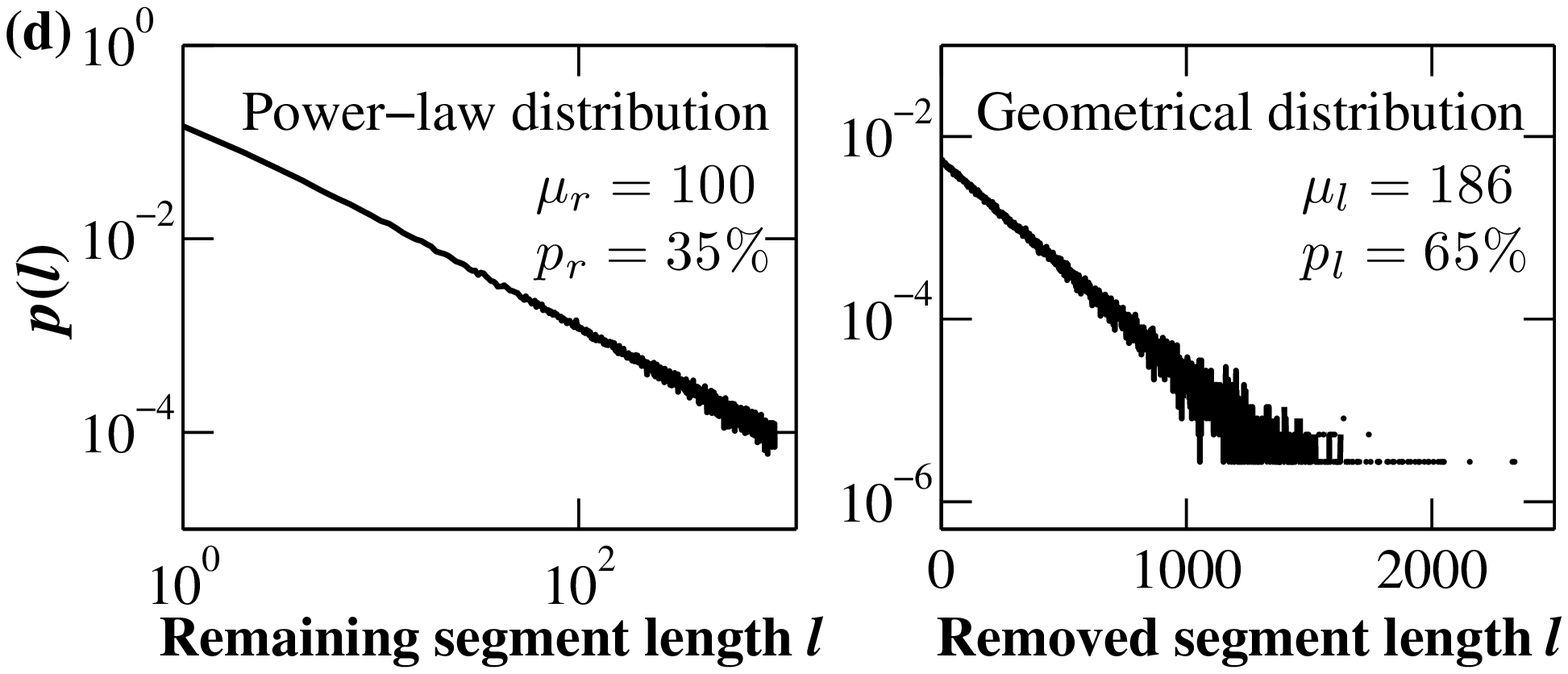}
}
\vspace{-0.5cm}
\caption{The distributions of remaining data segments (left column)
  and corresponding distributions of data loss segments (right
  column). The remaining data segments follow (a) exponential, (b)
  power-law, (c) Gaussian, and (d) $\delta$-distribution with average
  length $\mu_r$=100 and 35\% of data remaining. The data loss
  segments are always geometrically distributed independent of the
  distributions of remaining segments. Note that, the average lengths
  are practically the same as estimated from Eq.~\ref{equ-mu}.}
\label{fig-pdf-seg-gap}
\end{figure}

The approach to generate the appropriate surrogate signals with
different properties of remaining data segments is similar to the one
described in Sec.~\ref{secsegm}, except that now the binary series
$g(i)$ are obtained according to the parameters of the remaining data
segments, and the surrogate signals $\tilde{u}(i)$ are generated by
removing the $i$-th data point in the original signal $u(i)$ if
$g(i)=1$, and preserving the $i$-th data point if $g(i)=0$. The
relation between the average length of data loss segments ($\mu_{l}$)
and remaining data segments ($\mu_{r}$) can be derived as follows:

Let the length of the original signal be $N$. If $p_l$ is the percentage
of data loss, the amount of data loss is given by $N_l=p_lN$, and
the amount of remaining data is given by $N_r=p_rN=(1-p_l)N$. If
$\mu_l$ is the average length of the lost data segments, the number of
lost segments is approximately given by $n_l~\approx~N_l/\mu_l$. The
number of remaining data segments is approximately equal to the number
of data loss segments, i.e., $n_r~\approx~n_l$. Hence, the average
length of the remaining data segments is:
\begin{equation}
\mu_{r}~\approx~\frac{N_{r}}{n_{r}}~\approx~\frac{(1-p_l)}{p_l}\mu_{l}.
\label{equ-mu}
\end{equation}
Note that the lengths of data loss segments are always geometrically
distributed due to the shuffling procedure in our segmentation approach
(see Sec.~\ref{secsegm} and Fig.~\ref{fig-pdf-seg-gap}).

\begin{figure*}
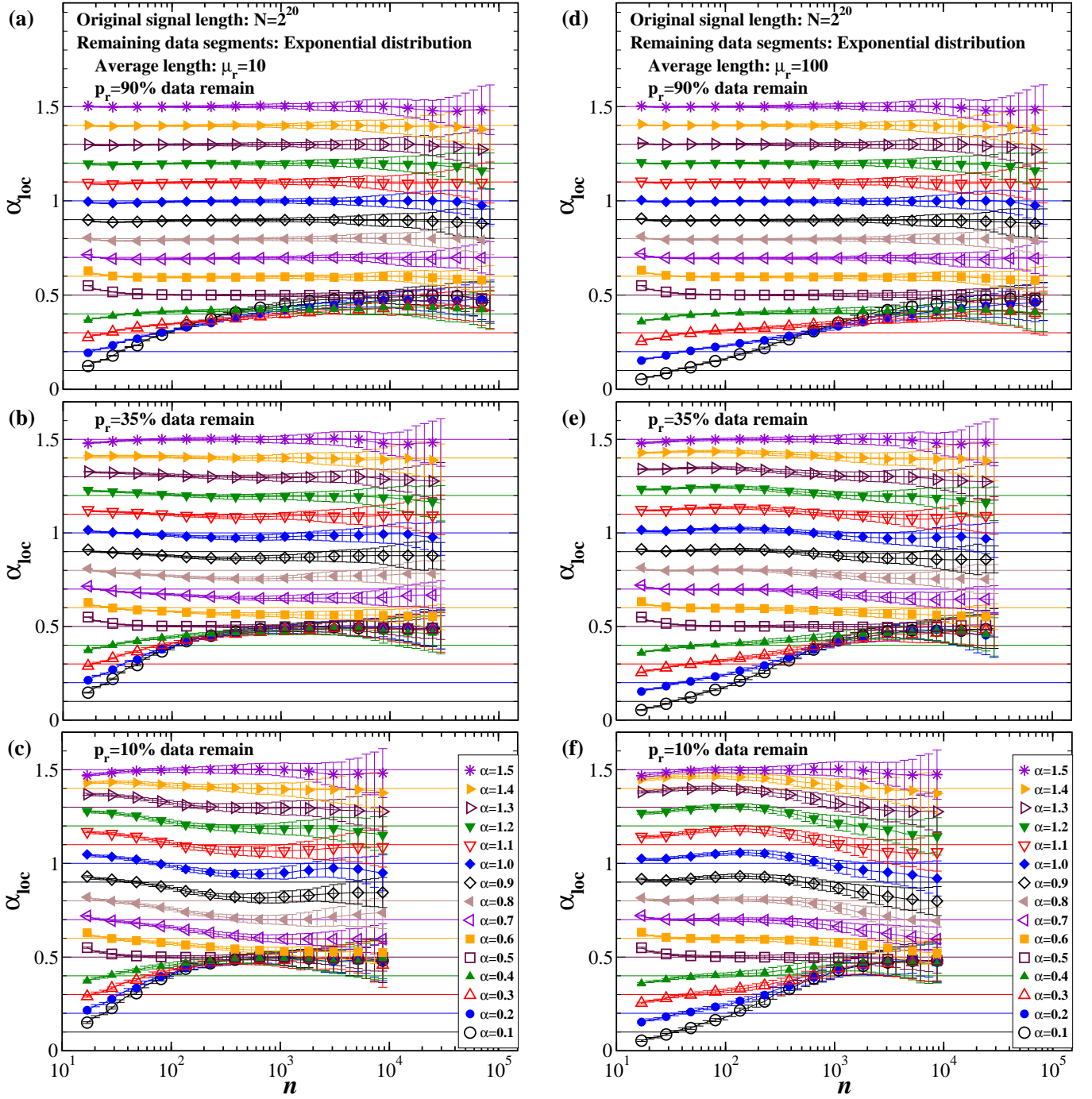

\centering{
\includegraphics[width=0.48\linewidth]{EPSfigures/aloc_seg_exp_t10_p90.eps}
\hspace{0.5cm}
\includegraphics[width=0.48\linewidth]{EPSfigures/aloc_seg_exp_t100_p90.eps}
}
\centering{
\includegraphics[width=0.48\linewidth]{EPSfigures/aloc_seg_exp_t10_p35.eps}
\hspace{0.5cm}
\includegraphics[width=0.48\linewidth]{EPSfigures/aloc_seg_exp_t100_p35.eps}
}
\centering{
\includegraphics[width=0.481\linewidth]{EPSfigures/aloc_seg_exp_t10_p10.eps}
\hspace{0.5cm}
\includegraphics[width=0.481\linewidth]{EPSfigures/aloc_seg_exp_t100_p10.eps}
}
\caption{(Color online) Effect of data loss on the local scaling
  behavior of long-range correlated signals. The lengths of the
  remaining data segments are exponentially distributed with average
  length $\mu_r=10$ ((a)-(c)) and $\mu_r=100$ ((d)-(f)). The symbols
  indicate average $\alpha_{loc}$ values obtained from 100 different
  realizations of surrogate signals with the same correlation exponent
  $\alpha$, and the error bars show the standard deviations. The more
  data are removed, the more the scaling exponent deviates from the
  original exponent. For anti-correlated signals, the removal of
  larger segments ($\mu_r=100$) has less effect on the scaling
  behavior. For positively correlated signals, the deviations vary
  across scales, showing both overestimated and underestimated
  regions.}
\label{fig-aloc-seg-exp-t10}
\end{figure*}

We find similar changes in the scaling behavior as observed in
Fig.~\ref{fig-aloc-exp-t10} where the distribution of removed segment
lengths was specified. As illustrated in
Fig.~\ref{fig-aloc-seg-exp-t10} where the lengths of remaining
segments are exponentially distributed, the local scaling behavior of
anti-correlated surrogate signals deviate monotonically from original
behavior towards uncorrelation at larger scales. While the local
scaling exponents of positively correlated surrogate signals vary
across scales, showing both overestimated and underestimated
regions. These regions as well as the scales at which the
anti-correlated signals reach $\alpha_{loc}=0.5$ are also shifted
towards larger scales when the average length of remaining segments
$\mu_r$ increases. However, in contrast to what was observed in
Fig.~\ref{fig-aloc-exp-t10}, there is no shift to smaller scales with
increasing percentage of data loss. Note that, according to
Eq.~\ref{equ-mu}, an average length $\mu_r=10$ of remaining segments
and a percentage $p_r=10\%$ of remaining data (as shown in
Fig.~\ref{fig-aloc-seg-exp-t10}c), corresponds to an average length
$\mu_l=90$ of removed segments and a percentage $p_l=90\%$ of removed
data. Thus the local scaling behavior observed in
Fig.~\ref{fig-aloc-seg-exp-t10}c is vary similar to
Fig.~\ref{fig-aloc-exp-t10}g (where $\mu_l=100$ and $p_l=90\%$), and
Fig.~\ref{fig-aloc-seg-exp-t10}d ($\mu_r=100$, $p_r=90\%$, $\mu_l=11$)
is similar to Fig.~\ref{fig-aloc-exp-t10}a ($\mu_l=10$, $p_l=10\%$).

\begin{figure*}
\centering{
\includegraphics[width=0.465\linewidth]{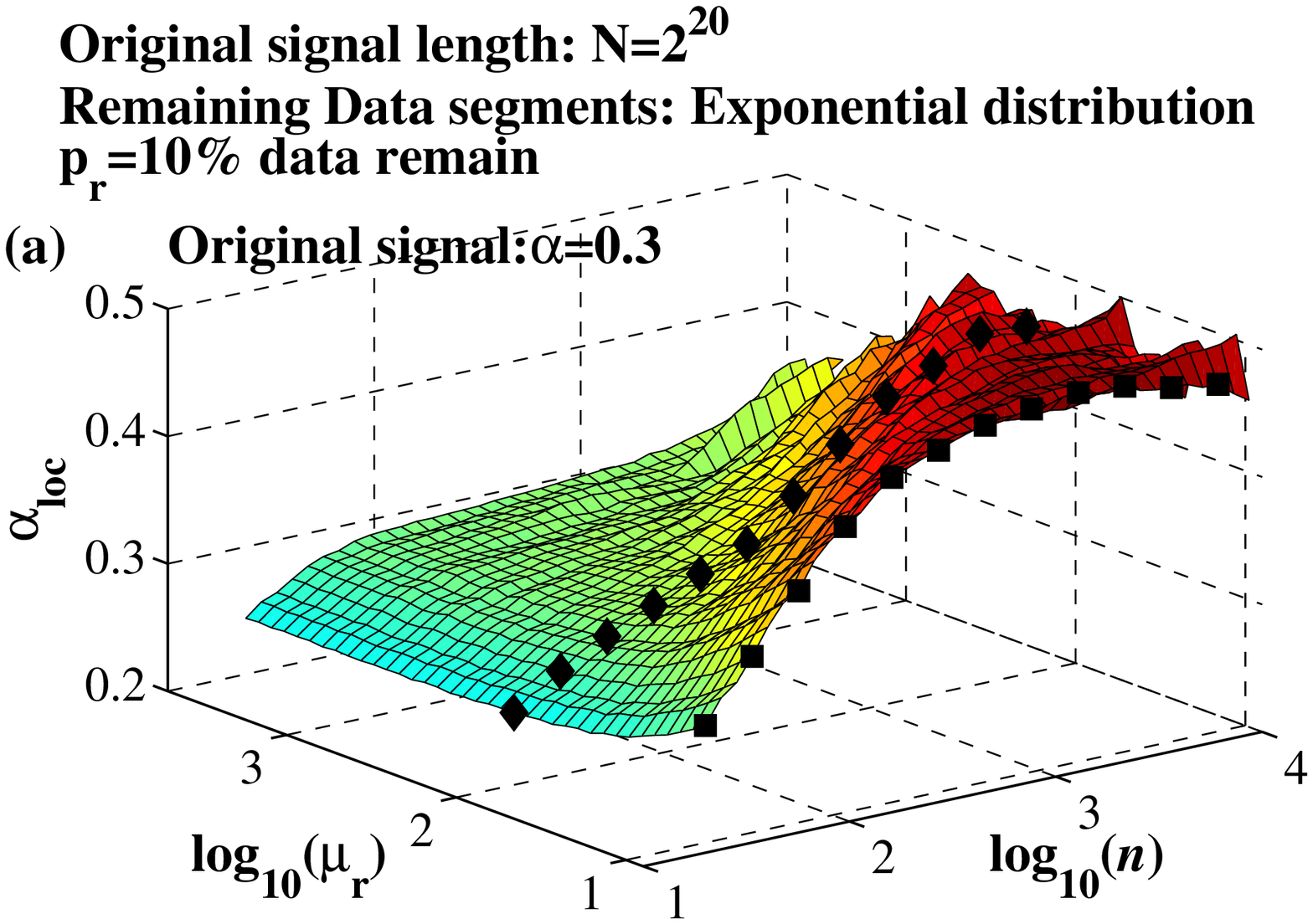}
\hspace{0.5cm}
\includegraphics[width=0.465\linewidth]{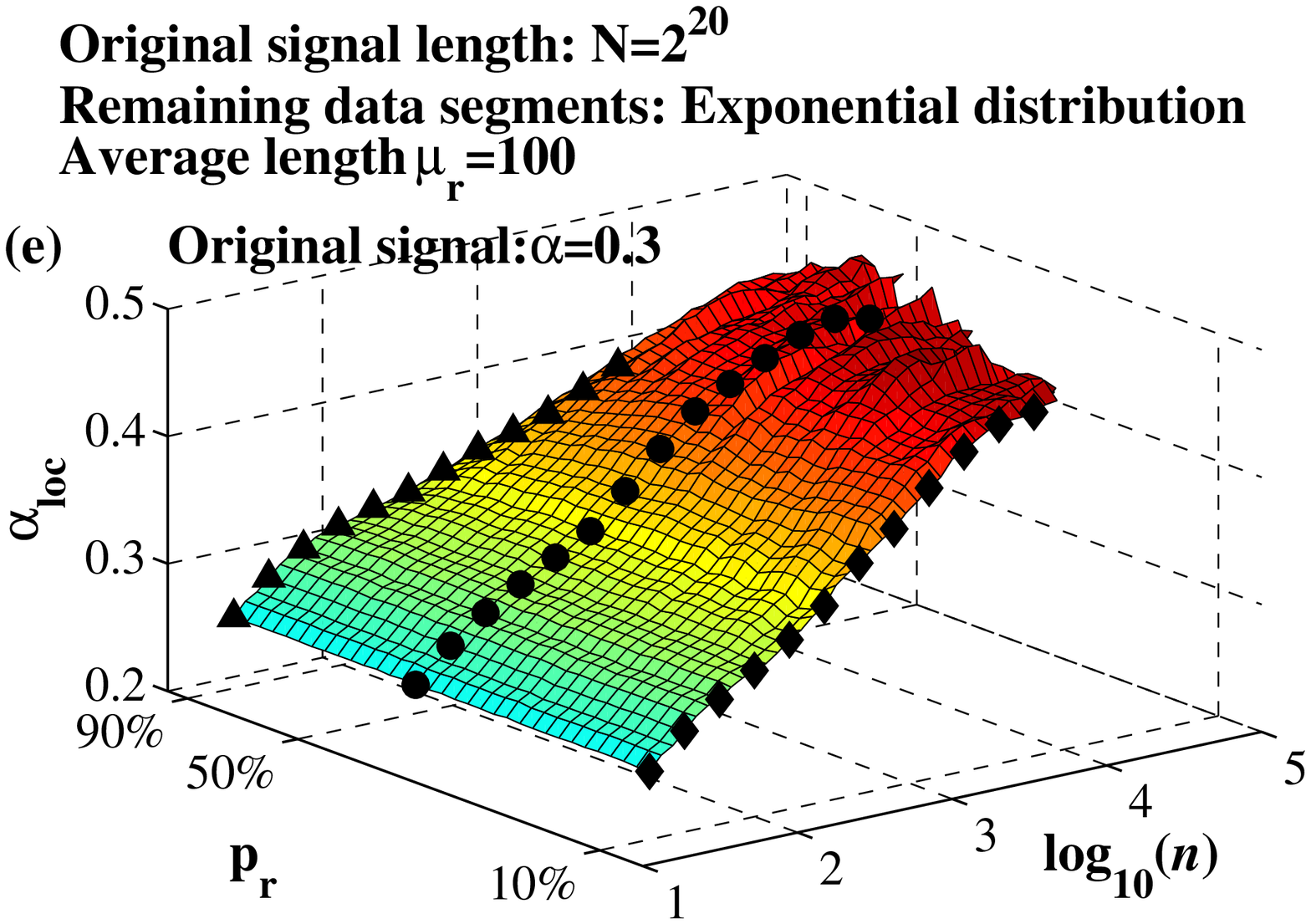}
}
\centering{
\includegraphics[width=0.465\linewidth]{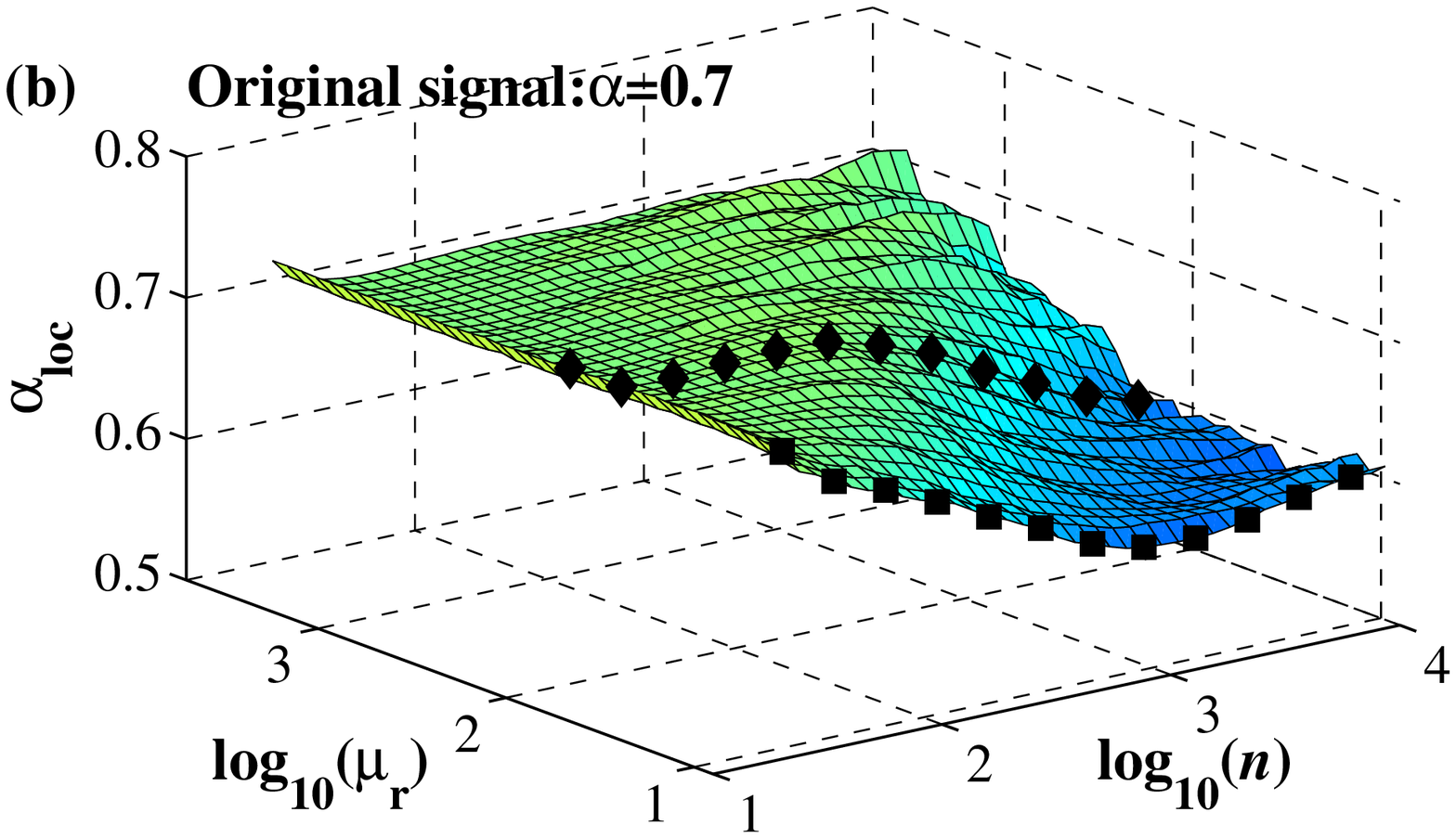}
\hspace{0.5cm}
\includegraphics[width=0.465\linewidth]{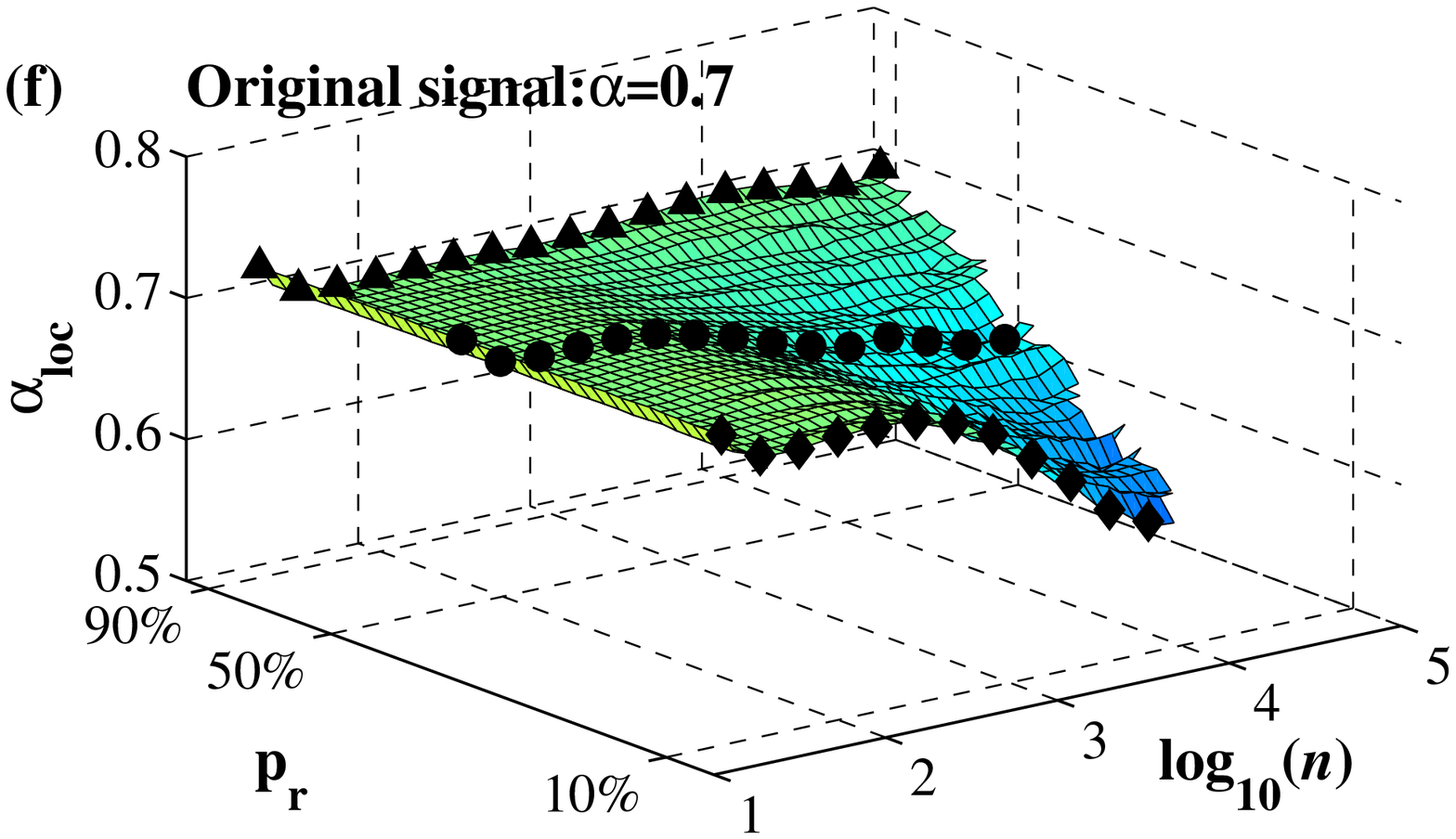}
}
\centering{
\includegraphics[width=0.465\linewidth]{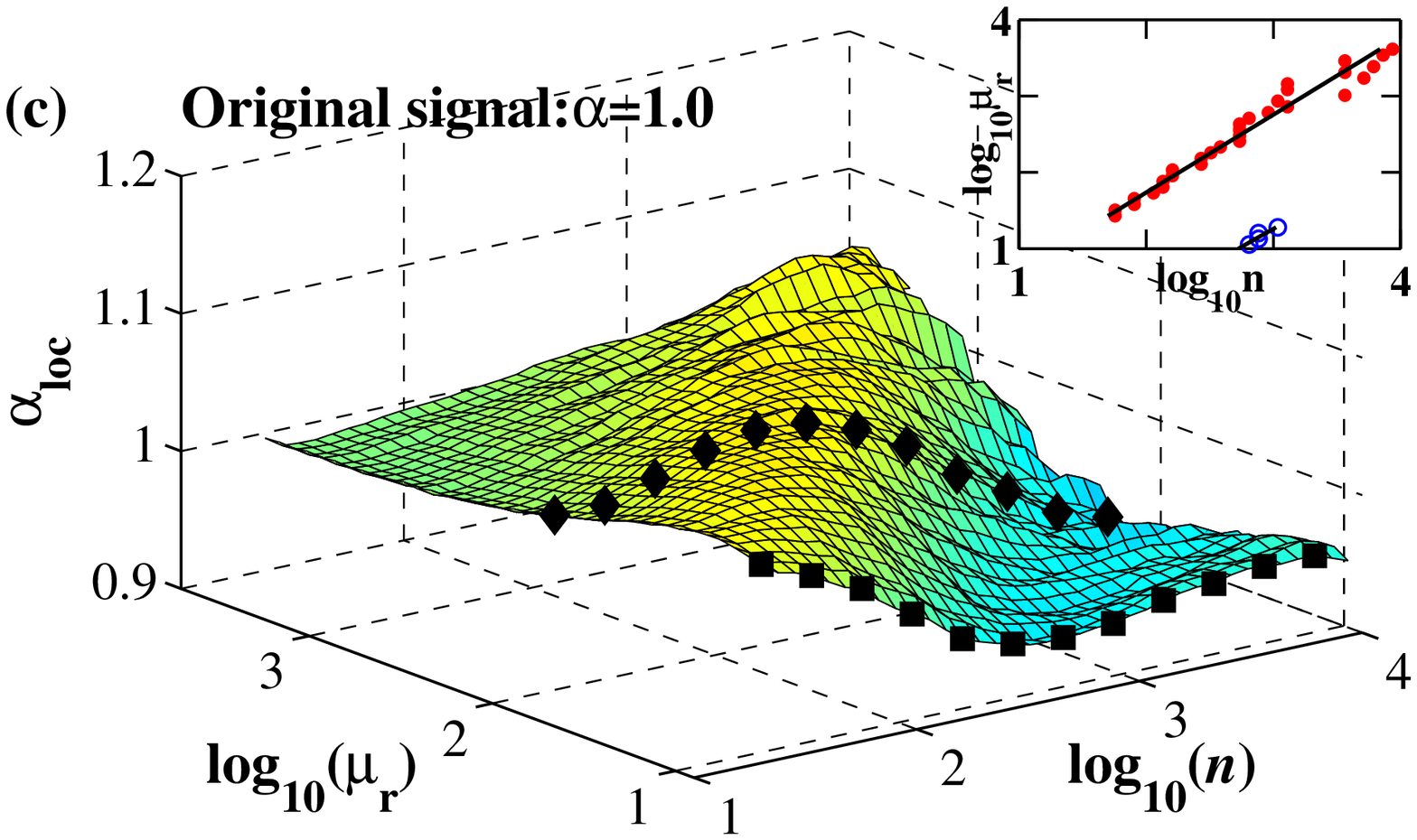}
\hspace{0.5cm}
\includegraphics[width=0.465\linewidth]{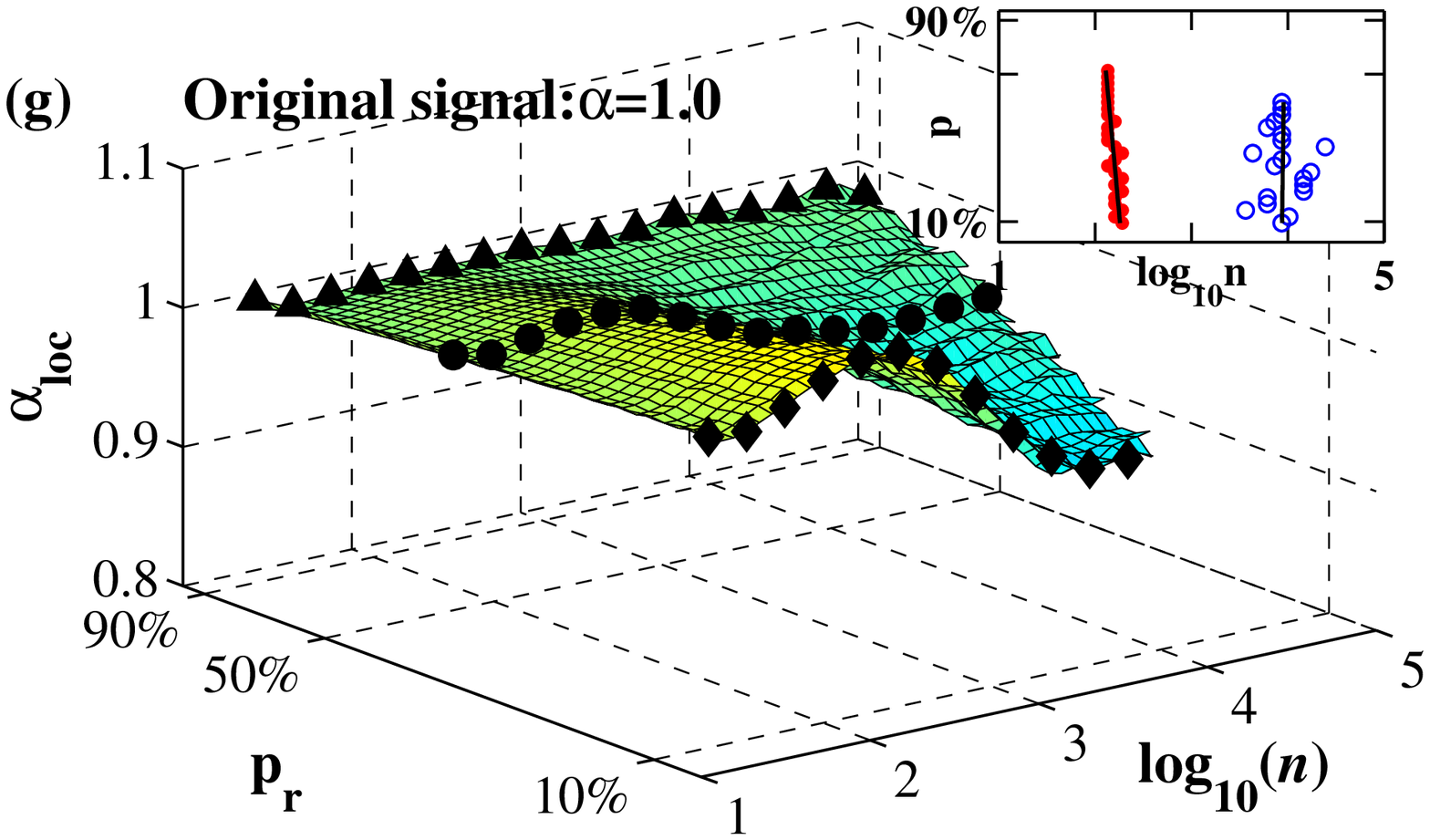}
}
\centering{
\includegraphics[width=0.465\linewidth]{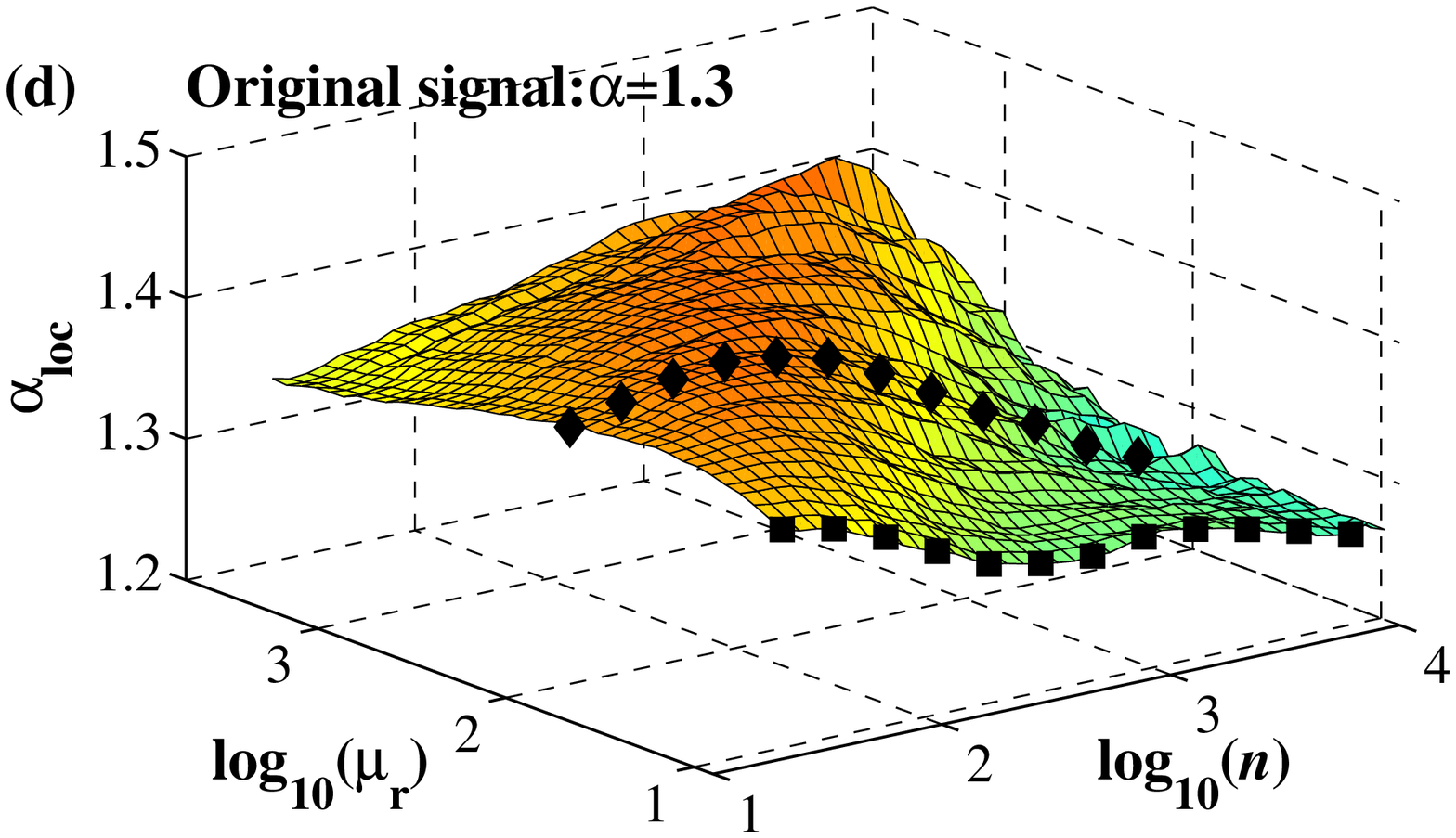}
\hspace{0.5cm}
\includegraphics[width=0.465\linewidth]{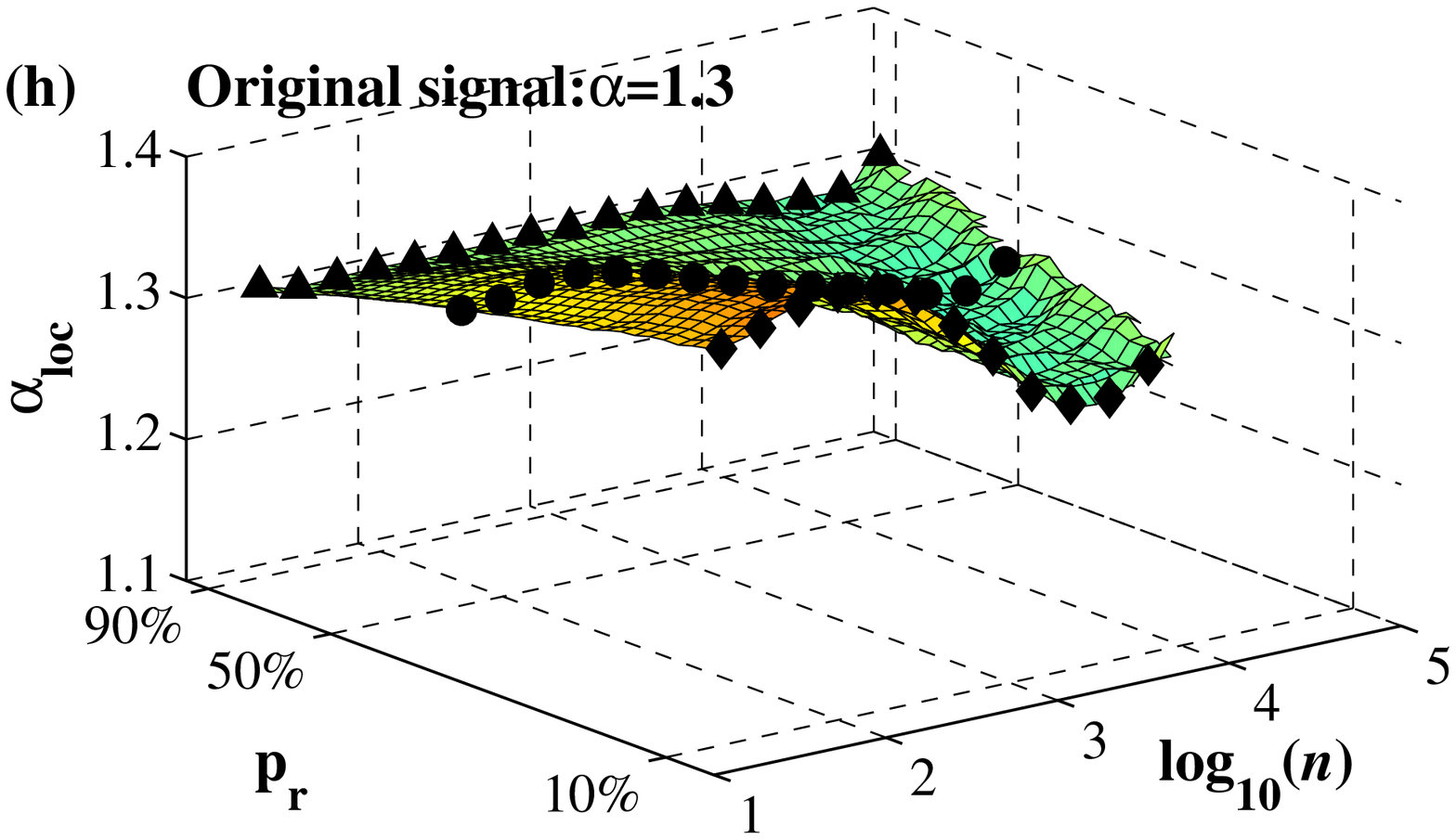}
}
\caption{(Color online) Effect of the average length $\mu_r$ of
  remaining data segments (a)-(d) and effect of the percentage $p_r$
  of remaining data (e)-(h) on the local scaling behavior in
  anti-correlated signals [(a), (e): $\alpha=0.3$] and positively
  correlated signals [(b), (f): $\alpha=0.7$; (c), (g): $\alpha=1.0$;
  (d), (h): $\alpha=1.3$]. For (a)-(d), $p_r=10\%$ of data are
  remained, and for (e)-(h), the average length of remaining segments
  $\mu_r=100$. In all the cases, the remaining segments are
  exponentially distributed, and the length of the original signals
  $N=2^{20}$.  The symbols in the inset figures in (c) and (g)
  indicate the positions where $\alpha_{loc}$ values reach a maximum
  (red closed circle) and a minimum (blue open circle), which show
  that the overestimated and underestimated regions are shifted to
  larger scales only with increasing $\mu_r$ and are not shifted with
  the percentage $p_r$ of remaining data changes. The local scaling
  curves highlighted by black symbols correspond to the curves shown
  in Fig.~\ref{fig-aloc-seg-exp-t10} (rectangle: $\mu_r=10$,
  $p_r=10\%$; diamond: $\mu_r=100$, $p_r=10\%$; circle: $\mu_r=100$,
  $p_r=35\%$; triangle: $\mu_r=100$; $p_r=90\%$).}
\label{fig-aloc3d-seg-exp-p10}
\end{figure*}

In Fig.~\ref{fig-aloc3d-seg-exp-p10}a-d, we show how the local scaling
behavior changes with the average length $\mu_r$ of remaining
segments. Similar to Fig.~\ref{fig-aloc3d-exp-p90}a-d where the
distribution of removed segments was specified, the variation of the
local scaling behavior of positively correlated signals also shows
overestimated regions at smaller scales followed by underestimated
regions at larger scales. Both regions are shifted to larger scales,
when the average length of remaining segments increases, forming a
power-law relationship between the shift in the local scaling behavior
and $\mu_r$ (Fig.~\ref{fig-aloc3d-seg-exp-p10}c). For anti-correlated
signals the local scaling behavior also shows a power-law relationship
between the scale at which $\alpha_{loc}$ reaches 0.5 and the average
length $\mu_r$.  Note that, according to Eq.~\ref{equ-mu}, the
$\alpha_{loc}$ curves from $\mu_r$=8 to 455 in
Fig.~\ref{fig-aloc3d-seg-exp-p10}a-d correspond to $\mu_l$=72 to 4095 in
Fig.~\ref{fig-aloc3d-exp-p90}a-d, thus the local scaling behavior in
these two regions are very similar.

With increasing percentage $p_r$ of remaining data, the deviation from
the original scaling behavior becomes smaller
(Fig.~\ref{fig-aloc3d-seg-exp-p10}e-h). However, for anti-correlated
signals, the scale at which $\alpha_{loc}$ reaches 0.5 does not depend
on the percentage of data loss (Fig.~\ref{fig-aloc3d-seg-exp-p10}e),
in contrast to Fig.~\ref{fig-aloc3d-exp-p90}e where removed data
segments were studied. Similarly, the overestimated regions in
positively correlated signals are also not shifted with the percentage
of data loss (Fig.~\ref{fig-aloc3d-seg-exp-p10}f-h, and compare to
Fig.~\ref{fig-aloc3d-exp-p90}f-h).

\begin{figure}
\centering{
\includegraphics[width=1.0\linewidth]{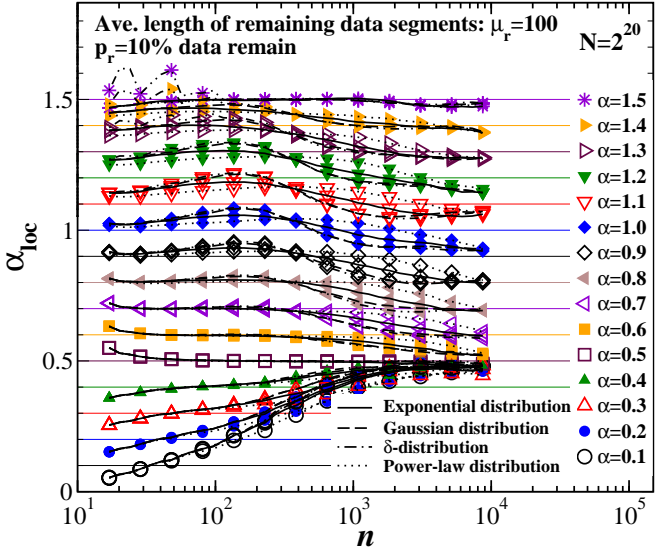}
}
\vspace{-0.5cm}
\caption{ Effect of different kinds of distributions of remaining data
  segments on the local scaling behavior. The Gaussian and
  $\delta$-distributions lead to identical and most pronounced
  deviations from the original scaling behavior for both
  anti-correlated and positively correlated signals. The power-law
  distribution leads to lowest deviations for anti-correlated signals
  and a smoother behavior of $\alpha_{loc}$ versus $\mu_r$, i.e., a
  less pronounced over- and underestimation of the original scaling
  behavior for positively correlated signals. Interestingly, for
  positively correlated signals, all four kinds of distributions yield
  the same local scaling exponent $\alpha_{loc}$ at certain scale
  ($n~\approx~300$ for $\mu_r=100$). Note that in case of the
  $\delta$-distribution, large jumps of $\alpha_{loc}$ values at small
  scales occur for original scaling exponents $\alpha=$1.3 to 1.5 (see
  text for more details).}
\label{fig-aloc-seg-vary-t100-p90}
\end{figure}

Next, we investigate how different kinds of distributions of remaining
data segments influence the local scaling behavior. As illustrate in
Fig.~\ref{fig-aloc-seg-vary-t100-p90}, the surrogate signals generated
by using Gaussian or $\delta$-distribution have almost identical local
scaling behavior and the most pronounced deviation from the original
local scaling behavior, and the power-law distribution shows the
smallest deviations. Note that, the local scaling exponent of
surrogate signals generated by a $\delta$-distribution jump to larger
$\alpha_{loc}$ values at certain small scales when the scaling
exponent of the original signal is 1.3, 1.4 and 1.5. This behavior is
caused by the discontinuities in the surrogate signal at the
transition points between remaining data segments, and since the
remaining segments are of fixed length, the transition points occur
periodically. If the segment length ($\mu=100$ in
Fig.~\ref{fig-aloc-seg-vary-t100-p90}) is an integral multiple of the
size of the fitting boxes (scales) in the DFA algorithm (e.g., $n=10,
20, 25, 50$), the transition points are not included in any fitting
box and thus the rms fluctuation functions of the surrogate signals
will be the same as in the original signals. In all other cases, the
discontinuities inside the fitting box will cause larger rms
fluctuation functions and lead to jumps in the local scaling exponents
at certain scales $n\le\mu_r$ as observed in
Fig.~\ref{fig-aloc-seg-vary-t100-p90}.

\begin{figure}
\centering{
\includegraphics[width=0.93\linewidth]{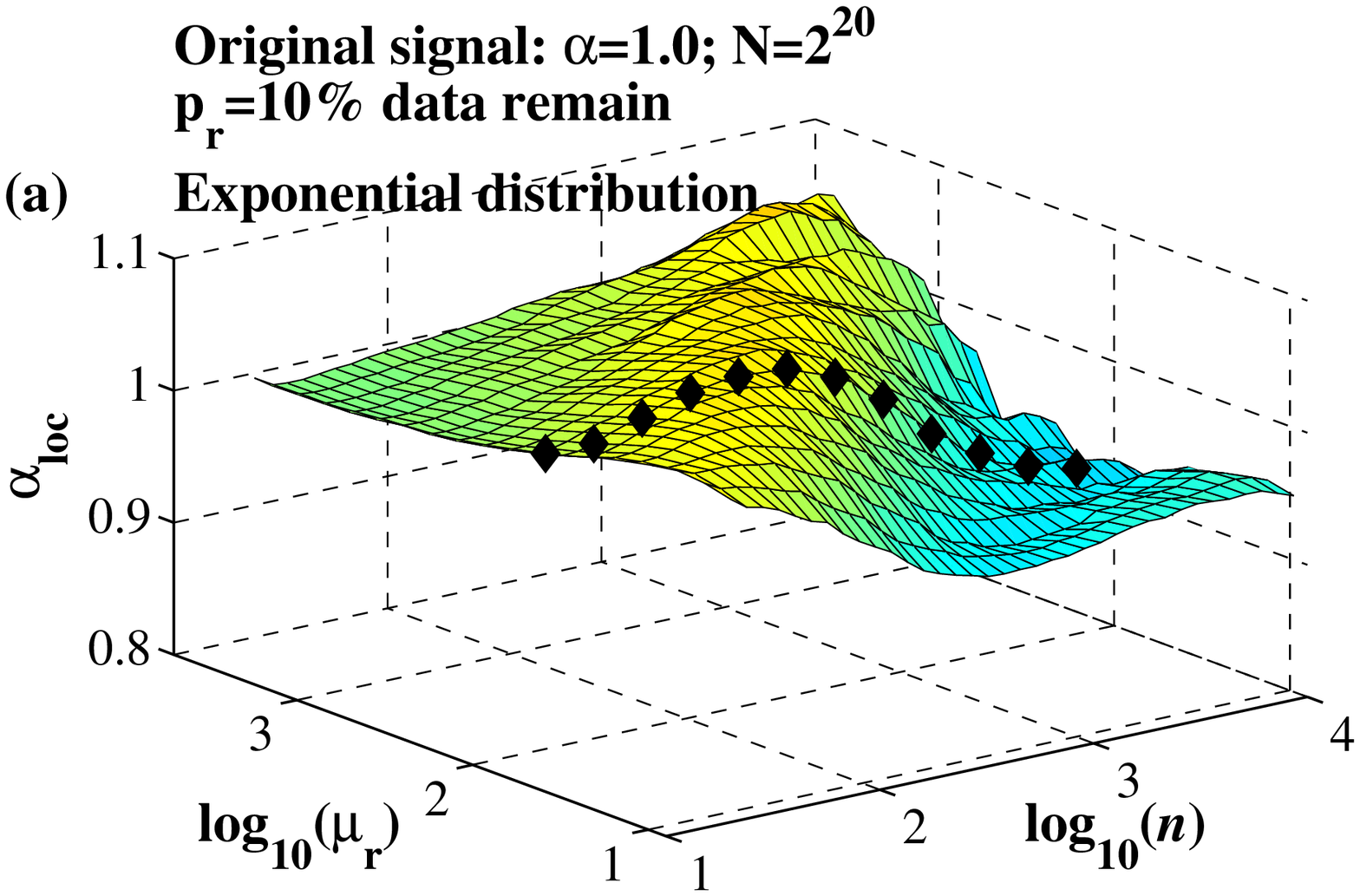}
}
\centering{
\includegraphics[width=0.93\linewidth]{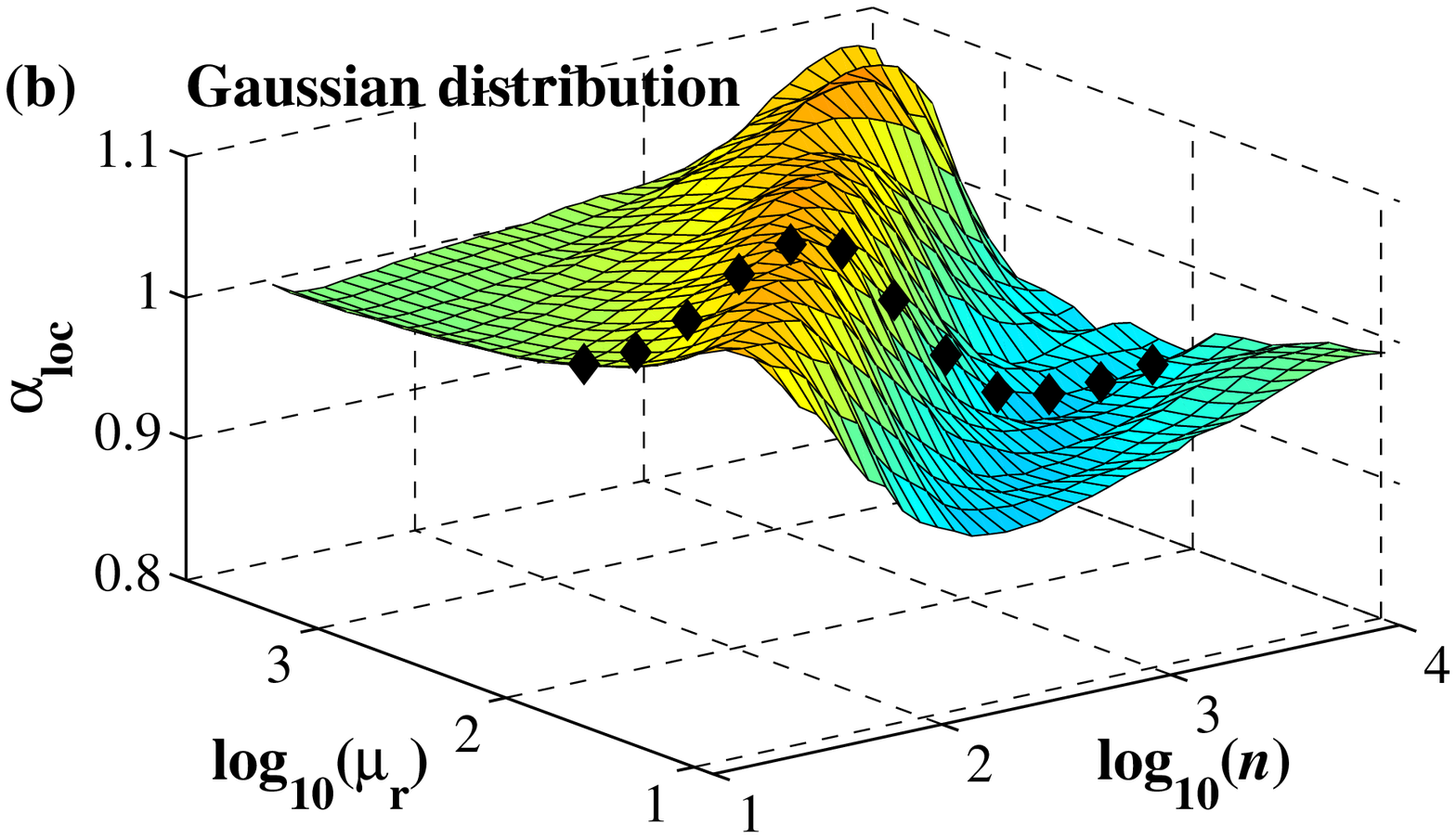}
}
\centering{
\includegraphics[width=0.93\linewidth]{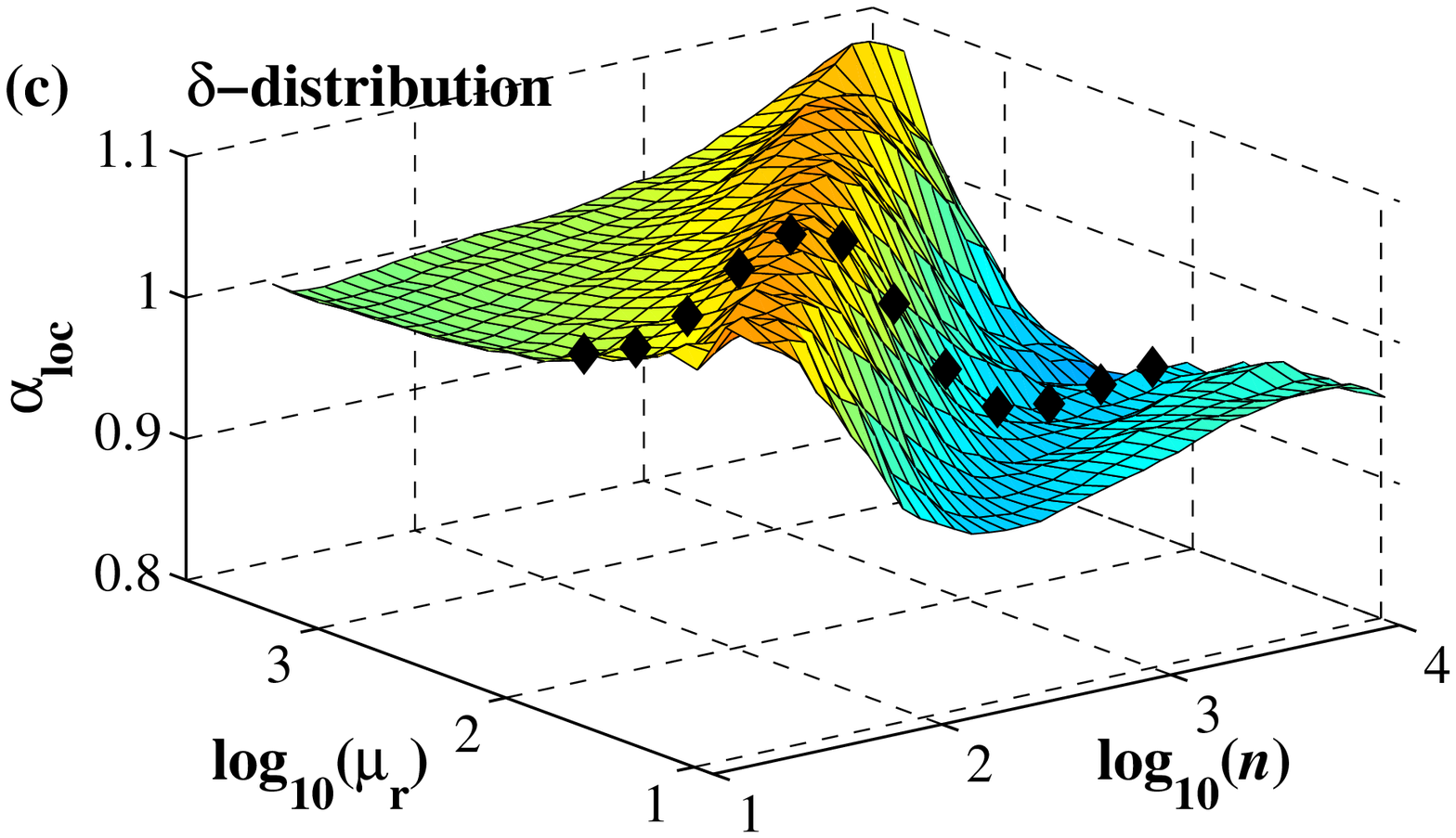}
}
\centering{
\includegraphics[width=0.93\linewidth]{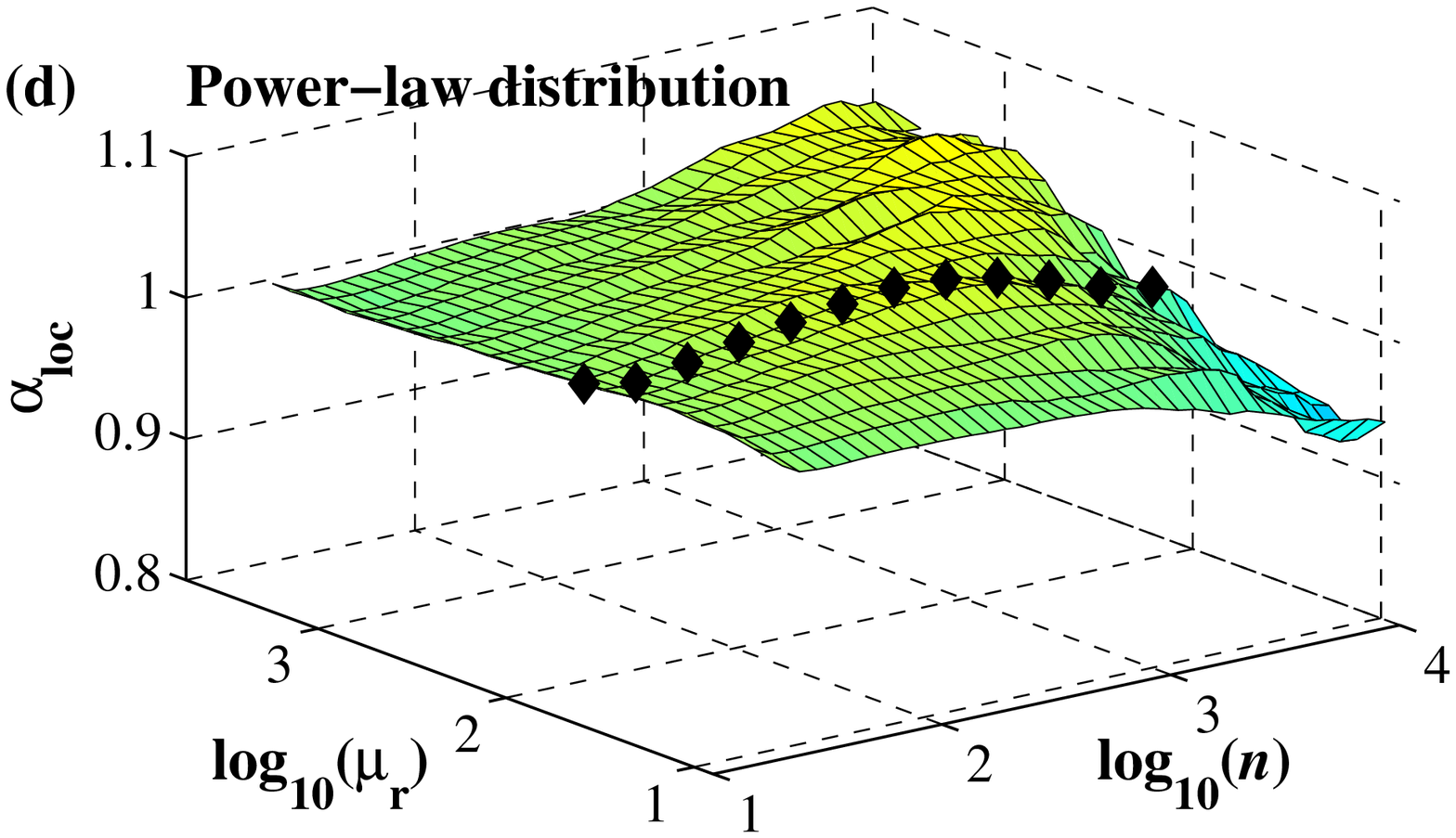}
}
\caption{Effect of different distributions and the average length
  $\mu_r$ of remaining data segments on the local scaling behavior.
  In all the cases, $p_r=10\%$ of data are remained, and the length of
  the original signals $N=2^{20}$. The Gaussian and
  $\delta$-distribution lead to very similar behavior with most
  pronounced $\alpha_{loc}$ deviations and a clear shift with
  $\mu_r$. In contrast, the power-law distribution shows no clear
  dependency of $\alpha_{loc}$ with $\mu_r$. The local scaling curves
  highlighted by black symbols correspond to the curves shown in
  Fig.~\ref{fig-aloc-seg-vary-t100-p90}.}
\label{fig-aloc3d-seg-vary-p10}
\end{figure}

In Fig.~\ref{fig-aloc3d-seg-vary-p10}, we show how the local scaling
curves of positively correlated signals change with the average length
$\mu_r$ of remaining segments, which follow an exponential
distribution (Fig.~\ref{fig-aloc3d-seg-vary-p10}a), a Gaussian
distribution (Fig.~\ref{fig-aloc3d-seg-vary-p10}b), a
$\delta$-distribution (Fig.~\ref{fig-aloc3d-seg-vary-p10}c), and a
power-law distribution (Fig.~\ref{fig-aloc3d-seg-vary-p10}d). The
Gaussian and $\delta$-distributions lead to a similar local scaling
behavior with regions of pronounced overestimation and underestimation
which are shifted to larger scales for increasing values of
$\mu_r$. This shift is also observed in the case of the exponential
distribution, however, the deviation from the original scaling
behavior (overestimation/underestimation) is less pronounced. In
contrast, the power-law distribution shows less variation of the local
scaling behavior and does not lead to such distinct regions of over-
and underestimated $\alpha_{loc}$ values. In addition, the local
scaling curves do not show a clear dependency (``shift'') with the
average length of remaining segments $\mu_r$. 

\begin{figure}
\centering{
\includegraphics[width=0.93\linewidth]{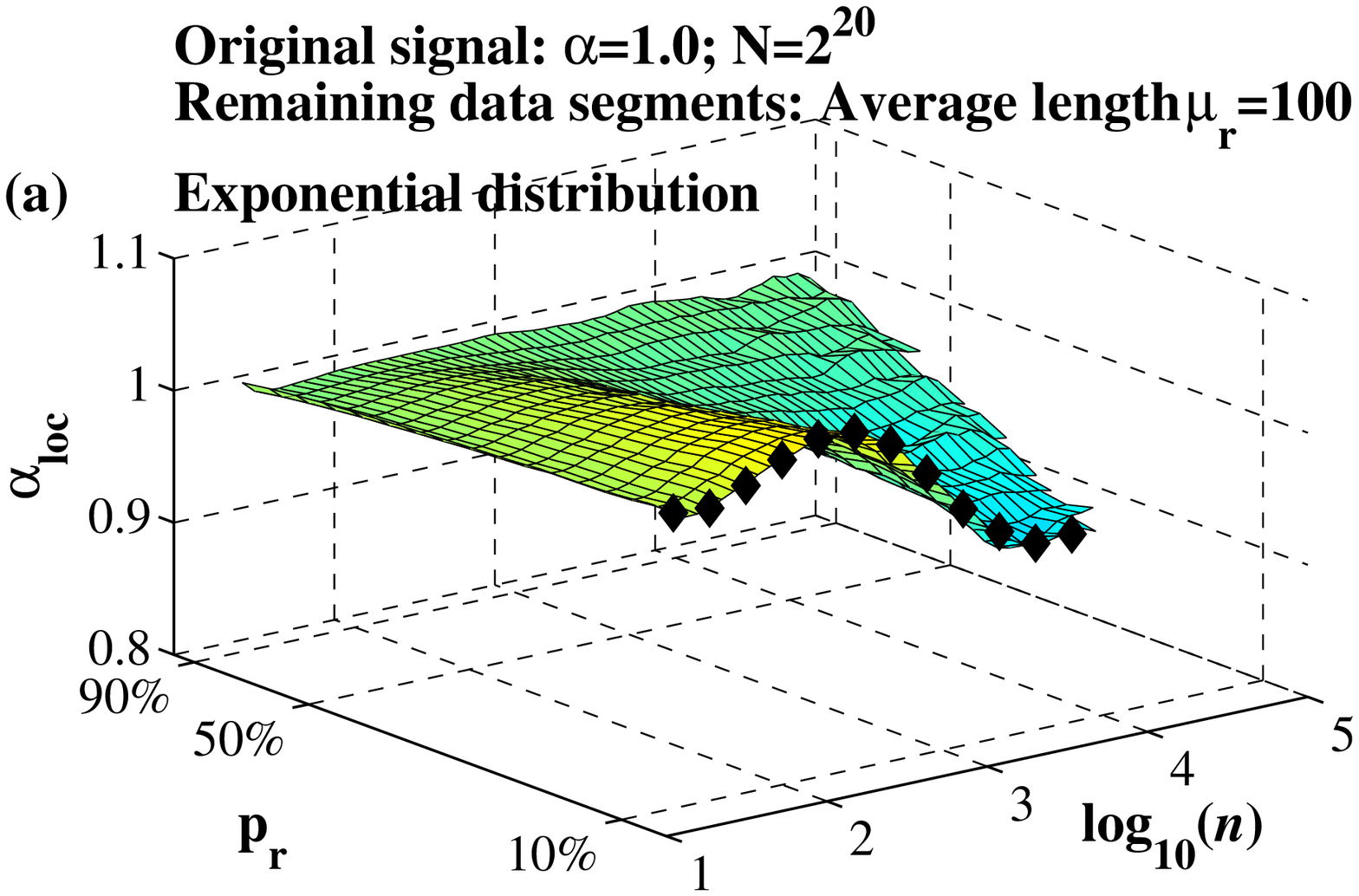}
}
\centering{
\includegraphics[width=0.93\linewidth]{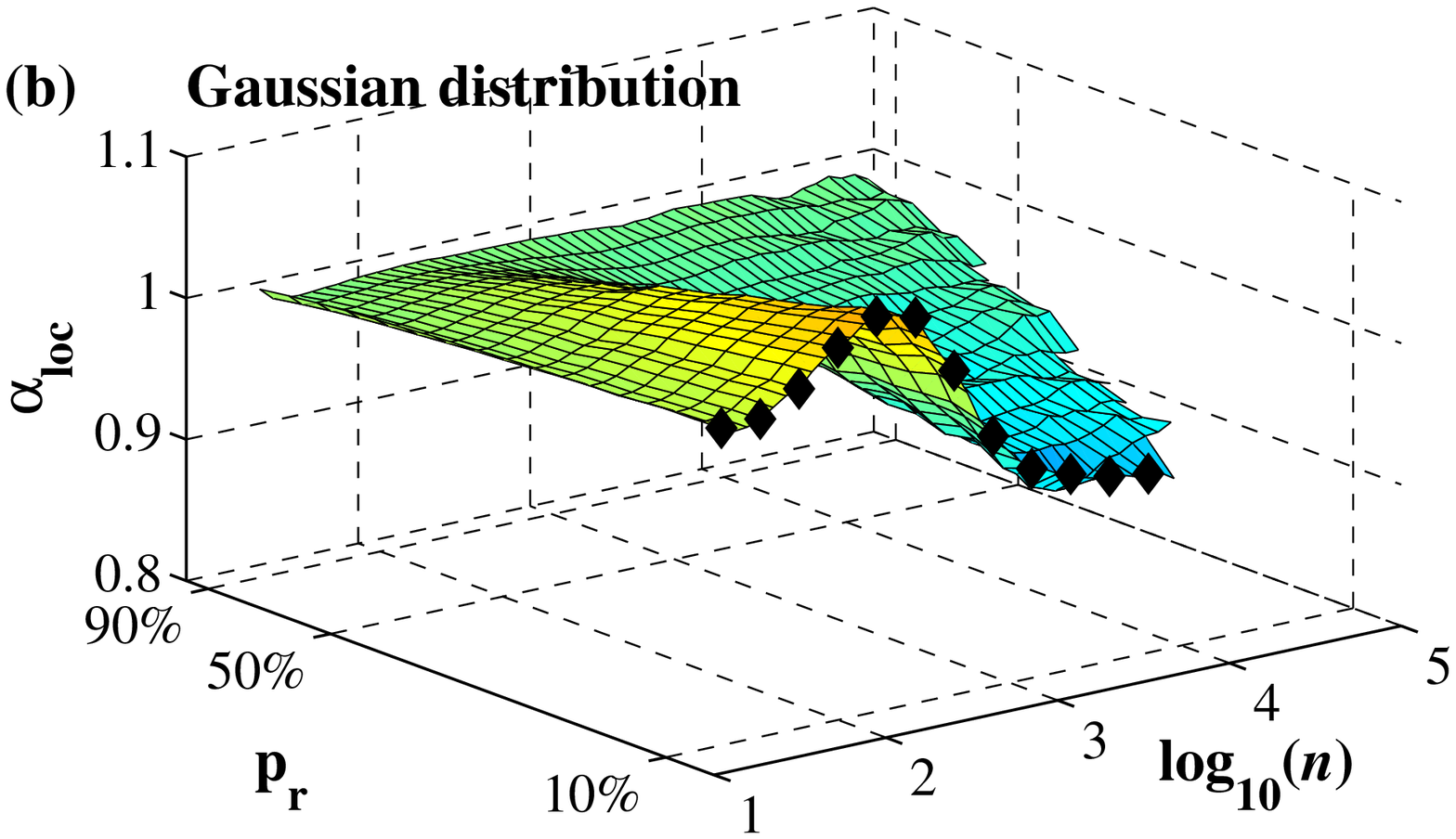}
}
\centering{
\includegraphics[width=0.93\linewidth]{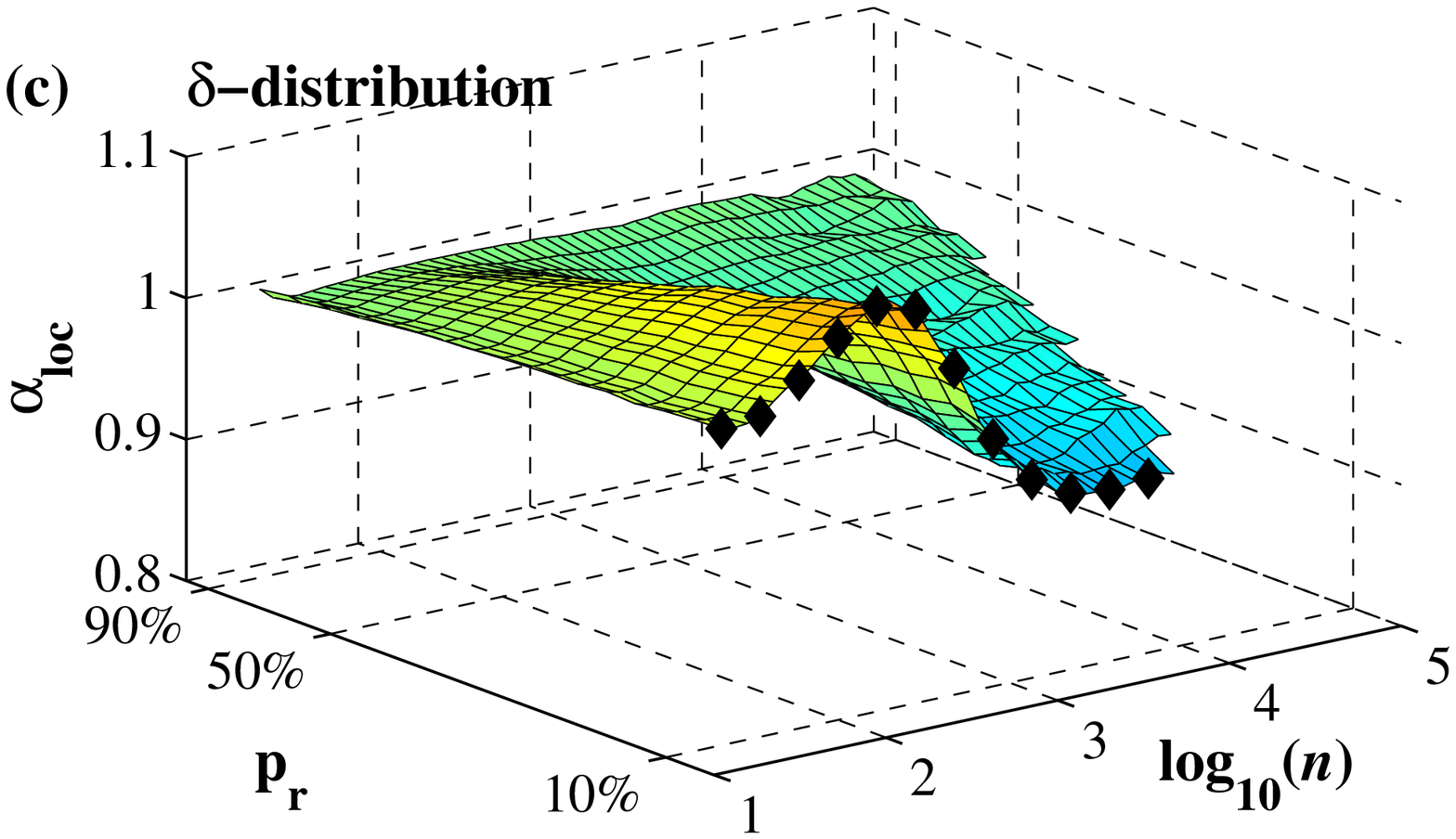}
}
\centering{
\includegraphics[width=0.93\linewidth]{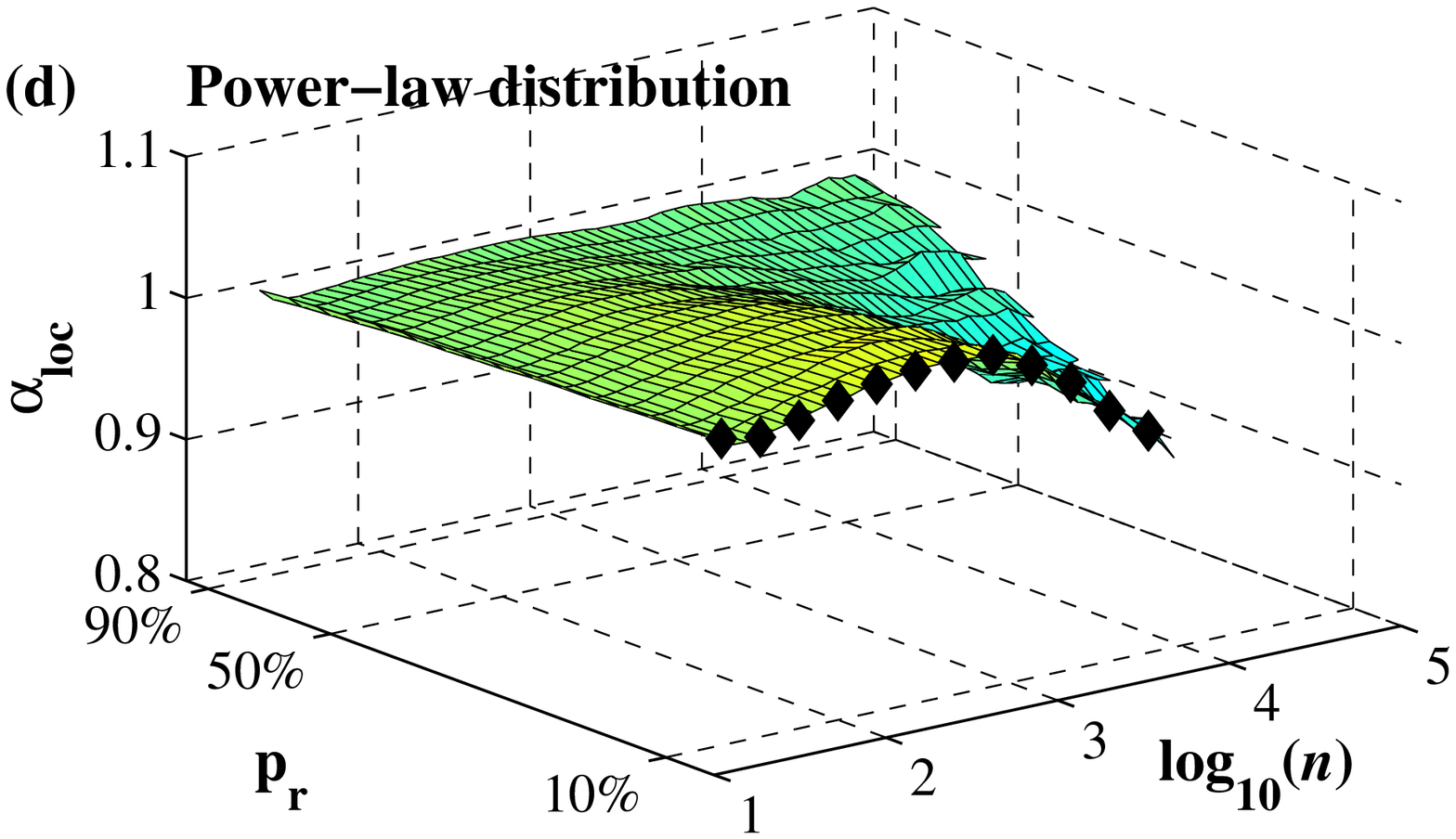}
}
\caption{Effect of different distributions of remaining data segments
  and the percentage $p_r$ of remaining data on the local scaling
  behavior. In all the cases, the average length of remaining segments
  $\mu_r=100$, and the length of the original signals $N=2^{20}$. The
  deviations from original scaling behavior are more pronounced for
  smaller percentages of remaining data. Note that the scale at which
  the most pronounced deviation is observed does not depend on
  $p_r$. The local scaling curves highlighted by black symbols
  correspond to the curves shown in
  Fig.~\ref{fig-aloc-seg-vary-t100-p90}.}
\label{fig-aloc3d-seg-vary-m100}
\end{figure}

The variation of the local scaling curves with the percentage $p_r$ of
remaining data for the four different distributions are presented in
Fig.~\ref{fig-aloc3d-seg-vary-m100}. Similar as shown in
Fig.~\ref{fig-aloc3d-seg-exp-p10}, the scale of most pronounced
deviation from the original scaling behavior is independent of the
percentage $p_r$ of remaining data.

\section{Summary and Conclusion}\label{secconclusion}

In this paper, we studied the effect of extreme data loss on the DFA
scaling behavior of long-range power-law correlated signals. In order
to simulate extreme data loss, often encountered in archaeological and
geological data, we developed a new segmentation approach to generate
correlated signals with randomly removed data segments. Using this
approach, surrogate signals can be generated for different percentages
of data loss, different average lengths and different distributions of
removed/remaining data segments. We compared the difference between
the DFA scaling behavior of original and surrogate signals by
systematically changing the percentage of data loss and the average
length of removed/remaining segments, and we also consider different
functional forms of the distributions of removed/remaining segment
lengths. We studied changes in the global scaling behavior as well as
in the local scaling exponents to reveal subtle deviations across
scales.

We find that anti-correlated signals are very sensitive to data
loss. Even if only 10\% of the data are removed, the scaling behavior
of the surrogate signals changes dramatically, showing uncorrelated
behavior at large scales. In contrast, positively correlated signals
are more robust to data loss and no significant changes in the {\it
  global} scaling behavior are observed for up to 90\% of data
loss. However, in case of extreme data loss, we find significant and
systematic deviations in the {\it local} scaling behavior which is
overestimated at small scales and underestimated at large
scales. Specifically, we find that for anti-correlated signals the
scale at which the local scaling exponent $\alpha_{loc}$ reaches 0.5
shifts to larger scales with increasing the average length $\mu_l$ (or
$\mu_r$) of the removed (or remaining) segments, following a power-law
relationship with $\mu_l$ (or $\mu_r$). For positively correlated
signals the regions of overestimation and underestimation of the local
scaling exponent are also shifted to larger scales following a
power-law with increasing $\mu_l$ (or $\mu_r$).

As expected, increasing the percentage of data loss leads to more
pronounced deviations in the local scaling behavior. However, the
variation of local scaling curves follows different rules if the
properties of either removed segments or remaining segments are
considered. When the average length $\mu_l$ of {\it removed} data
segments is kept constant, for increasing percentage $p_l$ of removed
data, the deviations of both anti-correlated and positively correlated
signals are shifted to smaller scales following a power-law with
$p_l$. When we focus on {\it remaining} data segments and keep their
average length $\mu_r$ constant, the deviations become more pronounced
with decreasing percentage $p_r$ of remaining data, however, the
deviations occur at the same scales.

This behavior can be explained by the relationship between removed and
remaining data. In case of a fixed percentage of removed or remaining
data, $\mu_l$ and $\mu_r$ are always directly proportional to each
other (Eq.~\ref{equ-mu}) and therefore the deviations (and the shift
of the most pronounced deviation) show a similar power-law relation
with $\mu_l$ and $\mu_r$, while fixing the average length of removed
or remaining segments leads to two different scenarios: (i) fixing
$\mu_l$ and changing $p_l$ leads to changes in $\mu_r$ proportional to
$p_l$; (ii) fixing $\mu_r$ and changing $p_r$ leads to changes in
$\mu_l$ proportional to $p_r$. Since the scale of the most pronounced
deviation from the original scaling behavior is shifted for scenario
(i) where $\mu_r$ is changing and $\mu_l$ is fixed, but not scenario
(ii) where $\mu_l$ is changing and $\mu_r$ is fixed, changes in
$\mu_l$ do not contribute to the observed shift. Thus, we suggest that
$\mu_r$ is the key parameter to determine the scales at which the
scaling behavior is mostly influenced, whereas the percentage of data
loss determines the extent of this influence.

Different distributions of the lengths of removed/remaining segments
affect the local scaling behavior differently. For Gaussian and
$\delta$-distributed segment lengths, deviations are most pronounced
and similar in extent, whereas power-law distributed segments show
smallest deviations and a very different overall behavior when compare
to exponential, Gaussian and $\delta$-distributed segments.

In conclusion, our study shows that it is important to consider not
only the percentage of data loss (removed/remaining data), but also
the average length of remaining segments to identify the scales at
which deviations from the original (``real'') DFA scaling behavior is
most pronounced. Therefore, when studying the scaling properties of
signals with extreme data loss, the DFA results should be carefully
interpreted to reveal the real scaling behavior.

\section*{Acknowledgments}
We thank the Brigham and Women's Hospital Biomedical Research
Institute Fund, the Spanish Junta de Andalucia (Grant
No. P06-FQM1858), Mitsubishi Chemical Corp., Japan, and the National
Natural Science Foundation of China (Grant No. 60501003) for support.

\end{document}